\documentclass[11pt]{article}
\pdfoutput=1 

\usepackage{jheppub} 
\usepackage[normalem]{ulem}
\usepackage[T1]{fontenc}
\usepackage{slashed,amsfonts,mathrsfs,bbm,bbding,pifont, fontawesome}
\usepackage{multicol}
\usepackage{multirow}
\usepackage{makecell}
\usepackage{booktabs,siunitx}
\usepackage{array,enumitem,booktabs,adjustbox}
\usepackage{units}
\usepackage{rotating}

\newcolumntype{M}[1]{>{\centering\arraybackslash}m{#1}}

\RequirePackage[colorlinks=true,urlcolor=blue,anchorcolor=blue,citecolor=blue,filecolor=blue,linkcolor=blue,menucolor=blue,linktocpage=true,pdfproducer=medialab,pagebackref=true]{hyperref}
 \hypersetup{
 colorlinks  = true,
 urlcolor    = blue,
 linkcolor   = blue,
 citecolor   = blue    
}

\graphicspath{{./Figures}}

\renewcommand{\dag}{\dagger}
\renewcommand{\to}{\rightarrow}

\renewcommand{\a}{\alpha}
\renewcommand{\b}{\beta}
\renewcommand{\d}{\delta}

\newcommand{\e}{\varepsilon}
\newcommand{\g}{\gamma}
\newcommand{\s}{\sigma}
\newcommand{\BBu}{B^{\mu\nu}}
\newcommand{\BBd}{B_{\mu\nu}}

\newcommand{\WWd}{W_{\mu\nu}}
\newcommand{\A}{\mathcal{A}}

\newcommand{\nn}{\nonumber}
\newcommand{\Dlr}{\overleftrightarrow{D_\mu}}
\newcommand{\Dlri}{\overleftrightarrow{D_\mu^i}}
\newcommand{\madgraph}{\textsc{Madgraph5\_aMC@NLO}}
\newcommand{\smeftsim}{\textsc{SMEFTsim}}
\newcommand{\Lag}{\mathscr{L}}

\title{ A sensitivity study of VBS and diboson WW to dimension-6 EFT operators at the LHC}

\author[d,e]{R. Bellan,}
\author[a,b]{G. Boldrini,}
\author[a]{D. Brambilla,}
\author[c]{I. Brivio,}
\author[a]{R. Brusa,}
\author[a,b]{F. Cetorelli,}
\author[a]{M. Chiusi,}
\author[d,e]{R. Covarelli,}
\author[h]{V. Del Tatto,}
\author[a,b]{P. Govoni,}
\author[b]{A. Massironi,}
\author[d]{L. Olivi,}
\author[d]{G. Ortona,}
\author[a]{G. Pizzati,}
\author[g]{A. Tarabini,}
\author[d,e]{A. Vagnerini,}
\author[a]{E. Vernazza,}
\author[a,f]{J. Xiao}

\affiliation[a]{Milano - Bicocca University, Piazza della Scienza 3, 20126 Milano, Italy}
\affiliation[b]{INFN Milano - Bicocca, Piazza della Scienza 3, 20126 Milano, Italy}
\affiliation[c]{Institut f\"ur Theoretische Physik, Universit\"at Heidelberg, \\Philosophenweg 16, 69120 Heidelberg, Germany}
\affiliation[d]{Torino University, Via Pietro Giuria 1, 10125 Torino, Italy}
\affiliation[e]{INFN Torino, Via Pietro Giuria 1, 10125 Torino, Italy}
\affiliation[f]{Peking University,  No.5 Yiheyuan Road Haidian District, Beijing, P.R.China 100871}
\affiliation[g]{Laboratoire Leprince-Ringuet, Inst. Polytech., Route de Saclay, 91128 Palaiseau, France}
\affiliation[h]{Trento University, Via Sommarive, 14, 38123 Trento, Italy}

\emailAdd{giacomo.boldrini@cern.ch}
\emailAdd{d.brambilla29@campus.unimib.it}
\emailAdd{brivio@thphys.uni-heidelberg.de}
\emailAdd{riccardo.bellan@unito.it}
\emailAdd{riccardo.brusa@cern.ch}
\emailAdd{f.cetorelli@campus.unimib.it}
\emailAdd{m.chiusi1@campus.unimib.it}
\emailAdd{e.vernazza1@campus.unimib.it}
\emailAdd{roberto.covarelli@unito.it}
\emailAdd{v.deltatto@campus.unimib.it}
\emailAdd{pietro.govoni@unimib.it}
\emailAdd{andrea.massironi@mib.infn.it}
\emailAdd{leonardo.olivi@edu.unito.it}
\emailAdd{ortona@to.infn.it}
\emailAdd{g.pizzati@campus.unimib.it}
\emailAdd{alessandro.tarabini@cern.ch}
\emailAdd{antonio.vagnerini@cern.ch}
\emailAdd{jie.xiao@cern.ch}

\abstract{
We present
a parton-level study of electro-weak production of vector-boson pairs at the Large Hadron Collider, establishing the sensitivity to a set of dimension-six operators
in the Standard Model Effective Field Theory (SMEFT).
Different final states are statistically combined, and we discuss how the orthogonality and interdependence of different analyses
must be considered to obtain the most stringent constraints.
The main novelties of our study are the inclusion of SMEFT effects in non-resonant diagrams and in irreducible QCD backgrounds,
and an exhaustive template analysis of optimal observables for each operator and process considered. We also assess for the first time the sensitivity of vector-boson-scattering searches in semileptonic final states. 

}

\begin{document} 
\maketitle

\section{Introduction}

While the CERN Large Hadron Collider (LHC) already collected an unprecedented amount of data, 
no significant deviations from the Standard Model (SM) predictions have been observed so far.
In the upcoming Runs, the LHC experiments will be able to look for physics Beyond the SM (BSM) in new ways, complementing the quest for patent signatures, such as new resonances, with searches for small deviations with respect to the SM expectations. This will be made possible by the increasing statistical precision granted by the available data sets,
together with the improved understanding of the detector behaviour
and the application of advanced data-analysis techniques 
 based on state-of-the-art machine learning algorithms.

The primary theoretical tool for BSM searches in precision measurements is the Standard Model Effective Field Theory (SMEFT)~\cite{Buchmuller:1985jz,Grzadkowski:2010es,Degrande:2012wf} (see~\cite{Brivio:2017vri} for a review). 
The SMEFT has been developed extensively in the past decade. State-of-the-art analyses are based on Higgs, diboson and top quark measurements 
and constrain about 20--30 SMEFT parameters simultaneously~\cite{daSilvaAlmeida:2018iqo,Biekoetter:2018ypq,Brivio:2019ius,Dawson:2020oco,Ellis:2020unq,Ethier:2021ydt,Ethier:2021bye,Almeida:2021asy}.
Going forward, the scope of these studies will be extended in several directions. For instance, there is interest in relaxing CP and flavour assumptions~\cite{Falkowski:2019hvp}, as well as in incorporating information from $B$-meson observables~\cite{Bissmann:2019gfc,Bissmann:2020mfi,Bruggisser:2021duo}.
The quality of the fit ingredients is also expected to improve in the future: on the theoretical side, more accurate SMEFT predictions are becoming available, that include 1-loop QCD or electro-weak (EW) corrections. On the experimental side, current measurements will improve substantially, achieving higher precision and allowing the extraction of differential information with higher resolution. Several processes will also become accessible at the LHC for the first time.

Vector Boson Scattering (VBS) ones are among the most interesting signatures in this perspective~\cite{Covarelli:2021gyz}.  The analysis of Run-II data by the ATLAS and CMS collaborations has recently led to the observation of VBS in the same-sign and opposite-sign WW~\cite{CMS:2017fhs,ATLAS:2019cbr,CMS-PAS-SMP-21-001}, WZ~\cite{ATLAS:2018mxa,CMS:2019uys} and ZZ~\cite{ATLAS:2020nlt,CMS:2020fqz} final states, and to strong evidence in the semileptonic WV final state~\cite{CMS-PAS-SMP-20-013}.
VBS processes with a photon, a heavy vector boson and two jets were recently observed as well~\cite{ATLAS:2019qhm,CMS:2020ioi,CMS:2020ypo}. This broad class of processes is mainly statistically limited at the LHC. Therefore the increased luminosity of its upcoming Runs will benefit their measurements in a particularly significant way,  allowing the extraction of new constraints on the EW sector, that are complementary to those from Higgs and diboson measurements.
Specifically, VBS processes provide tree-level sensitivity to effective operators inducing modifications of triple (TGC) and quartic gauge couplings (QGC), as well as of Higgs-gauge couplings away from the Higgs mass-shell, and even to contact interactions among four light quarks.

Mainly motivated by the QGC sensitivity, previous studies of EFT effects in VBS were often restricted to dimension-8 operators that generate genuine QGC corrections while leaving TGCs unaffected~\cite{PhysRevD.74.073005}, see e.g. Refs.~\cite{Perez:2018kav, Kalinowski:2018oxd, Brass:2018hfw,Bellan:2019xpr}.
The impact of dimension-6 operators was explored systematically only recently, driven by the interest in incorporating these measurements into global SMEFT analyses. 
Even though VBS measurements typically yield weaker constraints -- in absolute terms -- compared to EW precision observables, Higgs or inclusive diboson measurements, they can play a significant role in global analyses by constraining new directions in the parameter space and providing a link between the EW, Higgs and four-quarks sectors. 
Dedicated studies at dimension 6 were carried out for the ZZ~\cite{Gomez-Ambrosio:2018pnl} (see also \cite{Jager:2013iza}) and same-sign WW~\cite{Dedes:2020xmo} channels.
A first global analysis, that combines VBS and diboson constraints in several channels, was presented recently in~\cite{Ethier:2021ydt}, demonstrating that, while at present VBS has a visible but small impact in global EFT fits, in the future it will play a more significant role. 

In this work we examine the individual and combined sensitivity of five different VBS channels to a set of 14 dimension-6 operators in the Warsaw basis~\cite{Grzadkowski:2010es}. The primary goals of our study are to explore the complementarity among different VBS channels and to determine their relative projected sensitivities to different classes of effective operators. We also aim at assessing the relevance of SMEFT contributions that are quadratic in the Wilson coefficients, and of SMEFT corrections to the irreducible QCD backgrounds. 

We treat VBS processes as $2\to6$ scatterings, retaining EFT contributions to non-resonant diagrams. As these processes exhibit a great kinematic complexity, we perform a comparative study of several kinematic variables for each channel, to determine which observables are most sensitive to each dimension-6 operator.
The expected performance of the VBS channels is compared to that of diboson processes, which are traditionally the main playground for EFT analyses at dimension-6, due to their large production cross-sections and reasonable signal-over-background ratios. This is achieved by including inclusive WW diboson production~\cite{CMS:2020mxy,ATLAS:2019rob} in our analysis.

For simplicity, we work at parton-level and at leading-order (LO) in QCD, and we neglect all backgrounds other than diboson-plus-jets.\footnote{In most cases, diboson-plus-jets constitutes the dominant background. The main exception is $ZVjj$ production in the semileptonic channel, for which the dominant background is $Z+$jets. Further (reducible) background sources can come from particle mis-identification or similar detector effects, and cannot be reproduced in a parton-level analysis.} 
Due to these simplifications, that prevent us from reproducing a fully realistic analysis, we refrain from confronting our results with current measurements. Instead, we extract expected limits on the Wilson coefficients for integrated luminosities of $\unit[100]{fb^{-1}}$ or larger, that should be interpreted as indicative of the \emph{maximal} sensitivity of these processes to EFT effects.

The manuscript is organised as follows: the theory framework is defined in Section~\ref{sec.EFT}. The adopted strategy for the numerical and statistical analysis is presented in Sec.~\ref{sec:analysis}, while the specific properties of the individual processes are discussed in Sec.~\ref{sec:analyses}. We present our results in Sec.~\ref{sec:results} and in Sec.~\ref{sec:conclusions} we conclude. 

\section{Effective Field Theory framework}\label{sec.EFT}
We consider the SMEFT truncated at dimension-six:
\begin{equation}
 \Lag_{SMEFT} = \Lag_{SM} + \frac{1}{\Lambda^2}\sum_\alpha c_\alpha Q_\alpha + \mathcal{O}(\Lambda^{-4})
\end{equation} 
where $\Lambda$ is the cutoff scale of the EFT, $c_\alpha$ are the Wilson coefficients, and the index $\alpha$ runs over the labels of a complete and non-redundant set of operators that are invariant under the SM gauge symmetries. Here we work in the so-called Warsaw basis~\cite{Grzadkowski:2010es} and the Lagrangian is defined as in the \smeftsim\ package implementation~\cite{Brivio:2017btx,Brivio:2020onw}. In particular, we require the fermionic operators to be invariant (up to insertions of the Yukawa couplings) under a $U(3)^5$ flavour symmetry. We also assume CP conservation and take the Fermi constant ($G_F$) and the W and Z boson pole masses ($m_W, m_Z$) as input quantities for the EW sector~\cite{Brivio:2017bnu,lhceftwg_inputs}.

\begin{table}[t]
\centering
\begingroup
\renewcommand{\arraystretch}{1.5} 
\begin{tabular}{>{$}r<{$}@{  }>{$}l<{$}@{\hspace*{1cm}}>{$}r<{$}@{  }>{$}l<{$}}
\toprule
Q^{(1)}_{Hl}  &= (H^\dag i\Dlr H) (\bar l_p \g^\mu l_p)                                      
&
Q^{(3)}_{Hl}  &= (H^\dag i\Dlri H) (\bar l_p \s^i \g^\mu l_p) 
\\                           
Q^{(1)}_{Hq}  &= (H^\dag i\Dlr H) (\bar q_p \g^\mu q_p)  
&
Q^{(3)}_{Hq}  &= (H^\dag i\Dlri H) (\bar q_p \s^i \g^\mu q_p)
\\                
Q^{(1)}_{qq}  &= (\bar{q}_p\gamma_{\mu}q_p)(\bar q_r\gamma^{\mu} q_r)     
&    
Q^{(1,1)}_{qq}&= (\bar{q}_p\gamma_{\mu}q_r)(\bar q_r\gamma^{\mu} q_p)                  
\\       
Q^{(3)}_{qq}  &= (\bar{q}_p\gamma_{\mu}\s^i q_p)(\bar q_r\gamma^{\mu}\s^i q_r)       
& 
Q^{(3,1)}_{qq}&= (\bar{q}_p\gamma_{\mu}\s^i q_r)(\bar q_r\gamma^{\mu}\s^i q_p)                                  
\\      
Q_{HD}        &= (H^\dag D_\mu H) (H^\dag D^\mu H)                                    & 
Q_{H \square} &= (H^\dag H) \square (H^\dag H)                                            \\               
Q_{HWB}       &= (H^\dag \s^i H) \WWd^i\BBu                                             &  
Q_{HW}        &= (H^\dag H) \WWd^i W^{i\mu\nu}   
\\         
Q_{W}         &= \e^{ijk} W^{i\nu}_{\mu}W^{j\rho}_{\nu} W^{k\mu}_\rho                                  
&
Q^{(1)}_{ll}  &= (\bar l_{p}\g_\mu l_{r})(\bar l_{r} \g^\mu l_{p})                                                      \\
\bottomrule
\end{tabular}
\endgroup
\caption{
  The subset of Warsaw basis operators considered in this work. Repeated indices are understood to be summed over. $p,r$ are flavour indices, and a $U(3)^5$-invariant flavour structure is assumed. 
 \label{tab:d6operators} 
 }
\end{table}

The $SU(2)$ Higgs doublet is denoted by $H$, the gauge field strengths associated to the $SU(2)$ and $U(1)$ symmetries by $\WWd^i$ and $\BBd$ respectively, the left-handed lepton and quark doublets by~$l, q$ and the right-handed quark and charged-lepton fields by $u,d,e$.
The $SU(2)$ indices are indicated with $i,j,k$ and the Pauli matrices by $\sigma^i$. Flavour indices are indicated with $p,r$.
For further notational conventions we refer the reader to Ref.~\cite{Brivio:2020onw}.

For our numerical analysis, we consider only the subset of 14 operators given in Table~\ref{tab:d6operators}. It contains all the operators that enter via modifications of the EW input quantities ($Q_{Hl}^{(3)}, Q_{ll}^{(1)}, Q_{HD}, Q_{HWB}$), plus a set of dimension-6 operators that give significant contributions to all VBS processes, once the experimental selection cuts are applied.
These are mainly induced via modifications of $Vff$ ($Q_{Hl}^{(1)}, Q_{Hl}^{(3)}, Q_{Hq}^{(1)}, Q_{Hq}^{(3)}$), gauge ($Q_W$) and $HVV$ ($Q_{HD}, Q_{HW}, Q_{HWB}, Q_{H\square}$) couplings, or via four-quark contact terms ($Q_{qq}^{(1)},Q_{qq}^{(3)},Q_{qq}^{(1,1)},Q_{qq}^{(3,1)}$).
The set in Table~\ref{tab:d6operators} represents a convenient selection of operators for the purposes of this work, as it allows to examine all the mentioned categories of SMEFT effects and to explore the complementarity between VBS processes in constraining EFT parameters, while at the same time avoiding a very high-dimensional fit space.
Fermionic operators with right-handed fermions and bosonic operators such as $Q_{HB}$, that only enter a subset of VBS processes, as well as contact interactions between two quarks and two leptons, would need to be added for an exhaustive global analysis. These would amount to about 20 extra degrees of freedom, most of which are not expected to introduce new significant features to the fit. In this sense, we consider the set in Table~\ref{tab:d6operators} adequate for a study of the sensitivity of VBS processes to EFT effects and
we leave a more complete analysis for future work.

Working at order $\Lambda^{-2}$, a generic scattering amplitude has the form:
\begin{equation}
\label{2p1}
\A = \A_{\text{SM}} + \sum_\a\frac{c_{\a}}{\Lambda^{2}} \cdot \A_{Q_{\a}}\,,
\end{equation}
where $\A_{\text{SM}}$ is the SM amplitude and $\A_{Q_\a}$ is the total amplitude obtained with one insertion of the operator $Q_\a$. The latter scales linearly with $c_\a/\Lambda^2$, and this dependence has been made explicit in Eq.~\eqref{2p1}.
 
As a consequence, the expected number of events in a given phase-space region scales with the Wilson coefficients as:
\begin{align}
\label{eq:2p2}
    N&\,\propto\,|\A|^{2} = 
    |\A_{\text{SM}}|^2
    + 
    \sum_\a\frac{c_\a}{\Lambda^2} \cdot 2\Re(\A_{\text{SM}}\A_{Q_\a}^{\dagger})
    +   
    \sum_{\a,\beta}\frac{c_\a c_\beta}{\Lambda^4}  \cdot (\A_{Q_\a}\A_{Q_\beta}^\dagger)
    \\
    N&=N_{SM} + \sum_\a \left[\frac{c_\a}{\Lambda^2}N_\a^{int} +\frac{c_\a^2}{\Lambda^4}N_{\a}^{quad}\right]
    +
    \sum_{\a\neq\beta}\frac{c_\a c_\beta}{\Lambda^4}N_{\a,\b}^{mix}\,, 
\label{eq:2p2N}    
\end{align}
where in Eq.~\eqref{eq:2p2} both $\a$ and $\beta$ run over all relevant indices, and for $\a=\beta$ the last contribution reduces to $(c_\a^2/\Lambda^4)|\A_{Q_\a}|^2$.
The final result is therefore the sum of the SM prediction, a term that is linear in the Wilson coefficients and stems from the interference between SM and BSM amplitudes, and a pure BSM contribution, that is quadratic in the Wilson coefficients and is usually referred to as ``quadratic'' term. In Eq.~\eqref{eq:2p2N} the latter has been split into individual and mixed quadratic contributions for later convenience.

The result in Eq.~\eqref{eq:2p2N} applies both to integrated observables and bin-by-bin to differential ones.
For a given observable, the quantities $N_{SM}, N_\a^{int}, N_\a^{quad},N_{\a,\beta}^{mix}$ can be estimated numerically. For $n$ operators contributing, there are $n$ independent linear terms, $n$ individual squares and $n(n-1)/2$ mixed ones. Therefore, the whole set of EFT contributions can be determined by computing $N$ at a total of $n(n+3)/2$ (119 for $n=14$) independent points in the parameter space $\{c_\a\}$. In this work, these estimates are performed via Monte Carlo event simulations, as described in the next section. 

Note that insertions of a SMEFT operator into the scattering amplitude can take place both in vertices and in the propagators of unstable particles, via corrections to their masses and decay widths. 
As all masses are taken to be input quantities, only corrections to the decay widths of the W, Z and Higgs bosons are relevant to our calculation.   
Width corrections are polynomial in the Wilson coefficients and enter the denominator of the scattering amplitude. In order to recover the form in Eq.~\eqref{2p1} they are typically Taylor-expanded in~$c_\a$. The \emph{linear} corrections obtained in this way are well-defined~\cite{Helset:2017mlf} and
can be estimated directly with available Monte Carlo tools~\cite{Brivio:2020onw}. In practice, one has
\begin{equation}\label{eq.Nint_expand}
    N_\a^{int} = N_{\a,{\rm vert.}}^{int} + N_{\a,\d \Gamma_W}^{int} + N_{\a,\d \Gamma_Z}^{int} + N_{\a,\d \Gamma_H}^{int}\,.
\end{equation}
where  $N_{\a,{\rm vert.}}^{int}$ is the contribution from vertex insertions while the $N_{\a,\d\Gamma}^{int}$ terms come from insertions in the W, Z and H propagators respectively. In the Monte Carlo simulation, each of the terms in~\eqref{eq.Nint_expand} is estimated separately.

The propagator contributions to $N_\a^{quad}, N_{\a,\b}^{mix}$, on the other hand,
are ambiguous. The source of ambiguity is that the quadratic terms defined above provide partial predictions at order $\Lambda^{-4}$, as other contributions of the same order, such as insertions of two EFT operators in the same amplitude, are neglected. While for vertex corrections the distinction between double insertions and "standard quadratics" is well-defined, for propagator corrections it is less clear. For instance, the result obtained expanding first $\A$ to linear order in $c_\a$ and then taking a narrow-width approximation (NWA) on $|\A_{Q_\a}|^2$ is different from the one obtained taking first the NWA and then expanding the SMEFT corrections to $\Gamma$ in the $(\s_{\rm prod.} \Gamma_{\rm part.}/\Gamma)$ expression. Agreement between the two procedures is recovered if, in the first case, the square of the linear correction to the width $(\d\Gamma^{lin})^2$ is retained in the expanded $\A_{Q_\a}$. 

As, with currently available tools, it is not possible to produce complete simulations to quadratic order in the dimension-6 Wilson coefficients, we choose to omit all propagator corrections from our baseline analyses. Although admittedly not ideal, this solution ensures a consistent comparison between linear and quadratic results and avoids introducing arbitrariness in our predictions. Nevertheless, we do estimate the impact of propagator corrections on the purely linear analysis, as discussed in  Sec.~\ref{Individual_constraints}.

\section{Analysis procedure}
\label{sec:analysis}

We consider five distinct VBS channels, characterized by the presence of two well-separated hadronic jets in the final state, plus the inclusive production of opposite-sign W pairs, that serves as a representative of diboson processes. As indicated in Table~\ref{tab:mgcommands}, all VBS channels are defined as $2\to6$ scatterings, while 
inclusive WW diboson production is defined as a $2\to4$ process.
The channels considered are:
\begin{itemize}
    \item $W^\pm W^\pm$+2j: same-sign WW (SSWW) production with two same-sign leptons, missing transverse energy and two jets in the final state; 
    \item $W^+W^-$+2j: opposite-sign WW (OSWW) production with two opposite-sign leptons, missing transverse energy and two jets in the final state;
    \item $W^\pm Z$+2j: WZ production with three charged leptons, missing transverse energy and two jets in the final state; 
    \item $ZZ$+2j: ZZ production with four charged leptons and two jets in the final state;
    \item $ZV$+2j: ZV production (combining W$^\pm$Z and ZZ) with a semileptonic final state, where the Z boson decays into two charged leptons and the V one decays into two quarks, plus two tagging jets;
    \item $W^+W^-$: inclusive WW production with two opposite-sign leptons and missing transverse energy. As we work at LO in QCD, this channel has no additional jets.
\end{itemize}

All Feynman diagrams leading to the relevant final states are retained in our signal definitions, including those with non-resonant topologies. 
In the fully-leptonic channels, the final states include electrons and muons and, for simplicity, the two vector bosons are required to decay into lepton (or lepton-neutrino) pairs of different flavor. 
For all VBS channels except SSWW+2j, irreducible QCD backgrounds are present in the SM, that also receive SMEFT corrections. These contributions are taken into account in our study and their role in the SMEFT fit is discussed in Sec.~\ref{qcd_effect}.

\subsection{Event generation}\label{sec.event_gen}

For each channel we simulate events at parton- and tree-level with \madgraph\ (v.~2.6.5)~\cite{Alwall:2014hca},
interfaced to the \smeftsim\ package (v.~3)~\cite{Brivio:2017btx,Brivio:2020onw}.
The syntax used to generate each process is reported in Table~\ref{tab:mgcommands} for the SM case\footnote{Although higher order SM predictions are available for several of the channels considered~\cite{Jager:2006zc,Melia:2011dw,Jager:2011ms,Greiner:2012im,Jager:2013iza,Jager:2013mu,Rauch:2016upa,Biedermann:2016yds,Biedermann:2017bss,Ballestrero:2018anz,Jager:2018cyo,Denner:2019tmn,Chiesa:2019ulk,Denner:2020zit,Andersen:2021vnf,Denner:2021hsa,Denner:2022pwc}, we prefer to keep the SM terms at LO, in order to have all channels and all components of the predictions estimated consistently within a same generation framework. Given that we only provide expected limits, this approximation has only a limited impact on the fit results. }, together with a specification of how the syntax is modified for the extraction of the SMEFT contributions. 
The simulations are produced for $pp$ collisions at a centre-of-mass energy of 13~TeV, using the NNLO parton distribution functions provided by the NNPDF collaboration~\cite{Ball:2017nwa},
with $\alpha_S = 0.118$ and in the four-flavour scheme (LHAPDF identification code 325500).
The renormalization and factorization scale choices are determined by the Monte Carlo generator as the transverse mass of the $2\rightarrow 2$ scattering after $k_T$ clustering. 

\begin{table}[t]
  \centering
  \begingroup
  \renewcommand{\arraystretch}{1.6} 
  {\footnotesize
\begin{tabular}{>{\bf}c|>{\tt}l}
\toprule
WW & generate p p > e+ ve mu- vm\~{} SMHLOOP=0
\\ 
SSWW+2j EW & generate p p > e+ ve mu+ vm j j QCD=0 SMHLOOP=0           
\\              
OSWW+2j EW & generate p p > e+ ve mu- vm\~{} j j QCD=0 SMHLOOP=0              
\\           
WZ+2j EW & generate p p > e+ e- mu+ vm  j j QCD=0 SMHLOOP=0                        
\\  
ZZ+2j EW & generate p p > e+ e- mu+ mu- j j QCD=0 SMHLOOP=0       
\\
ZV+2j EW & generate p p > z w+(w-,z) j j QCD=0 SMHLOOP=0, z > l+ l-,  w+(w-,z) > j j
\\ 
OSWW+2j QCD & generate p p > e+ ve mu- vm\~{} j j QCD==2 SMHLOOP=0
\\ 
WZ+2j QCD & generate p p > e+ e- mu+ vm j j QCD==2 SMHLOOP=0
\\ 
ZZ+2j QCD & generate p p > e+ e- mu+ mu- j j QCD==2 SMHLOOP=0
\\ 
ZV+2j QCD & generate p p > z w+(w-,z) j j QCD==2 SMHLOOP=0, z > l+ l-, w+(w-,z) > j j
\\ 
\bottomrule
$\mathbf{N_{\a,{\rm vert}}^{int}}$ & + NPprop=0 NP=1 NP\char`^2==1
\\
$\mathbf{N_{\a,\d\Gamma}^{int}}$ & + NP=0 NPprop=2 NPprop\char`^2==2
\\
$\mathbf{N_{\a}^{quad},N_{\a\b}^{mix}}$ & + NPprop=0 NP=1 NP\char`^2==2
\\
\bottomrule
  \end{tabular}
  }
  \endgroup
  \caption{Upper block: \madgraph\ strings used to generate the SM components of the processes of interest.  Where relevant, the charged conjugate processes were included in the generation. The charged lepton shortcut {\tt l-} stands for electron or muon.
  Lower block: strings added to the SM ones in order to generate the three main classes of SMEFT corrections, following the \smeftsim\ conventions.
 \label{tab:mgcommands} 
  }
\end{table}

For each process, the SM, interference and quadratic EFT components (Eq.~\eqref{eq:2p2N}) are extracted using two alternative techniques:
\begin{enumerate}[label=\alph*),itemsep=0pt,topsep=.5em,leftmargin=15pt]
 \item performing an event simulation for each contribution. The SM signal ($N_{SM}$), the pure interference ($N_{\a,{\rm vert.}}^{int}$ and $N_{\a,\d\Gamma}^{int}$) and quadratic term ($N_\a^{quad}$) for individual operators, as well as the mixed quadratic contributions $N_{\a,\beta}^{mix}$ can all be generated directly exploiting the interaction-order syntax in \madgraph\ and \smeftsim.
 
 \item exploiting the reweighting method in \madgraph~\cite{Mattelaer_2016}. In this case the events are generated only once, for one point in parameter space, and are subsequently reweighted to match a different set of values of the Wilson coefficients. Because this procedure is based on the calculation of the matrix element at fixed phase-space points, the dependence of the weights on the Wilson coefficients is computed exactly. 
\end{enumerate}
While $N_\a^{int}$, $N_\a^{quad}$ can be determined directly, isolating the quantities $N_{\a,\b}^{mix}$ often requires algebraic combinations of results obtained at 2 or 3 different points. 
Besides being faster, the reweighting technique has the advantage of reducing significantly the statistical uncertainties in this operation. On the other hand, it can give unreliable results in regions of phase space that are poorly populated in the original event generation. For this reason, both techniques were used for estimating and cross-checking the signal dependence on the Wilson coefficients. 

Linear corrections stemming from SMEFT insertions in vertices and in the propagators of the W, Z and Higgs bosons were simulated separately, with the syntax indicated in Table~\ref{tab:mgcommands}. Note that propagator corrections have identical shapes for all operators contributing. For instance, for any observable:
\begin{align}
N_{c_{Hq}^{(3)},\d\Gamma_W}^{int} &=    -N_{c_{Hl}^{(3)},\d\Gamma_W}^{int} 
=\frac{4}{3}N_{c_{ll}^{(1)},\d\Gamma_W}^{int} 
\end{align}
and similarly for the other $N_{\a,\d\Gamma}^{int}$ terms. In this way, all $N_{\a,\d\Gamma}^{int}$ can be estimated with one simulation per process and heavy boson (W/Z/H).
For later convenience, we report here the numerical expressions of the linearized width corrections (retaining only the operators in Tab.~\ref{tab:d6operators}):
\begin{align}
\label{eq.dW_expressions}
\frac{\Lambda^2}{v^2}
\frac{\d\Gamma_W}{\Gamma_W^{SM}} &= \frac{4}{3} c_{Hq}^{(3)} - \frac{4}{3} c_{Hl}^{(3)} - c_{ll}^{(1)}\,,
\\
\frac{\Lambda^2}{v^2}\frac{\d\Gamma_Z}{\Gamma_Z^{SM}} &= 
  1.61 c_{Hq}^{(3)} 
- 1.37c_{Hl}^{(3)} 
+ c_{ll}^{(1)}
+ 0.47 c_{Hq}^{(1)} 
- 0.18 c_{Hl}^{(1)}
-0.07 c_{HD} 
+ 0.46 c_{HWB} 
\,,\nn
\\
\frac{\Lambda^2}{v^2}
\frac{\d\Gamma_H}{\Gamma_H^{SM}} &= 
  0.36 c_{Hq}^{(3)} 
- 2.62 c_{Hl}^{(3)} 
+ 1.40 c_{ll}^{(1)}
+ 1.83 c_{H\square} 
- 0.46 c_{HD} 
- 1.26 c_{HW} 
+ 1.23 c_{HWB} 
\,.\nn
\end{align}

Table~\ref{tab:3} summarizes which dimension-six operators contribute to each of the processes under investigation, when they are considered individually.
As mentioned above, all operators considered here contribute to all VBS channels, both at interference and quadratic level. However, some of them do not enter certain QCD-induced backgrounds or inclusive diboson (WW) production. Also, certain operators only contribute to non-resonant diagrams, i.e. where the lepton pairs in final state do not come from an intermediate gauge boson. These are indicated with brackets in the table.
A priori, contributions from the W, Z and Higgs propagators can introduce a dependence on operators that do not enter via vertices. However, we have verified that this is never the case for the processes and operators considered.

\begin{table}[htbp]
\renewcommand{\arraystretch}{1.5}\centering
\setlength{\tabcolsep}{10pt}
\begin{adjustbox}{angle=0,scale=.95}
{\small
\begin{tabular}{|l|{c} | *{5}{c} | *{4}{c} |} 
 \toprule
    & \rotatebox{90}{\textbf{WW}} & 
    \rotatebox{90}{\textbf{SSWW+2j EW}} & 
    \rotatebox{90}{\textbf{OSWW+2j EW}} &
    \rotatebox{90}{\textbf{WZ+2j EW}} & 
    \rotatebox{90}{\textbf{ZZ+2j EW}} & 
    \rotatebox{90}{\textbf{ZV+2j EW}} &
    \rotatebox{90}{\textbf{OSWW+2j QCD}} &
    \rotatebox{90}{\textbf{WZ+2j QCD}} &
    \rotatebox{90}{\textbf{ZZ+2j QCD}} &
    \rotatebox{90}{\textbf{ZV+2j QCD}} \\
    \midrule
    $Q_{HD}$ & \ding{51} & \ding{51} & \ding{51} & \ding{51} & \ding{51} & \ding{51} & \ding{51} & \ding{51} & \ding{51} & \ding{51} \\
    $Q_{H\square}$ & \ding{51} & \ding{51} & \ding{51} & \ding{51} & \ding{51} & \ding{51} & & & &\\ 
    $Q_{HW}$ & & \ding{51} & \ding{51} & \ding{51} & \ding{51} & \ding{51} & & & &\\
    $Q_{HWB}$ & \ding{51}& \ding{51} & \ding{51} & \ding{51} & \ding{51} & \ding{51} & \ding{51} & \ding{51} & \ding{51} & \ding{51}\\
    $Q_{W}$ & \ding{51} & \ding{51} & \ding{51} & \ding{51} & \ding{51} & \ding{51} & \ding{51} & \ding{51} &  & \ding{51}\\    
    $Q_{Hq}^{(1)}$ & \ding{51}& \ding{51} & \ding{51} & \ding{51} & \ding{51} & \ding{51} & \ding{51} & \ding{51}& \ding{51} & \ding{51}\\
    $Q_{Hq}^{(3)}$ & \ding{51} & \ding{51} & \ding{51} & \ding{51} & \ding{51} & \ding{51} & \ding{51} & \ding{51}& \ding{51} & \ding{51}\\
    $Q_{Hl}^{(1)}$ & (\ding{51}) & (\ding{51}) & \ding{51} & \ding{51} & \ding{51} & \ding{51} & \ding{51} & \ding{51}& \ding{51} & \ding{51}\\
    $Q_{Hl}^{(3)}$ & \ding{51} & \ding{51} & \ding{51} & \ding{51} & \ding{51}  & \ding{51} & \ding{51} & \ding{51}& \ding{51} & \ding{51}\\
    $Q_{ll}^{(1)}$ & \ding{51} & \ding{51} & \ding{51} & \ding{51} & \ding{51} & \ding{51} & \ding{51} & \ding{51} & \ding{51} & \ding{51}\\
    $Q_{qq}^{(3)}$ & & \ding{51} & \ding{51} & \ding{51} & \ding{51} & \ding{51} & & & &\\
    $Q_{qq}^{(3,1)}$ & & \ding{51} & \ding{51} & \ding{51} & \ding{51} & \ding{51} & & & & \\
    $Q_{qq}^{(1,1)}$ & & \ding{51} & \ding{51} & \ding{51} & \ding{51} & \ding{51} & & & & \\
    $Q_{qq}^{(1)}$ & & \ding{51} & \ding{51} & \ding{51} & \ding{51} & \ding{51} & & & & \\
   \bottomrule
\end{tabular}
}
\end{adjustbox}
\caption{Summary table for the dependence of the processes under investigation on the 14 benchmark EFT operators. Empty cells indicate that there are no diagrams for that operator for a given process. The brackets indicate that the operator only enters non-resonant diagrams.}
\label{tab:3}
\end{table}

\subsection{Event selection and analysis strategy}\label{sec:analysis}

For each process under study, the simulated events are first filtered  applying kinematic cuts that reproduce typical acceptance regions for lepton and jet reconstruction in LHC experiments, and isolate a phase space region where VBS production is strongly enhanced over the background. For this study, we refrain from applying any unitarisation procedure or clipping of the high-energy distribution bins. 
Following a procedure similar to Refs.~\cite{Higgs_Combination_note, CMS_combine_2015, ATLAS:2016neq},
we perform a template analysis based on the distributions of the SM, linear and quadratic signal components as a function of a set of final-state-dependent leptonic and partonic observables. 
The full list of variables considered is reported in Table~\ref{tab:variables}, along with the selection cuts applied to each final state, and they are defined as follows:
\begin{itemize}
\item $p_{T,l^i}$, the transverse momentum of the $i$-th charged lepton (sorted by their $p_T$, from largest to smallest) with respect to the beam axis, $\eta_{l^i}$ its pseudo-rapidity;

\item $m_{ll}$, the invariant mass of the two leading-$p_T$ leptons (independently of their charge and flavor);

\item MET, the missing transverse energy, calculated as the transverse component of the vector sum of all neutrino momenta;

\item $p_{T,j^i}$, the transverse momentum of the $i$-th outgoing parton (sorted by $p_T$), $\eta_{j^i}$, $\phi_{j^i}$ its pseudo-rapidity and azimuthal angle respectively;

\item $m_{jj}$, the invariant mass of the two highest-$p_T$ outgoing partons, $\Delta\eta_{jj}$ their pseudo-rapidity separation and $\Delta\phi_{jj}$ their angular separation in the azimuthal plane;

\item $\Delta R(l,j)$, the lepton-parton separation in the distance parameter $R$, defined as the sum in quadrature of $\phi$ and $\eta$ distances.

\end{itemize}
For the remaining observables we use an analogous notation, so their definitions can be easily inferred. The angular variables $\Phi_{planes}$, $\theta_{lW}, \theta_{lZ}$ and $\theta^*$ for the WZ+2j channel are defined in Sec.~\ref{sec.WZ} below. For the semileptonic ZV+2j channel, two dijet pairs are formed based on the highest invariant mass as described in Sec.~\ref{sec.ZV} and the observables associated to such systems are labelled as "mjj max" or "mjj nomax" with clear connection to the result of the tagging procedure.
For each process and EFT operator, only the variable that gives the strongest constraint on the Wilson coefficient (determined a posteriori) is employed in the construction of the likelihood used for the results extraction, as described below.

\begin{table}[!htbp]
\vspace*{-10mm}
\renewcommand{\arraystretch}{1.9}
\centering\small
\begin{tabular}{|>{$}p{23mm}<{$} >{$}p{58mm}<{$} >{$}p{32mm}<{$} >{\raggedleft}p{1.2cm}@{ }p{.5cm}|} 
\toprule
\textbf{Process} & \text{\bf Variables of interest} & \text{\bf Selections}  & \multicolumn{2}{c|}{\bf Expected events}  \\
\hline

\textbf{WW} \newline (pp \rightarrow 2l2\nu) 
& 
{\rm MET},\, m_{ll},\, p_{T,l^{i}},\, p_{T,ll},\, \eta_{l^{i}} 
& 
  {\rm MET} > \unit[30]{GeV} \newline 
     m_{ll} > \unit[60]{GeV} \newline 
p_{T,l^{1}} > \unit[25]{GeV} \newline 
p_{T,l^{2}} > \unit[20]{GeV} \newline 
|\eta_{l^i}|< 2.5 
&

(EW)~~& 30600
\\   
\hline

\textbf{SSWW+2j} \newline (pp \rightarrow 2l2\nu jj) 
\newline \newline 
\textbf{OSWW+2j} \newline (pp \rightarrow 2l2\nu jj) 
\newline \newline 
\textbf{WZ+2j} \newline (pp \rightarrow 2e\mu\nu jj) 
& 
{\rm MET},\, m_{jj},\, m_{ll},\, \phi_{j^{i}},\, p_{T,j^{i}} \newline 
p_{T,l^{i}},\, p_{T,ll},\, \Delta\eta_{jj},\, \Delta\phi_{jj},\, \eta_{j^{i}},\, \eta_{l^{i}}
\newline \newline \newline \newline \newline 
{\rm MET},\, m_{jj},\, m_{ll},\, \phi_{j^{i}},\, p_{T,j^{i}},\, p_{T,l^{i}} \newline
p_{T,ll},\, \Delta\eta_{jj},\, \Delta\phi_{jj},\, \eta_{j^{i}},\, \eta_{l^{i}},\, m_{3l} \newline 
p_{T,3l},\, m_{WZ},\, \delta\eta_{WZ},\, \delta\phi_{WZ},\, \Phi_{planes} \newline
\theta_{lW},\, \theta_{lZ},\, \theta^* 
& 
{\rm MET} > \unit[30]{GeV}    \newline 
     m_{jj} > \unit[500]{GeV} \newline
     m_{ll} > \unit[20]{GeV}  \newline 
p_{T,l^{1}} > \unit[25]{GeV}  \newline 
p_{T,l^{2}} > \unit[20]{GeV}  \newline 
p_{T,j^{i}} > \unit[30]{GeV}  \newline 
\Delta\eta_{jj} > 2.5         \newline 
 |\eta_{j^{i}}| < 5           \newline 
 |\eta_{l^i}|   < 2.5  
 & 
(EW)
\newline 
\newline 
\newline 
(EW)
\newline 
(QCD)
\newline 
\newline 
(EW)
\newline 
(QCD)~~~~
&
197
\newline 
\newline 
\newline 
493
\newline 
1967
\newline 
\newline 
35
\newline 
90
\\   
\hline

\textbf{ZZ+2j} \newline (pp \rightarrow 2e2\mu2j) 
&
m_{jj},\,  m_{l^{1}l^{2}},\, m_{ll},\, m_{4l},\, \phi_{j^{i}},\, p_{T,j^{i}},\,  p_{T,l^{i}},\,  \newline 
p_{T,l^{1}l^{2}},\, p_{T,l^{\pm}l^{\pm}},\, p_{T,l^{\pm}l^{\mp}},\, p_{T,Z},\, \Delta\phi_{jj},\, \newline \Delta\eta_{jj},\, \eta_{j^{i}},\, \eta_{l^{i}}      
& 
       m_{jj} > \unit[400]{GeV} \newline
  60 < m_{ll} < \unit[120]{GeV} \newline 
       m_{4l} > \unit[180]{GeV} \newline 
  p_{T,l^{1}} > \unit[20]{GeV}  \newline 
  p_{T,l^{2}} > \unit[10]{GeV}  \newline 
  p_{T,l^{i}} > \unit[5]{GeV}   \newline 
p_{T,j^{1,2}} > \unit[30]{GeV}  \newline 
      \Delta\eta_{jj} > 2.4     \newline 
       |\eta_{j^{i}}| < 4.7     \newline 
       |\eta_{l^i}|   < 2.5     \newline 
\Delta R(l^{i},j^{k}) > 0.4
& 

(EW)
\newline 
(QCD)~~~~
&
11
\newline 
176
\\   
\hline

\textbf{ZV+2j} \newline (pp \rightarrow 2ljjjj) 
& 
m_{jj}^{max},\, m_{jj}^{nomax},\, m_{ll},\, \phi_{j^{i}},\, p_{T,j^{i}},\, p_{T,l^{i}} \newline
p_{T,ll},\, \Delta\eta_{jj},\,  \Delta\eta_{jj}^{nomax},\, \Delta\phi_{jj}^{max}\newline \Delta\phi_{jj}^{nomax},\, \eta_{j^{i}} \eta_{l^{i}} 
&
        m_{jj} > \unit[1500]{GeV} \newline 
60 < m_{jj}^{V} < \unit[110]{GeV} \newline 
    85 < m_{ll} < \unit[95]{GeV}  \newline 
    p_{T,l^{1}} > \unit[25]{GeV}  \newline 
    p_{T,l^{2}} > \unit[20]{GeV}  \newline 
    p_{T,j^{i}} > \unit[100]{GeV} \newline 
\Delta\eta_{jj} > 3.5             \newline 
 |\eta_{j^{i}}| < 5               \newline 
 |\eta_{l^i}|   < 2.5 
 & 
(EW)
 \newline
(QCD)~~~~
 &
142
 \newline
50
\\   
\bottomrule
\end{tabular}
\caption{Summary table for processes, variables and selections considered in this work. The second column lists the observables examined, while the third one summarises the parton level phase space definition used in this analysis. The last column reports the expected SM event yields of the EW and QCD-induced processes after the analysis selections, for an integrated luminosity of $\unit[100]{fb^{-1}}$. The charged lepton shortcut $l$ stands for electron or muon.
}
\label{tab:variables}
\end{table}

\subsection{Likelihood construction}\label{sec:likelihood}

The likelihood function $\mathcal{L}$ for the data to match the EFT model is built
using Eq.~\eqref{eq:2p2N} to determine the mean
of the Poisson statistics ($N_k(\textbf{c})$) describing the expected number of events
surviving the data analysis selections,
either in total or in the $k$-th distribution bin:
\begin{equation}
\mathcal{L}(\textbf{c}) = \prod_k \frac{\left( N_k(\textbf{c})\right)^{n_k}}{n_k!}e^{-N_k(\textbf{c})}\,,
\end{equation}
where {\bf c} is the vector of free Wilson coefficients in the fit, $N_k$ the expected number of events defined as in Eq.~\eqref{eq:2p2N} and $n_k\equiv N_k(\textbf{0})$ is the expected number of events in the SM.  Where relevant, the QCD SM and EFT components are accounted for in the expression for $N_k$ and summed to their EW counterparts. 
The number of expected events is always normalised to an integrated luminosity of 100~fb$^{-1}$, unless otherwise stated. 
No systematic uncertainty is taken into account, except for a correlated 2\%
variation between all yields, samples, and bins, as a proxy of a typical
LHC luminosity uncertainty.
While actual data analyses will be more severely affected by systematics and theory uncertainties,
we refrain from providing estimates for the latter, as they can vary significantly depending on the channel and on the analysis details. 
This choice is meant to avoid introducing this arbitrariness into the sensitivity estimates, and implies that our results should be interpreted as optimistic estimates of the sensitivity of VBS processes to EFT effects.

For each process considered, only one of the kinematic distributions listed in Table~\ref{tab:variables} is employed at each time to construct the likelihood.
For one-dimensional fits, we choose the variable that gives the strongest limit at  $68\%$~confidence level (c.l.) on the Wilson coefficient of interest. Note that, due to this procedure,  constraints on different Wilson coefficients are generally derived with different, optimized likelihoods. 
The ranking procedure is applied separately for individual fits where the quadratic terms in the signal parameterization are retained or discarded. It is also repeated for each two-dimensional fit, retaining the observable that gives the smallest area inside the $68\%$~c.l. contour in the bi-dimensional plane. 
For the 14-dimensional fit, the profiled constraint on each Wilson coefficient is derived from a likelihood that implements, for each process, the same observable as in the corresponding 1D case.
The full lists of optimal variables employed for each fit are reported in Tables~\ref{tab:sensitivity_ranking}--\ref{tab.sensitivity_2d_inWW} in Appendix~\ref{app:ranking}.

The expected sensitivity to the Wilson coefficients is estimated based on the likelihood profile:
for single Wilson coefficient scans, 
the 68\% and 95\%~c.l. intervals in the coefficient estimates 
are determined by requiring $-2\Delta \log \mathcal{L} < 1$
and $-2\Delta \log \mathcal{L} < 3.84$ respectively,
being $\Delta \log \mathcal{L}$ the variation of the likelihood logarithm with respect to its maximum value.
For bi-dimensional scans, the intervals are instead $-2\Delta\log\mathcal{L} < 2.30$, $-2\Delta\log\mathcal{L} < 5.99$~\cite{pdg}.

\section{Processes considered}\label{sec:analyses}

\begin{figure}[tbp]
  \centering 
    \includegraphics[width=.49\textwidth]{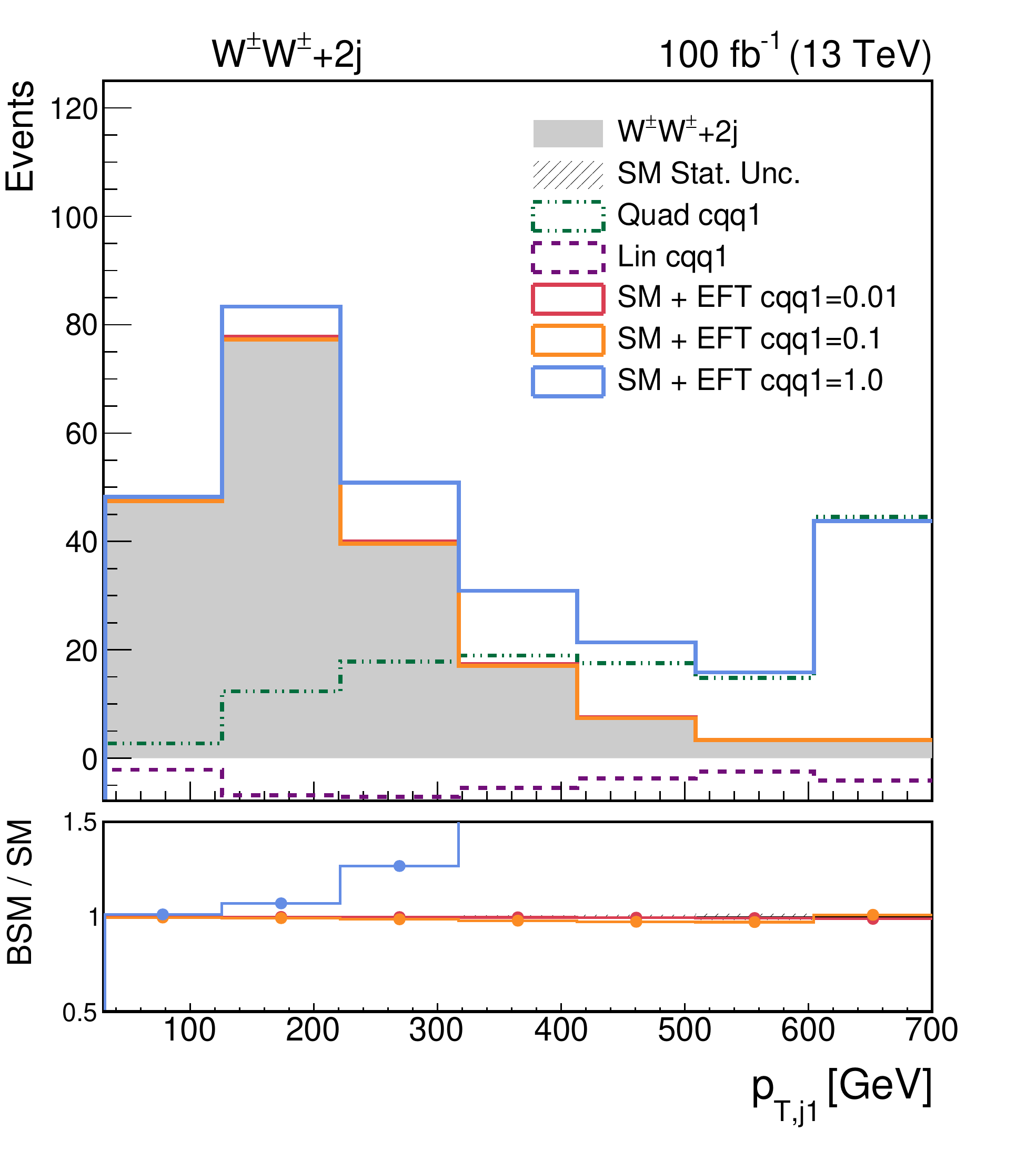}
    \includegraphics[width=.49\textwidth]{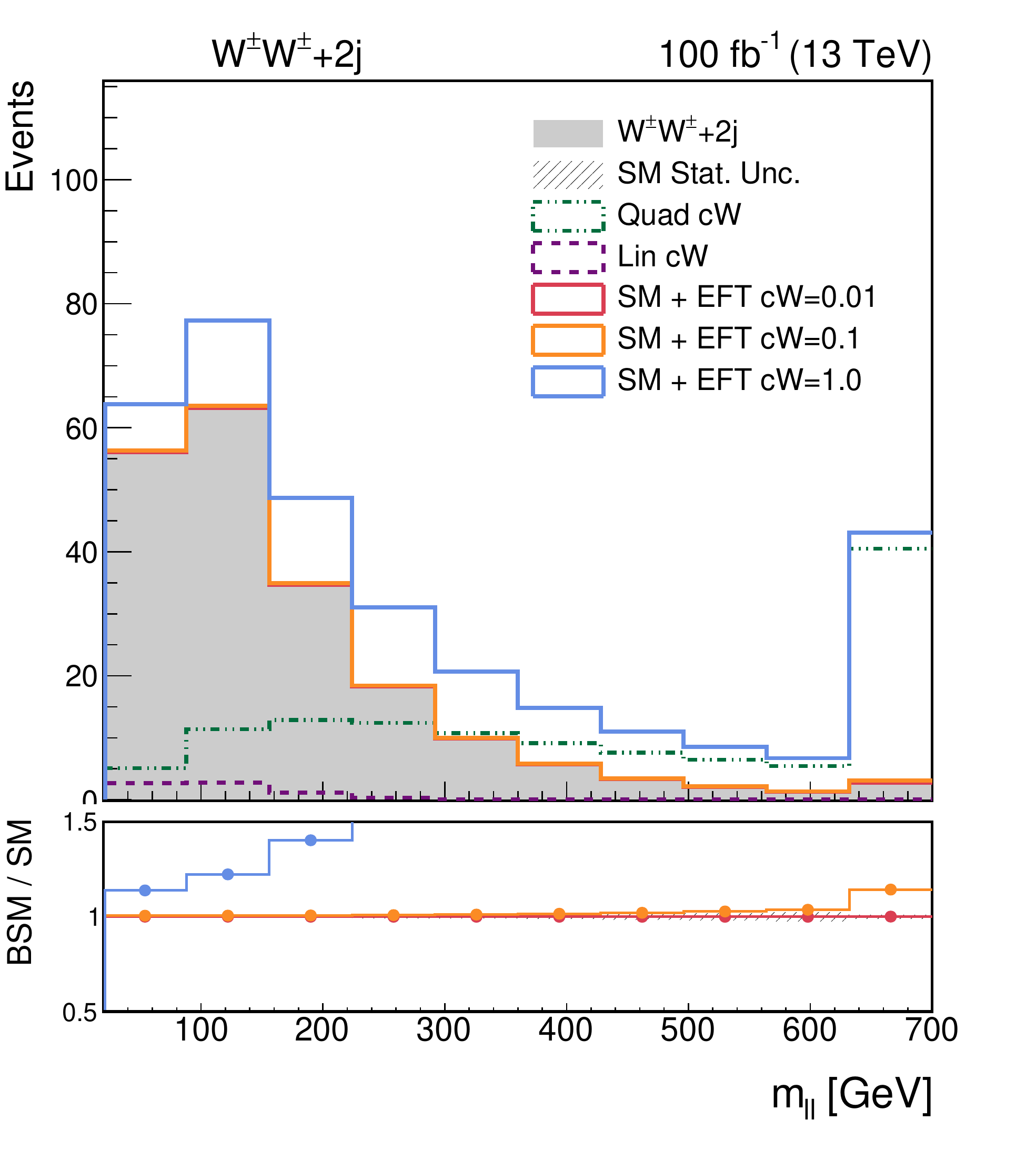}
  \caption{Impact of two Wilson coefficients on representative kinematic distributions in the SSWW+2j process.
  Solid lines show the total prediction for one Wilson coefficient at a time, with $c_\a/\Lambda^2 = 0.01$ (red), $0.1$ (orange) or $\unit[1]{TeV^{-2}}$ (blue).
  The pure interference (quadratic) EFT component, normalized to $c_\a/\Lambda^2=\unit[1]{TeV^{-2}}$, is indicated with a purple (green) dashed line. The SM prediction is shown in solid grey.
  The last bin comprises all the overflow events.\label{fig:Distributions_maintext} }  
\end{figure}

In this section we discuss the main properties of the processes considered and their role in the SMEFT analysis.  
The number of simulated SM events passing the selections, reported in Table~\ref{tab:variables}, provides an indication of the signal cross sections and of the typical signal-to-background ratios. An example of the impact of dimension-six operators on the kinematic distributions is shown for the SSWW+2j channel (see Sec.~\ref{sec.ssww}) in Fig.~\ref{fig:Distributions_maintext}. Further representative kinematic distributions are provided for all channels in Appendix~\ref{app:distributions}.

\subsection{Inclusive \texorpdfstring{$W^+W^-$}{WW}}

The inclusive W$^{+}$W$^{-}$ production is well-studied experimentally by the ATLAS and CMS Collaborations, see e.g.~\cite{ATLAS:2019rob,CMS:2020mxy} for the most recent results.  
The large cross-section of this process makes it an ideal environment for the study of new phenomena at high energies. Indeed, integrated and differential measurements of the WW production are included in most global analyses of dimension-six EFT operators, see Refs.~\cite{Biekoetter:2018ypq,daSilvaAlmeida:2018iqo,Ellis:2020unq,Ethier:2021ydt,Ethier:2021bye,Almeida:2021asy} for recent examples. 
This process is also a crucial component of the first combined EFT interpretations of EW/Higgs measurements that were recently published by the ATLAS Collaboration~\cite{ATL-PHYS-PUB-2021-010,ATL-PHYS-PUB-2021-022} and pave the way for larger global EFT fits within the LHC experiments. 
In our analysis we neglect all backgrounds to this process as,  in the phase space defined in Table~\ref{tab:variables}, their yield is approximately a half compared to that of the signal.

EFT corrections to the inclusive WW production at the LHC are very well-studied in the literature and NLO QCD corrections are known, see e.g. Refs.~\cite{Berthier:2016tkq,Falkowski:2016cxu,Baglio:2017bfe,Grojean:2018dqj,Baglio:2019uty,Baglio:2020oqu}. Here we simulate this process at LO in QCD, where WW only occurs with a $q\bar q$ initial state, via the annihilation of the $q\bar{q}$ pair or by a $t$-channel exchange of a parton. Within this approximation the main EFT effects are modifications of triple gauge couplings ($c_W$) and of couplings of the weak bosons to fermions ($c_{Hq}^{(1)},c_{Hq}^{(3)},c_{Hl}^{(3)}$), as well as input shift corrections ($c_{HD},c_{HWB},c_{ll}^{(1)}$). For most of these coefficients, the WW production gives more stringent constraints than VBS (in fully-leptonic final states), mainly due to the larger cross section. The only exception is the custodial-violating $c_{HD}$, that is best probed in processes where the dominant Feynman diagrams involve Z or photon couplings to fermions.
$c_{Hl}^{(1)}$ is poorly constrained in this final state, because it only enters non resonant diagrams with a Z boson in $s$-channel. 
Higgs ($c_{H\square},c_{HW}$) and four-fermion operators only enter at 1-loop in QCD, and are therefore neglected here.
Distributions for interesting variables for some operators this process is sensitive to, after the selections, are shown in Figure~\ref{fig:Distributions_inWW}.

\subsection{\texorpdfstring{$W^\pm W^\pm$}{WW}+2j}\label{sec.ssww}
The analysis of the EW production of a same-sign W boson pair (SSWW+2j) selects the leptonic final state in association with two jets, where the W bosons decay in an electron-muon pair plus the corresponding neutrinos. This process has been observed by the LHC Collaborations in the fully-leptonic final state~\cite{ATLAS:2019cbr,CMS:2017fhs,CMS:2020gfh}.  With two leptons of the same charge, moderate MET, and two jets with a large rapidity separation and a large dijet mass, the EW W$^{\pm}$W$^{\pm}$ process presents a clean signature in the detector. The LO QCD-induced background, with diagrams with exactly two QCD vertices, is small compared to the EW-induced production and can be kinematically separated from the signal. Thus it can be safely neglected in our analysis. The main source of reducible background for this channel are jet-induced fake leptons, mostly stemming from $b$ quarks in $t\bar t$ events. They can be neglected as their impact is very suppressed in the high-energy tails, where we have most of the EFT sensitivity. 

As shown in Fig.~\ref{fig:Combined_LL_Profiles_15}, this analysis is particularly sensitive to 4-quark operators, that induce very marked shape distortions both in energy and angular variables. At quadratic level, SSWW+2j has the highest discriminating power among their Wilson coefficients, see e.g. Fig.~\ref{fig:2DCombined}. On the other hand, it yields the weakest bounds on $c_{HD}$, as Z boson contributions are subdominant, and on $c_{HWB}$: the latter dominantly enters via corrections to the weak mixing angle, to which SSWW has limited sensitivity.
 
Distributions for interesting variables for some operators this process is sensitive to, after applying the selection cuts, are shown in Figure~\ref{fig:Distributions_SSWW}.

\subsection{\texorpdfstring{$W^+W^-$}{WW}+2j}

Despite possessing the largest cross section among the VBS processes, the purely EW production of a pair of opposite-sign W bosons (OSWW+2j) has been observed experimentally only very recently by the CMS Collaboration~\cite{CMS-PAS-SMP-21-001}. The main challenge in accessing this process is the very large irreducible QCD background (cfr. Tab.~\ref{tab:variables}).
We analyze the fully leptonic final state, characterized by the presence of two jets, with high energy and a large separation in pseudo-rapidity,  two opposite-charge leptons and missing transverse energy. 
The same-flavour final states ($e^+e^-\nu\nu$, $\mu^+ \mu^-\nu\nu$) are overwhelmed by Drell-Yan events, so here we restrict to different-flavour combinations ($e^{\pm}\mu^{\mp}\nu\nu)$, that have smaller backgrounds and hence a higher sensitivity.
The QCD induced production of W$^+$W$^-$+2j (order $\alpha^{4}_{EW}\alpha^{2}_{QCD}$) is sizeable and taken into account in the analysis. 
The selections in Table~\ref{tab:variables} reduce the background contamination, which mainly arises from $t\bar{t}$ production.

The Vector Boson Fusion Higgs contribution  is a subdominant component of the VBS signal, once the selections are applied. Nevertheless, some sensitivity to Higgs operators is retained. In fact, thanks to the presence of $s$-channel diagrams and to the relatively large cross section, OSWW is the most sensitive channel to anomalous Higgs couplings ($c_{H\square},c_{HW}$), among those considered in this work.
Distributions for interesting variables for some operators this process is sensitive to, after the selections, are shown in Figure~\ref{fig:Distributions_OSWW}.

\subsection{\texorpdfstring{$W^\pm Z$}{WZ}+2j}\label{sec.WZ}

We consider the EW WZ production in the fully leptonic channel plus two jets, with the Z boson decaying into electrons and the W boson decaying into a ($\mu,\nu_\mu$) pair. 
One of the advantages of this decay mode is the high purity of the multi-leptonic final state.
Additionally, the choice of different flavour decays for the Z and the W bosons provides an efficient discrimination between the two bosons.
The WZ+2j channel has a small cross-section but was observed at Run II by both ATLAS and CMS~\cite{CMS:2019uys,ATLAS:2018mxa}. 
The dominant QCD background for the WZ+2j state is the QCD radiation of partons from an incoming quark or gluon, and it is accounted for in the analysis.

The presence of a single neutrino in the final state allows to kinematically reconstruct its momentum component along the $z$-axis, by imposing a $m_W$ constraint on the invariant mass of the W boson decay products. Once the four-momenta of all the final state particles are known, one can extract the invariant mass of the WZ system $m_{WZ}$, the angular separations between the vector bosons $\delta\eta_{WZ}$, $\delta\phi_{WZ}$ and the separation between their decay planes $\Phi_{planes}$. 
Furthermore, knowing the collision centre-of-mass boost along the beam direction, more sophisticated angular observables can be constructed, such as 
the emission polar angles of the leptons with respect to the direction of the decaying bosons in the rest frame of the latter, $\theta_{lW}$ and $\theta_{lZ}$, and the vector bosons emission angle in the centre-of-mass reference frame, $\theta^{*}$.  

Among the channels considered, WZ+2j is the one with highest sensitivity to $c_{Hl}^{(1)}$ (together with ZZ+2j). The sensitivity to $c_{HD}$ and $c_{HWB}$ is also quite good: as mentioned above, the explicit presence of a Z boson (and also of photons) in the dominant diagrams enhances the sensitivity to these two operators, compensating for the smaller cross section compared to other VBS channels and even compared to the inclusive WW.

Distributions for interesting variables for some operators this process is sensitive to, after the selections, are shown in Figure~\ref{fig:Distributions_WZ}.

\subsection{\texorpdfstring{$ZZ$}{ZZ}+2j}

The observation of the EW ZZ production was achieved by the ATLAS Collaboration using the full Run-II data set~\cite{ATLAS:2020nlt}, while the CMS Collaboration has recently reported strong evidence for this process~\cite{CMS:2017zmo,CMS:2020fqz}. 
This VBS channel has the smallest cross-section among all the processes considered,
and is one of the rarest SM processes observed to date. Since the QCD production associated to two jets is a dominant background with respect to the VBS signal, the search for this process is very challenging, despite a very clean experimental
signature.

Both the ATLAS and CMS Collaborations use multivariate analyses to isolate the EW signal over the large QCD background, after some fiducial cuts. Here we perform a much simpler analysis using the selections listed in Table~\ref{tab:variables} and considering exclusively the $2e2\mu$ final state. The variables of interest include, besides those in common with other VBS channels, the invariant mass and total transverse momentum of the lepton pair with the largest transverse momentum $m_{l^1l^2}$, $p_{T,l^1l^2}$, the invariant mass of the four-lepton system $m_{4l}$, the total transverse momentum of the same-sign lepton pair $p_{T,e^{\pm}\mu^{\pm}}$ and the transverse momentum of the dilepton system $e^{+}e^{-}$ or $\mu^{+}\mu^{-}$ with invariant mass closest to $m_Z$
, taken as a proxy for $p_{T,Z}$.

A detailed EFT study of this particular process (restricting to resonant diagrams, which in this case is a good approximation) was presented in Ref.~\cite{Gomez-Ambrosio:2018pnl}. The sensitivity of ZZ+2j to EFT effects is quite limited due to the low total cross section and small signal-to-background ratio.  Nevertheless, this channel is relevant for constraining operators that affect specifically Z interactions, such as $c_{Hl}^{(1)}$. It is also competitive with the other processes for $c_{HD},c_{HWB}$.

Distributions for interesting variables for some operators this process is sensitive to, after the selections, are shown in Figure~\ref{fig:Distributions_ZZ}.

\subsection{\texorpdfstring{$ZV$}{ZV}+2j}\label{sec.ZV}

Even though the LHC collaborations have mainly studied VBS processes in the fully leptonic final states, semileptonic ZV production also ranks among the processes of greatest interest, as it benefits from the larger branching ratio of $V\to \text{hadrons}$ (about 67\% for W bosons and 70\% $Z$ bosons). For instance, ZV+2j was shown to be among the most sensitive channels to dimension-8 EFT effects \cite{VBS_recent_develop_2018,CMS:2019qfk}. 

We consider final states with four jets and a pair of electrons or muons,
addressing the challenge to identify properly the two tagging jets by identifying the jet pair with the highest invariant mass as the one produced by the scattering partons. 
This method correctly matches final state partons to the corresponding vector boson or scattering parton for at least 75\% of the events under investigation. Events where the final state is produced through a triple vector boson emission are vetoed in order to select the EW VBS channel.

We simulate the irreducible QCD-induced $l^{+}l^{-}+4j$ sample and include its EFT dependence in the results. Another major background for this channel is $Z$+jets, which, however, is not included here due to the significant computational challenges posed by its Monte Carlo simulation. To prevent this choice from introducing a bias in the global analysis, the ZV+2j results will be analysed separately from the other channels, see Sec.~\ref{Individual_constraints}. 
Up to this caveat, the semileptonic ZV+2j channel is found to be very competitive in terms of sensitivity to EFT effects, as it  combines the presence of Z bosons in dominant diagrams with a large cross section. For this reason, it yields the strongest constraints on $c_{HD},c_{Hl}^{(1)}$ and $c_{HWB}$. It is also competitive with the inclusive WW on $c_{Hl}^{(3)}, c_{Hq}^{(3)}$ and $c_{ll}^{(1)}$, and with OSWW+2j on $c_{HW}$.

Distributions for interesting variables for some operators this process is sensitive to, after the selections, are shown in Figure~\ref{fig:Distributions_VZ}.

\section{Results}\label{sec:results}

In this section we report the results of the likelihood scans with different configurations,
corresponding to the statistics expected after an integrated luminosity of $\unit[100]{fb^{-1}}$, collected by one given experiment at the LHC.

\subsection{One-dimensional constraints}\label{Individual_constraints}

\begin{figure}[htbp]
  \centering 
  \hspace{.24\textwidth}
  \includegraphics[width=.24\textwidth,trim={9mm 6mm 2.5cm 1.4cm},clip]{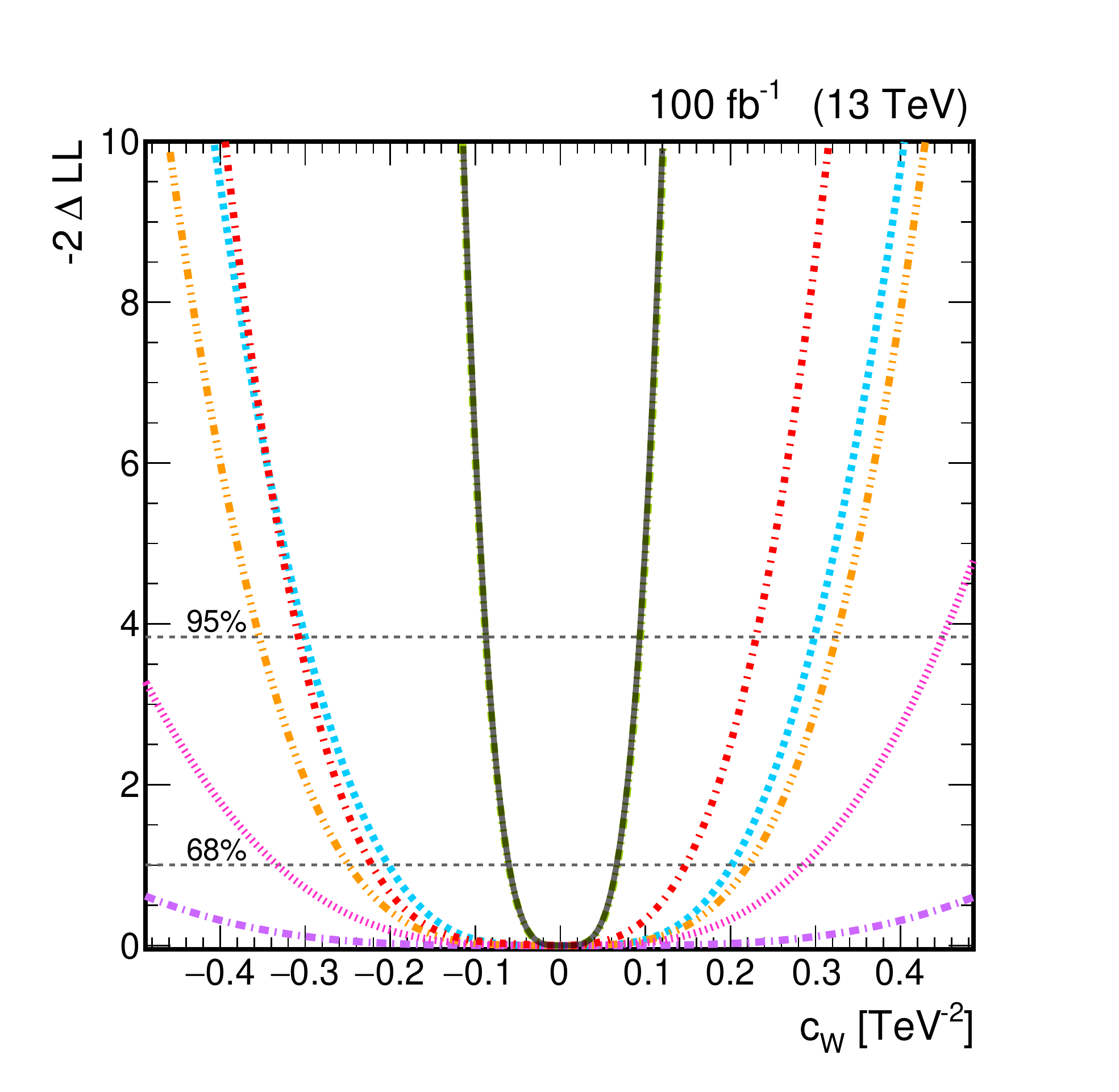}
  \includegraphics[width=.24\textwidth,trim={9mm 6mm 2.5cm 1.4cm},clip]{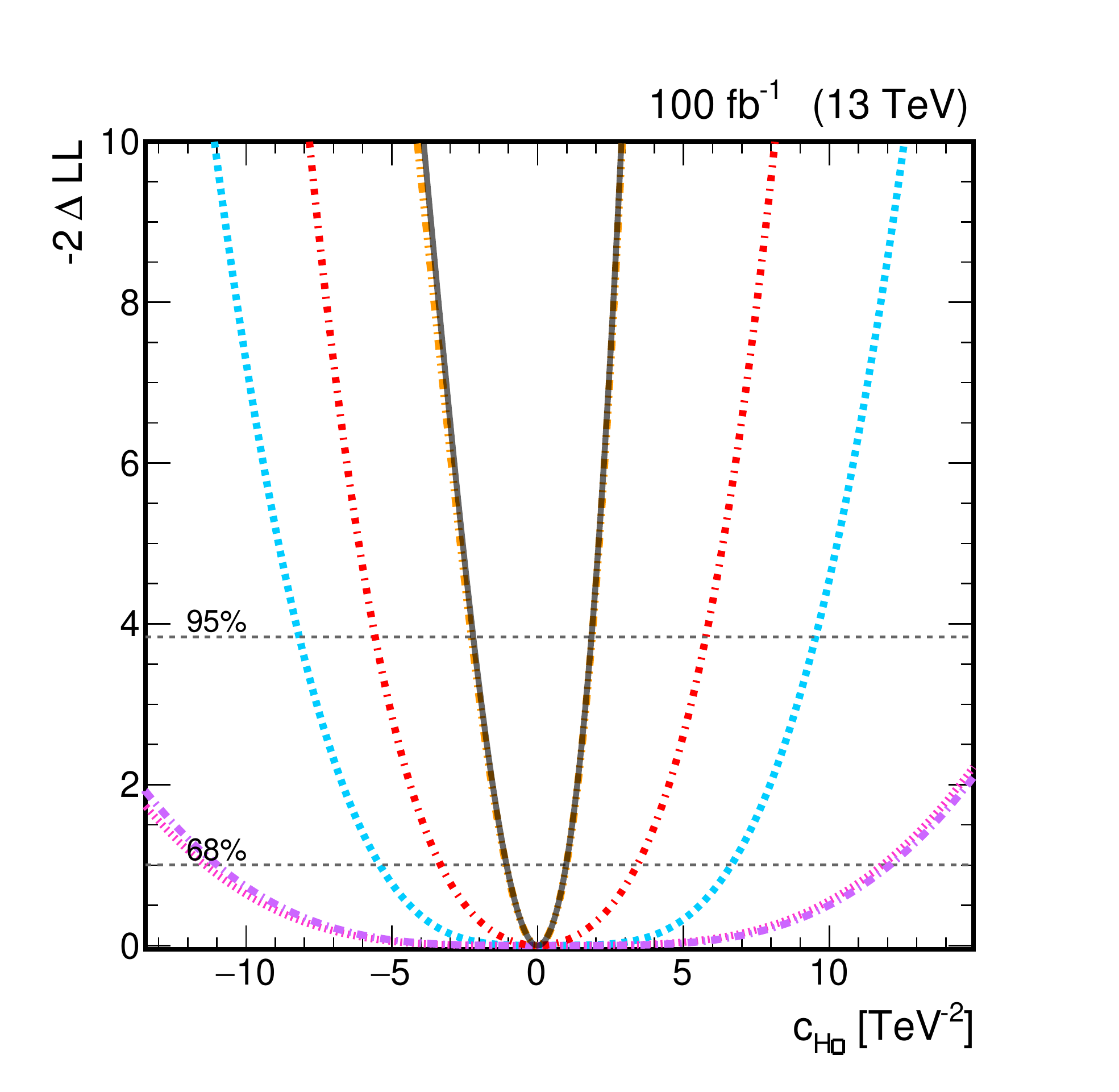}
  \includegraphics[width=.24\textwidth,trim={9mm 6mm 2.5cm 1.4cm},clip]{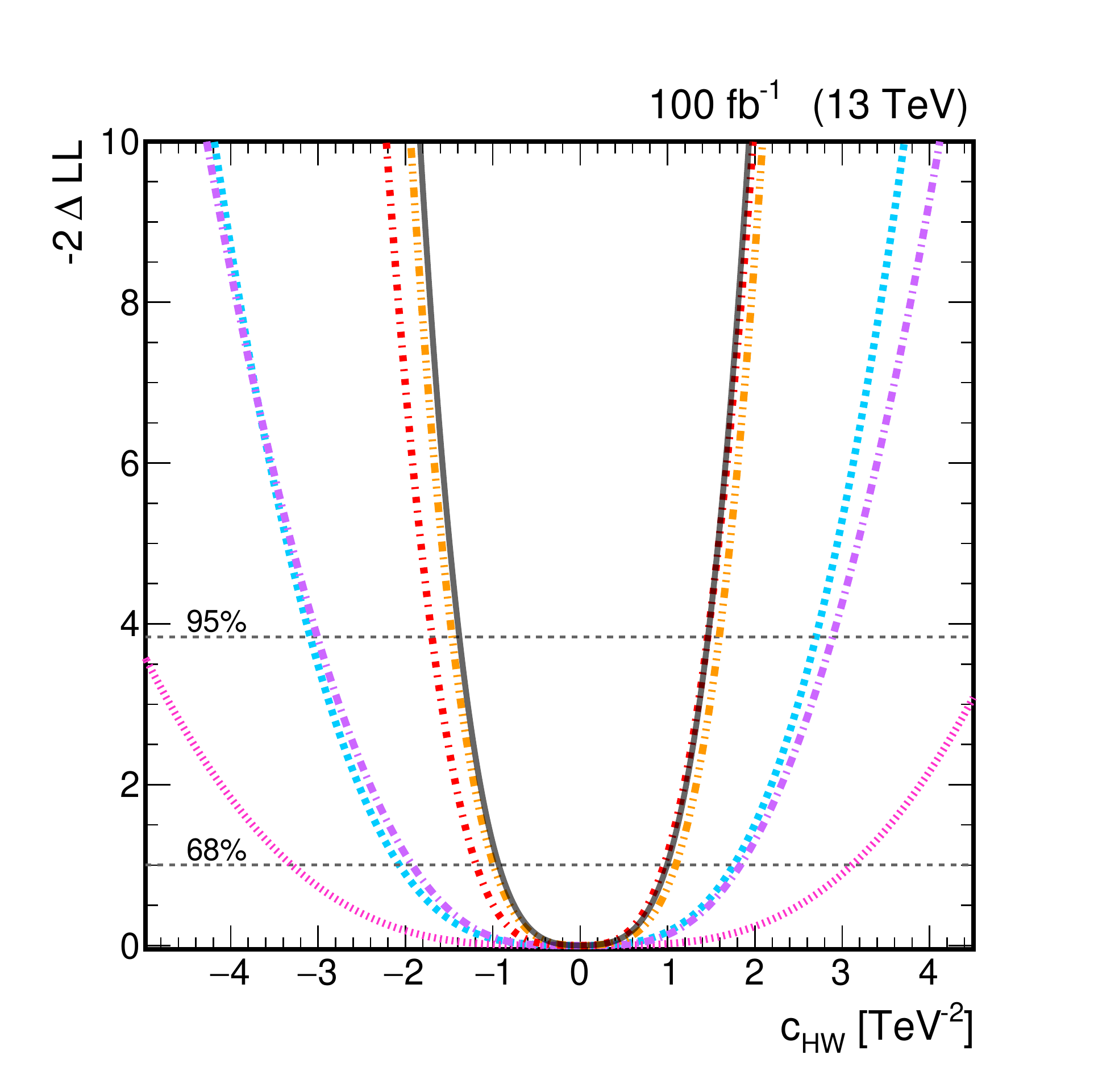}\\[2mm]
    
   \includegraphics[width=.24\textwidth,trim={9mm 6mm 2.5cm 1.4cm},clip]{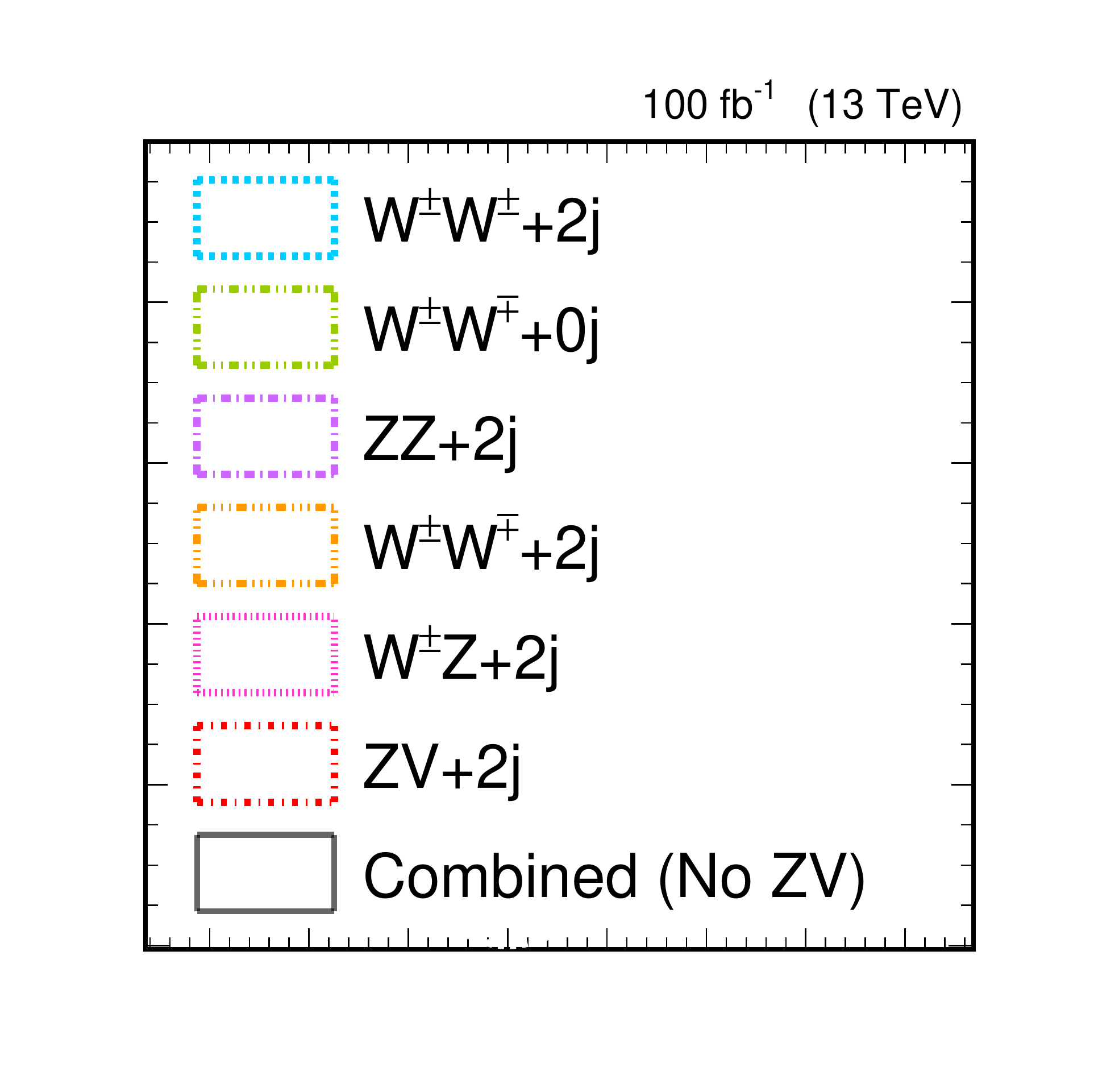}
  \includegraphics[width=.24\textwidth,trim={9mm 6mm 2.5cm 1.4cm},clip]{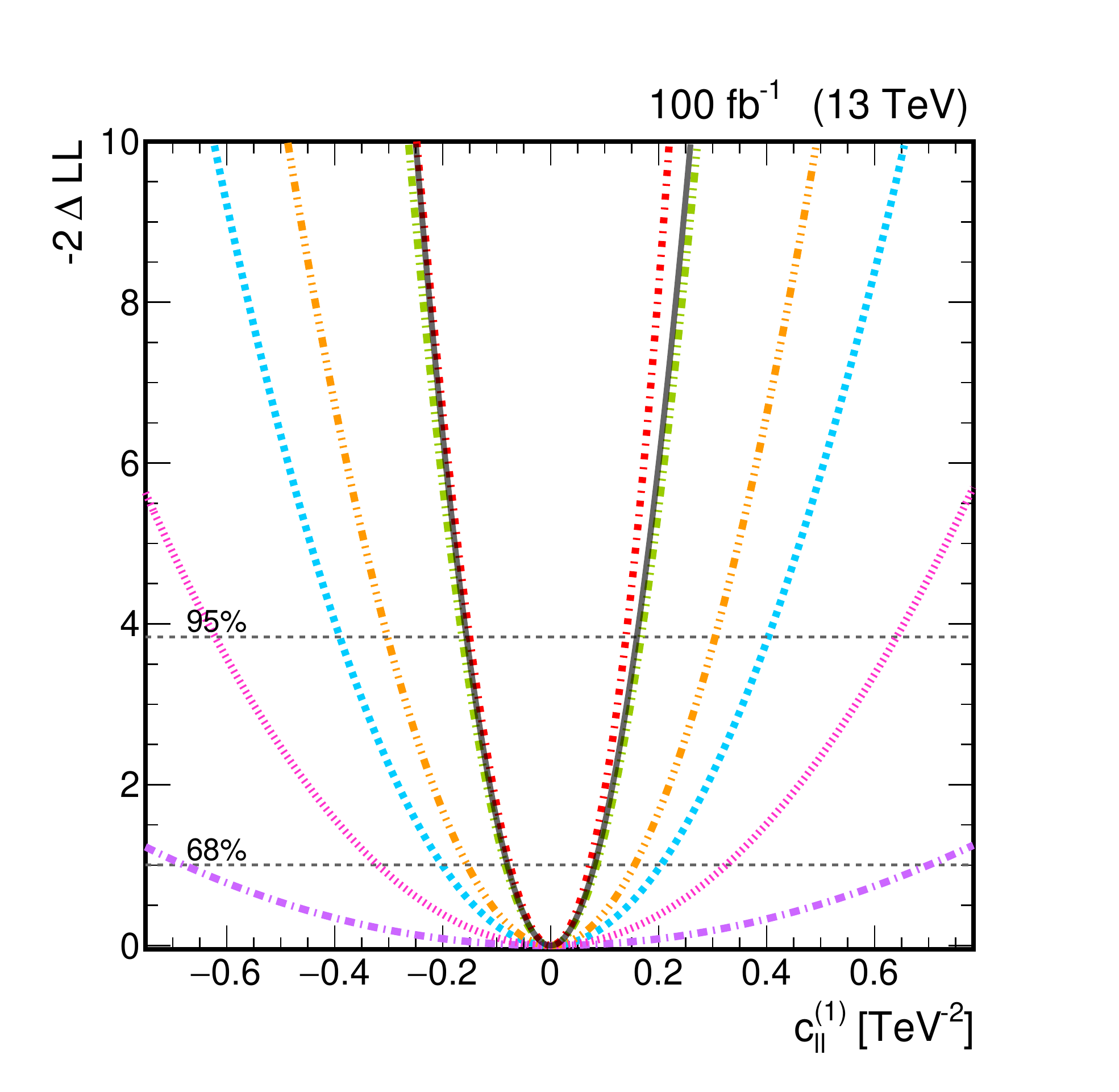}
  \includegraphics[width=.24\textwidth,trim={9mm 6mm 2.5cm 1.4cm},clip]{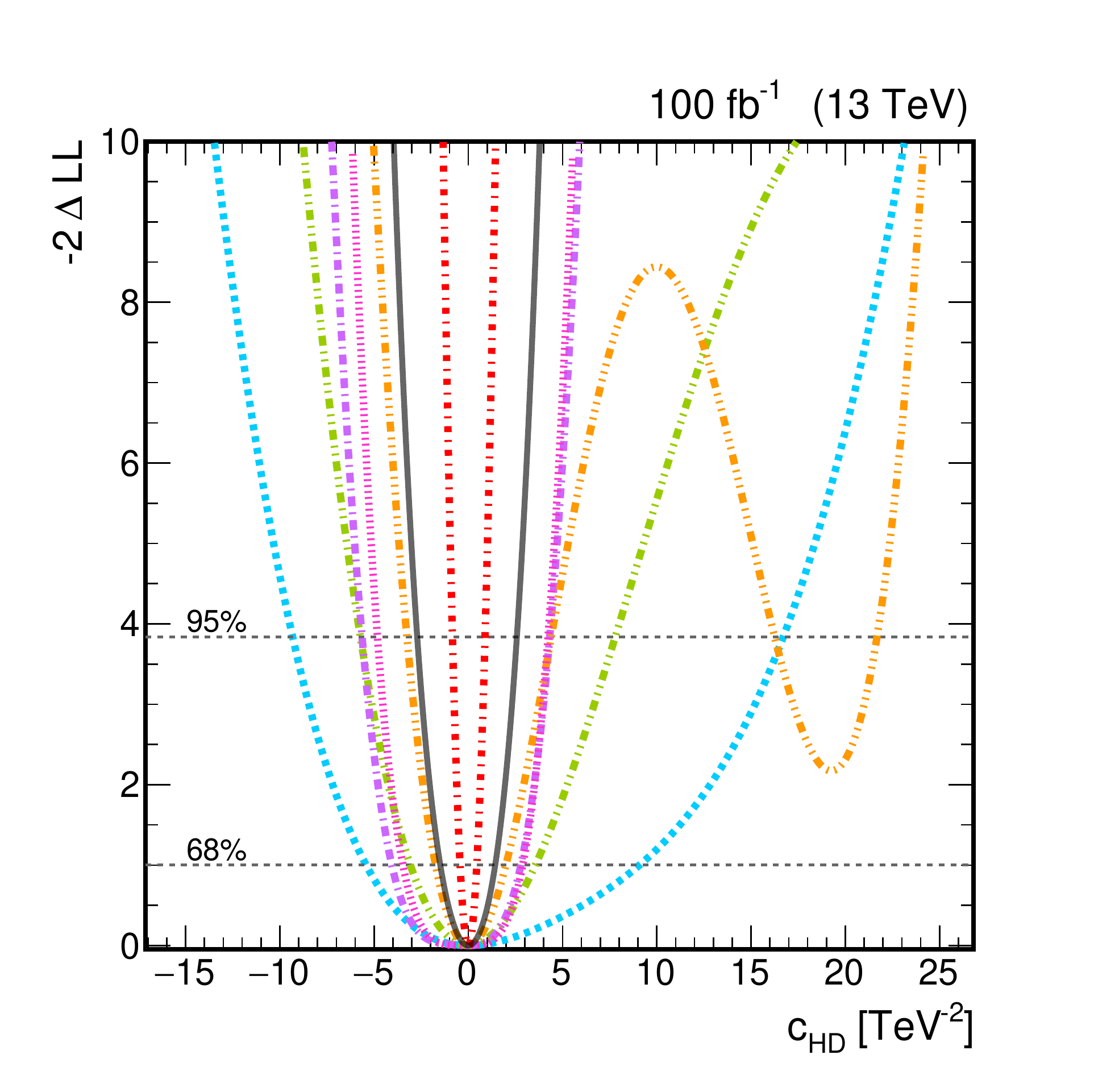}
  \includegraphics[width=.24\textwidth,trim={9mm 6mm 2.5cm 1.4cm},clip]{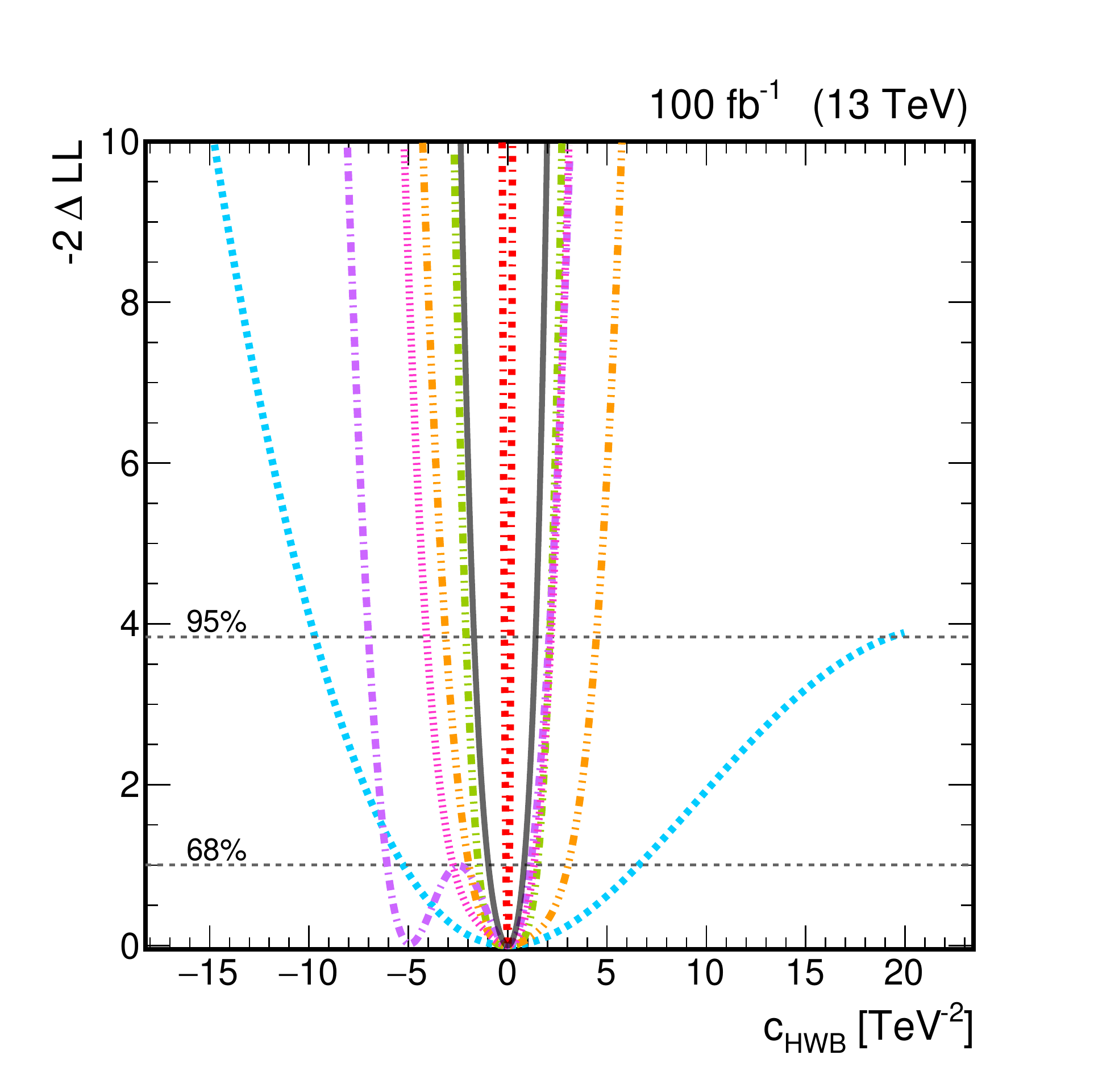}\\[2mm]
  
  \includegraphics[width=.24\textwidth,trim={9mm 6mm 2.5cm 1.4cm},clip]{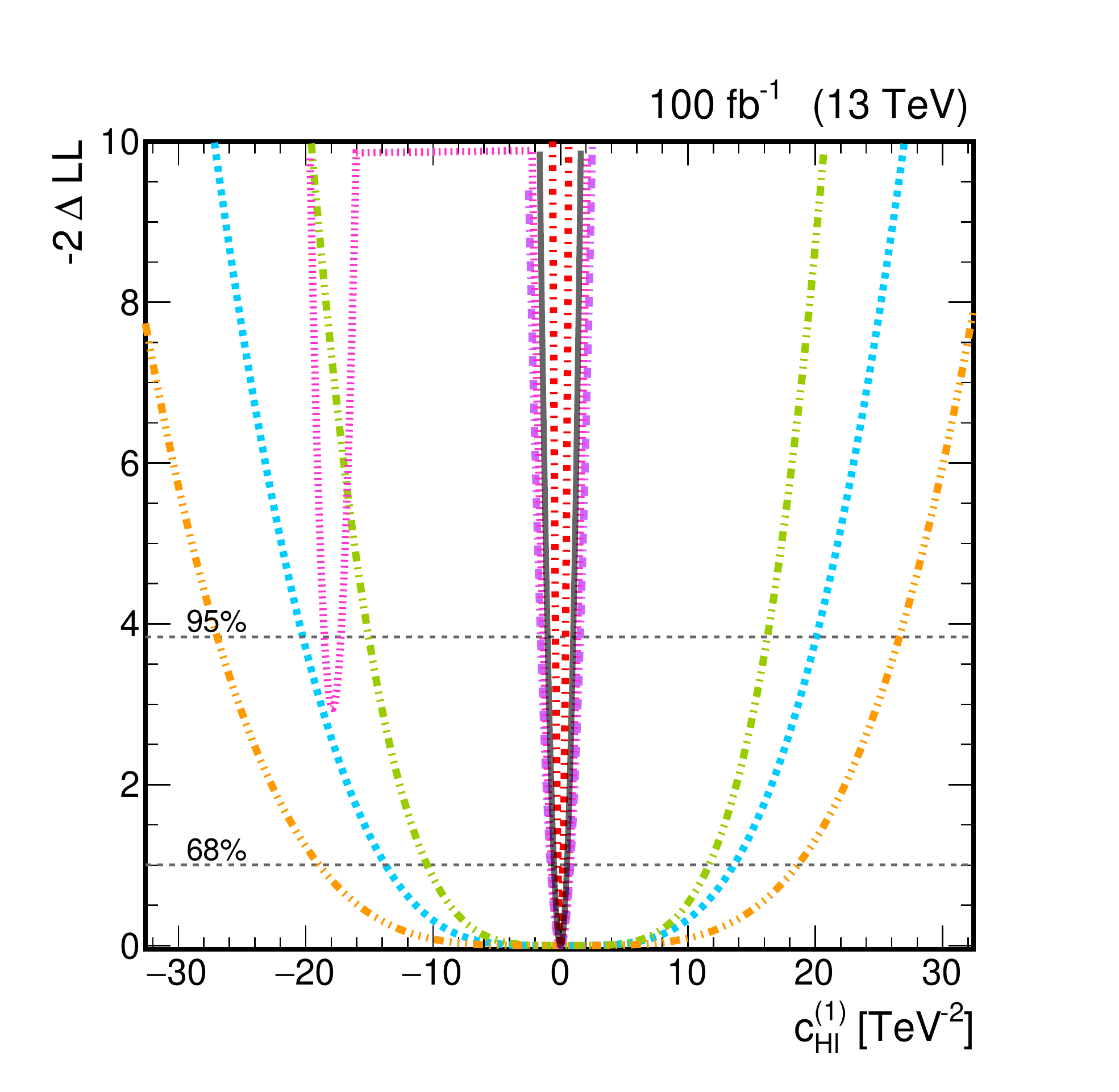}
  \includegraphics[width=.24\textwidth,trim={9mm 6mm 2.5cm 1.4cm},clip]{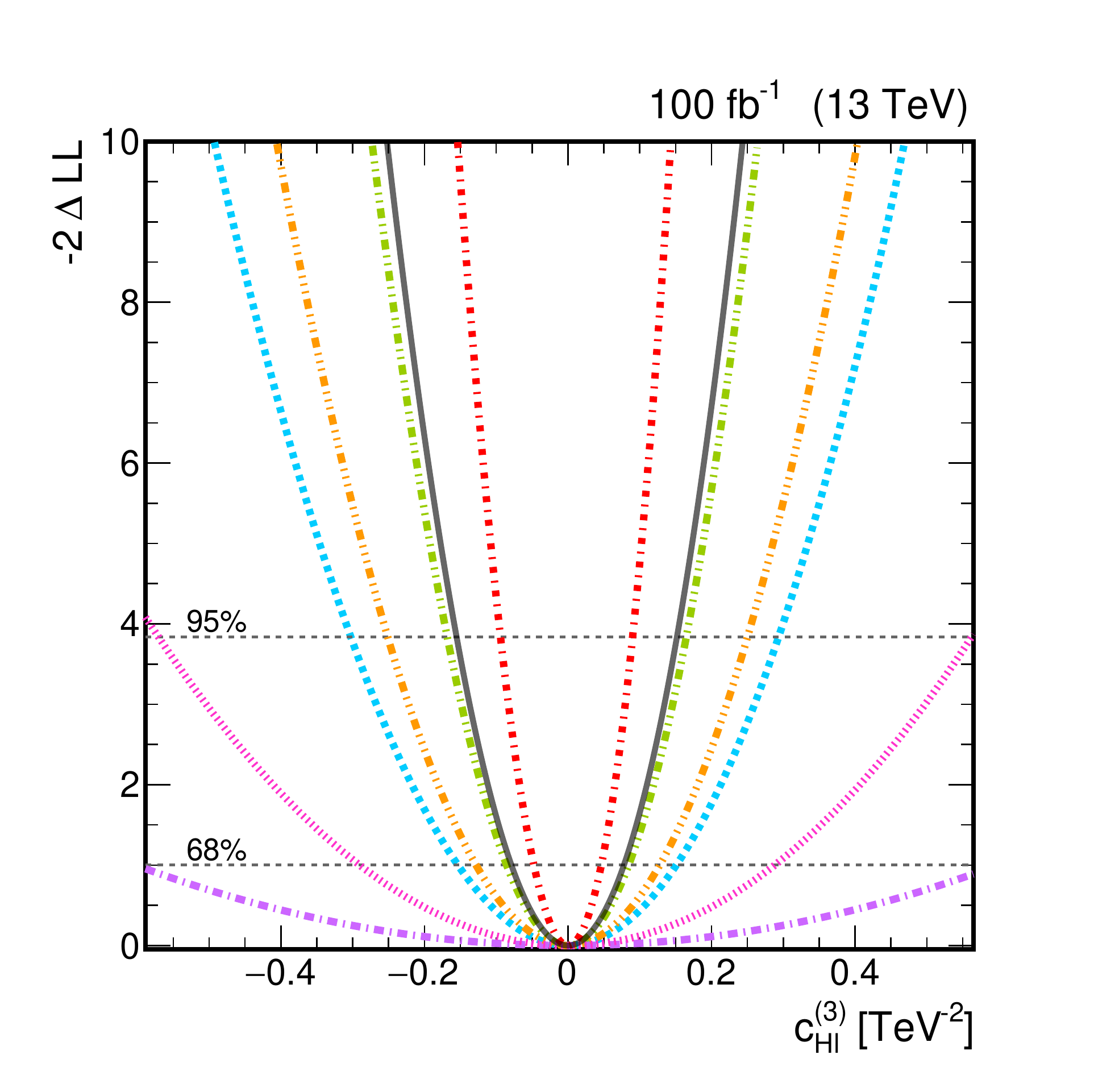}
  \includegraphics[width=.24\textwidth,trim={9mm 6mm 2.5cm 1.4cm},clip]{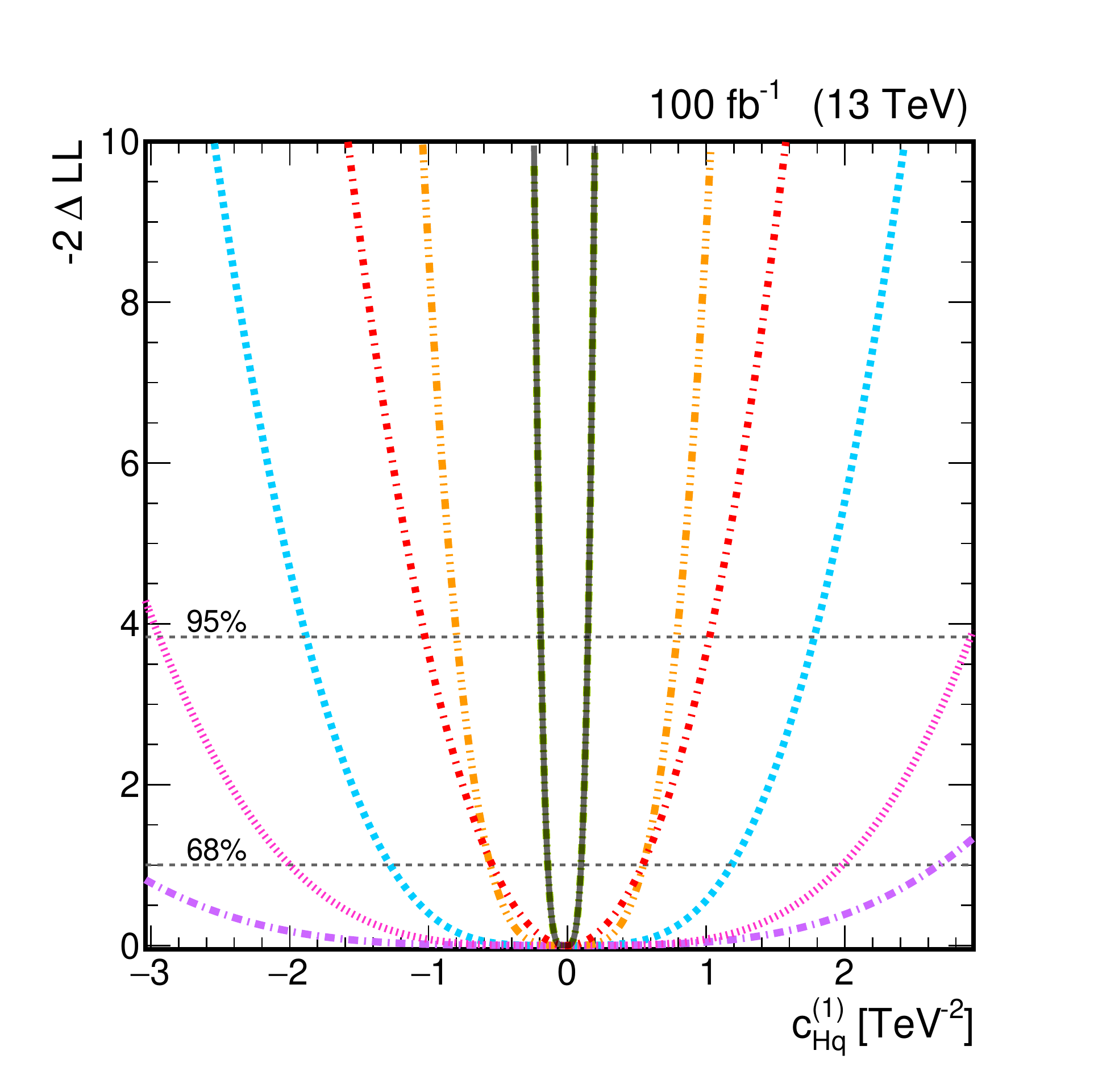}
  \includegraphics[width=.24\textwidth,trim={9mm 6mm 2.5cm 1.4cm},clip]{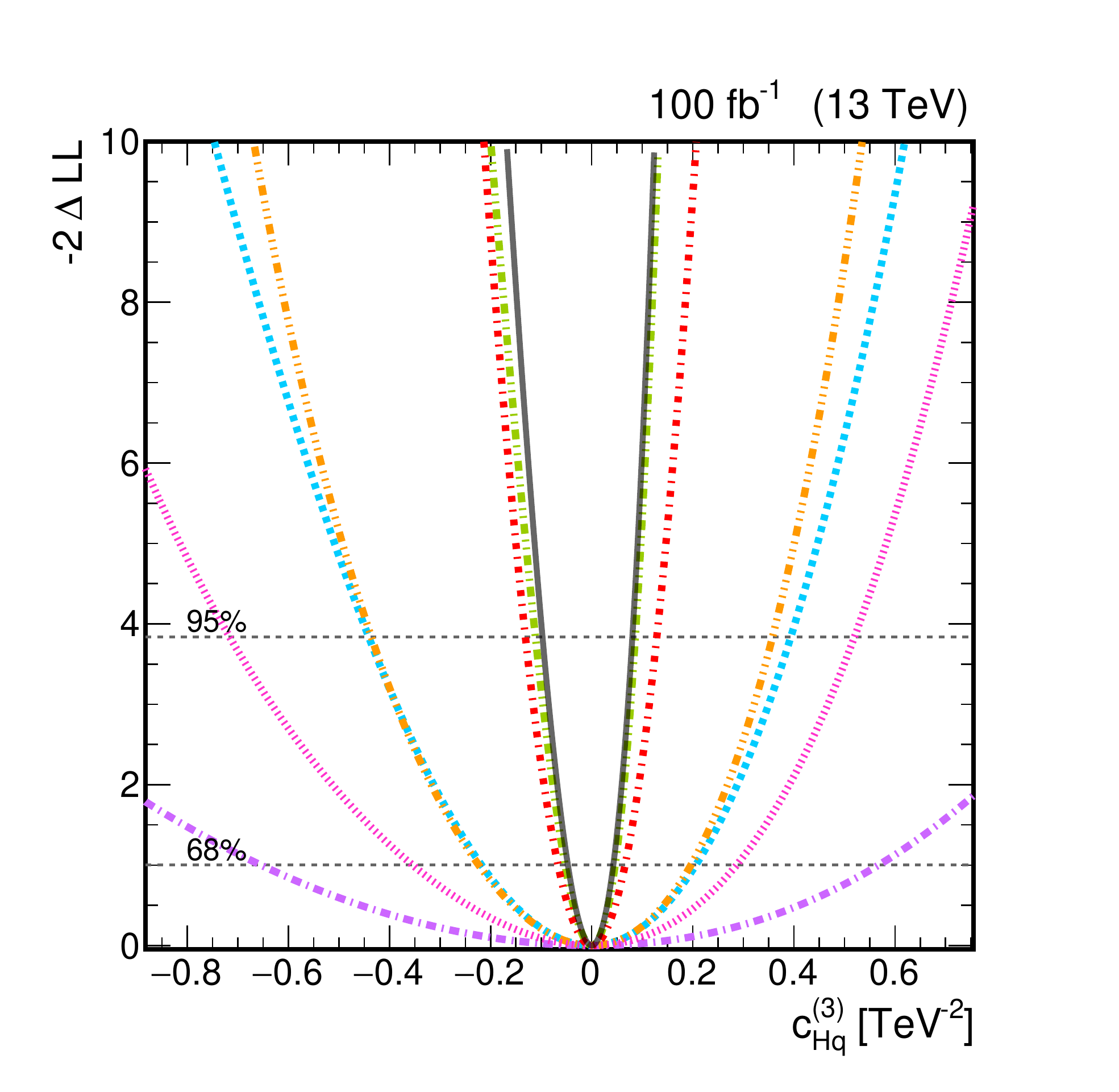}\\[2mm]
  
  \includegraphics[width=.24\textwidth,trim={9mm 6mm 2.5cm 1.4cm},clip]{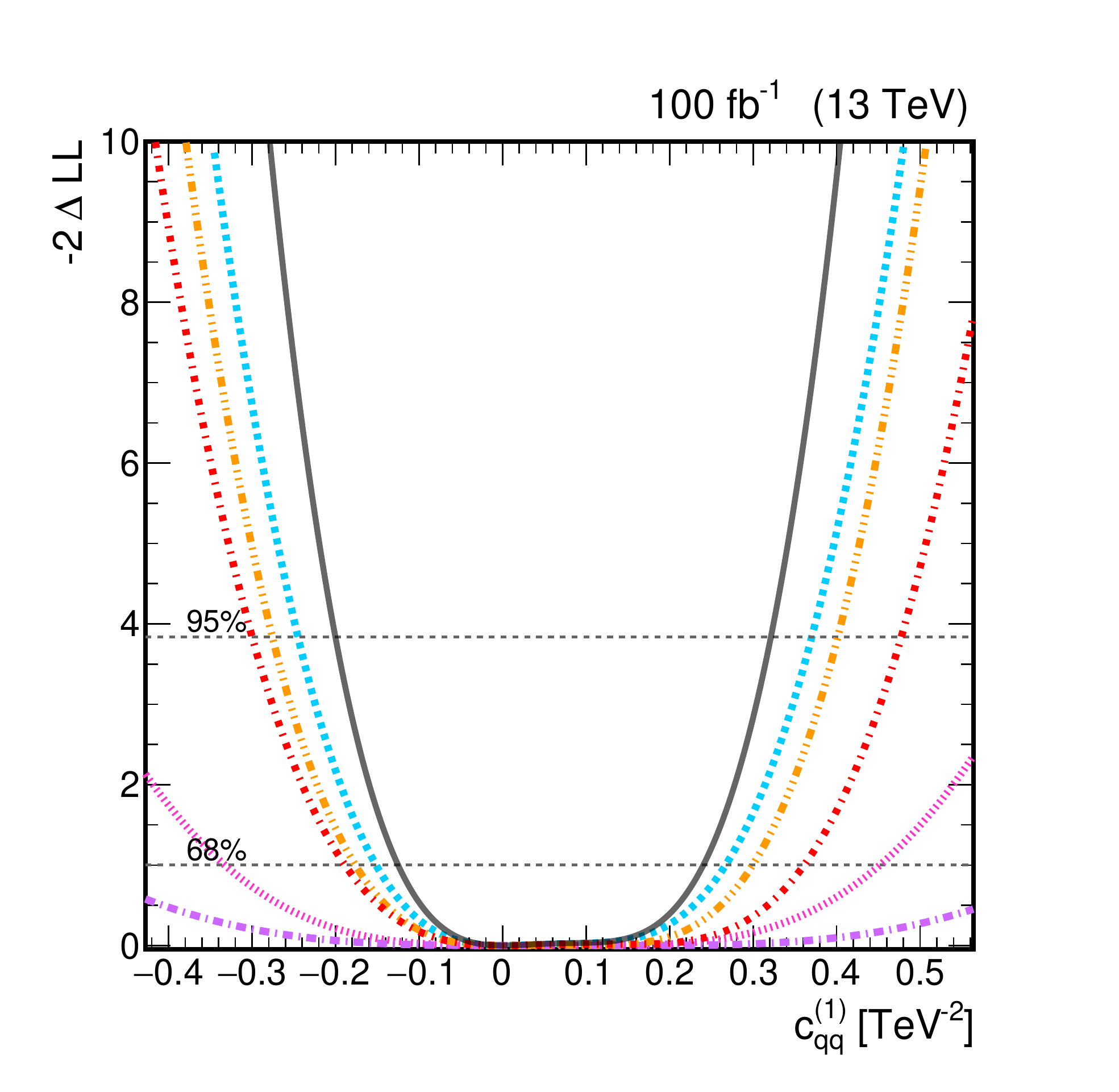}
  \includegraphics[width=.24\textwidth,trim={9mm 6mm 2.5cm 1.4cm},clip]{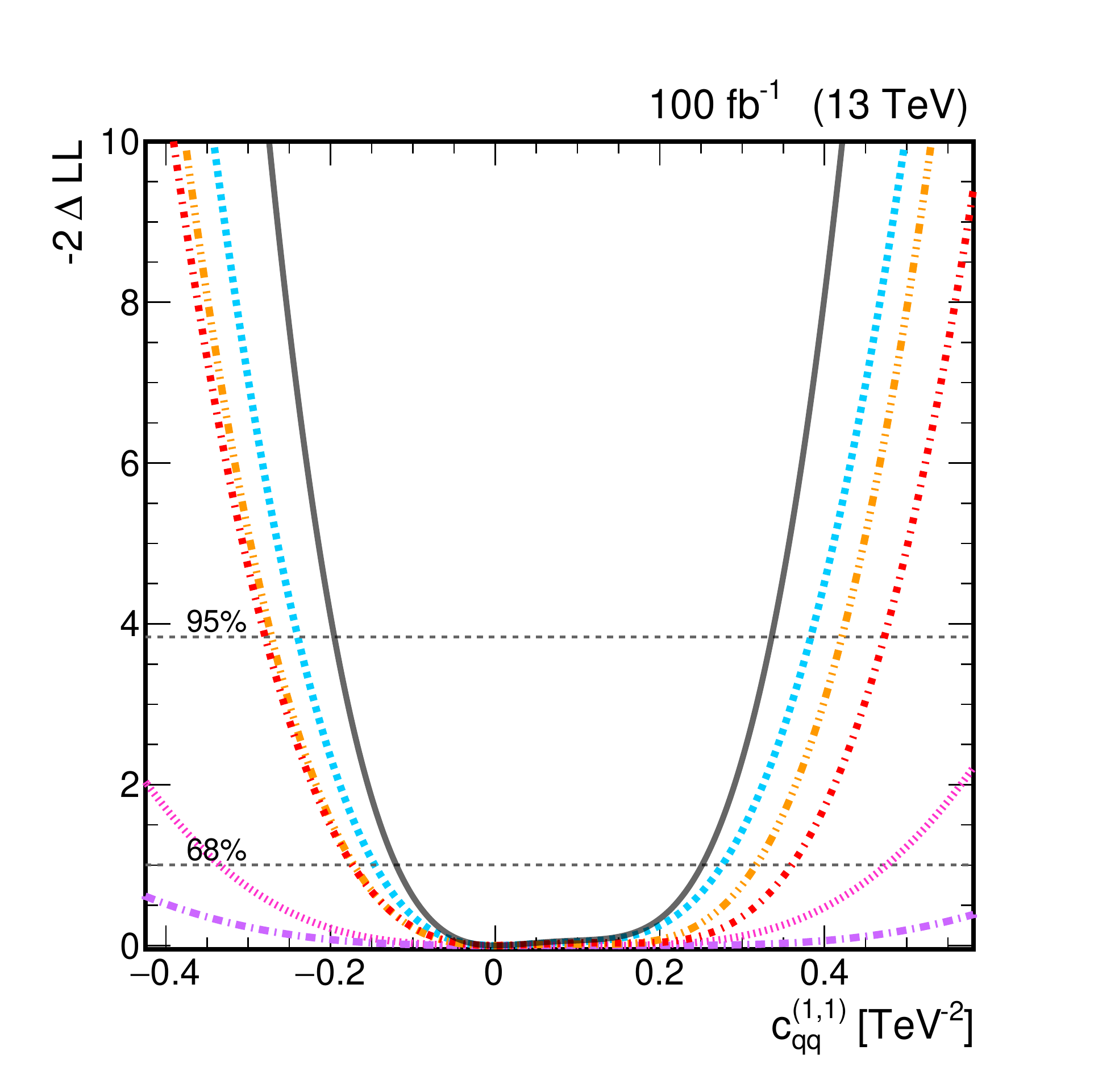}
  \includegraphics[width=.24\textwidth,trim={9mm 6mm 2.5cm 1.4cm},clip]{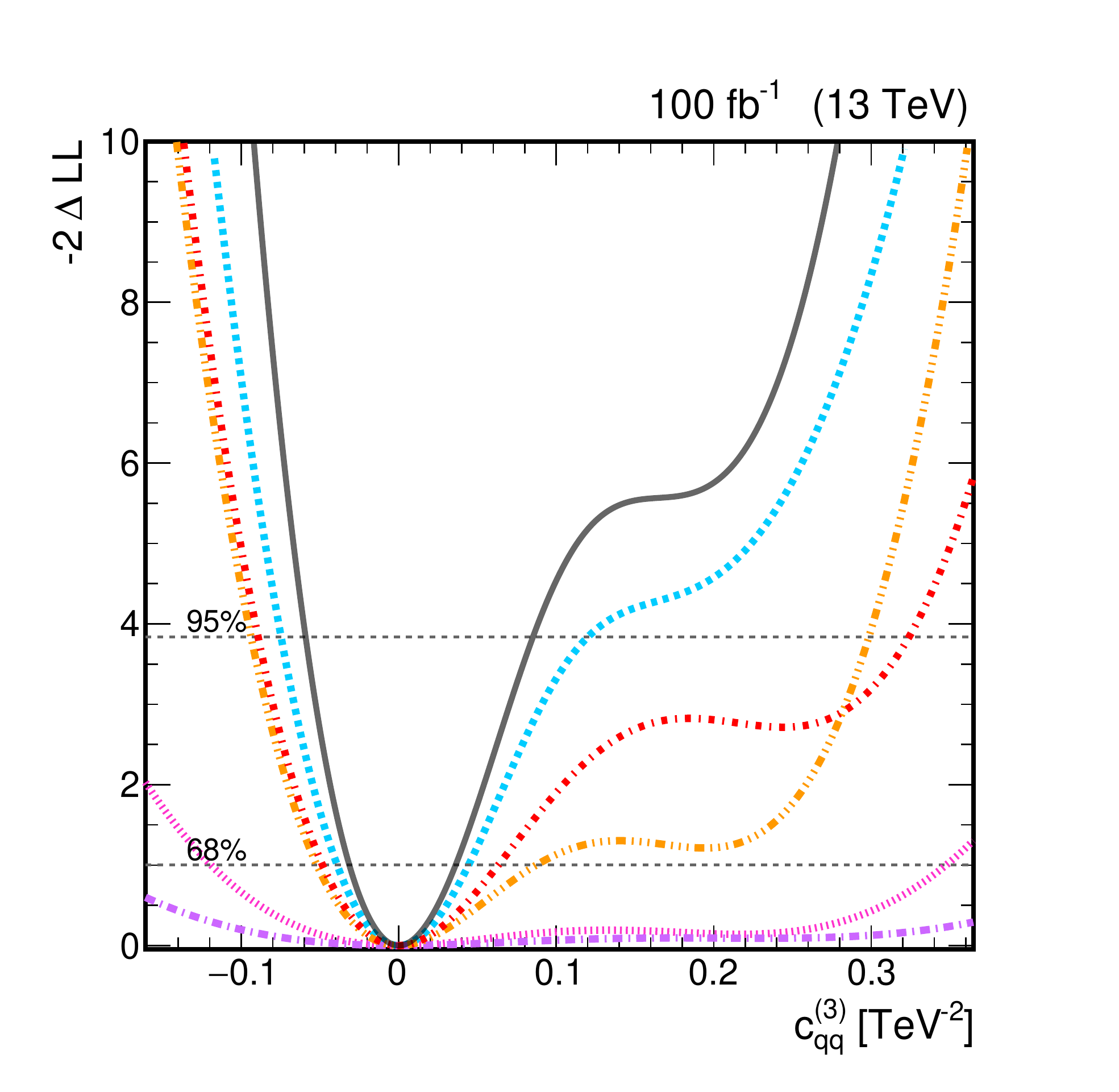}
  \includegraphics[width=.24\textwidth,trim={9mm 6mm 2.5cm 1.4cm},clip]{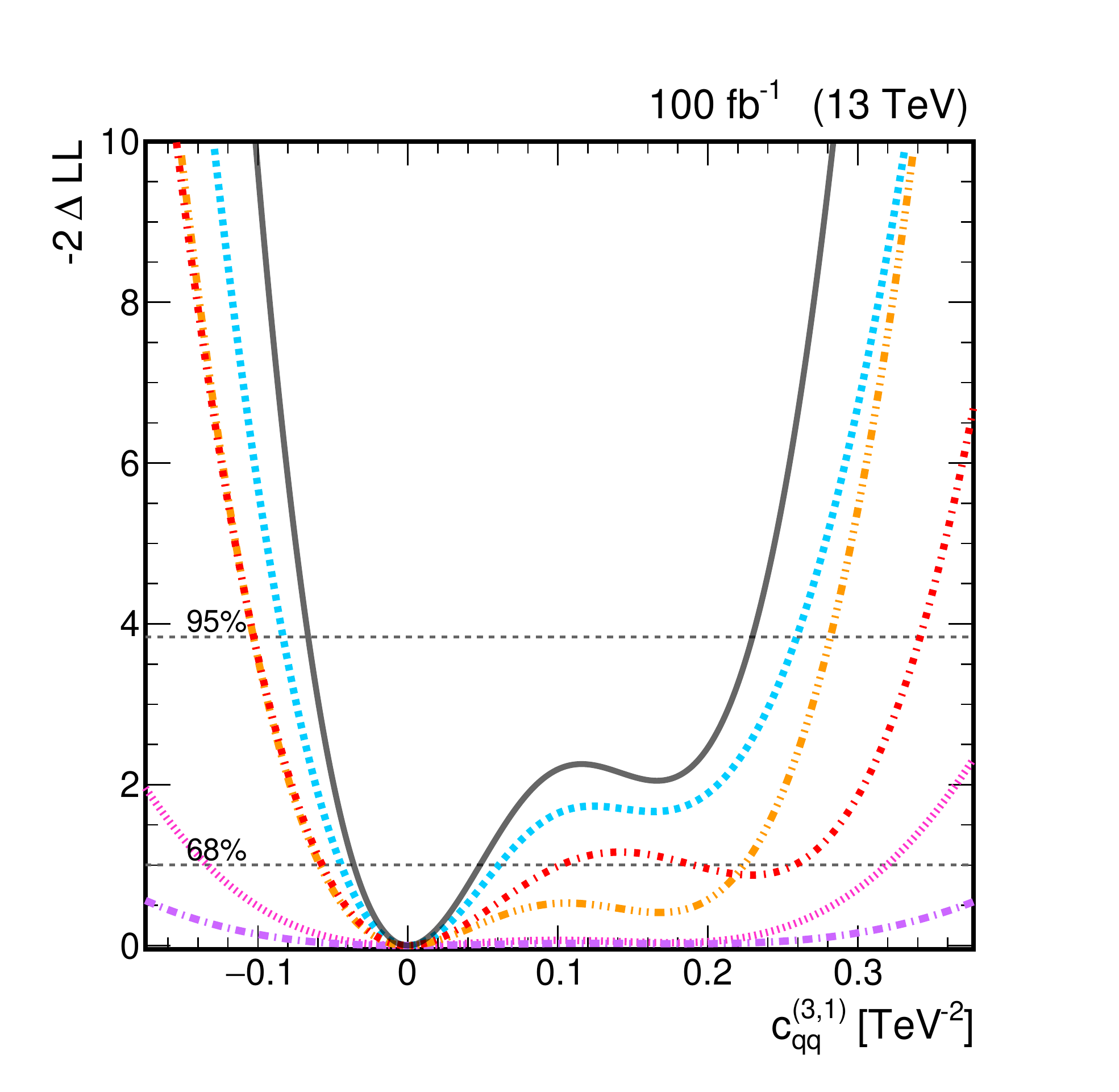}\\[2mm]
  
  \caption{Profiles of $-2\Delta\log \mathcal{L}$ reported for the individual channels (colored lines) and their combination (solid black line, excluding ZV+2j) as a function of the Wilson coefficients. Horizontal dashed lines mark the 68\% and 95\% confidence levels respectively, taken at $-2\Delta\log \mathcal{L} = 1$ and $-2\Delta\log \mathcal{L} = 3.84$. For each coefficient, the likelihood was built taking, for each channel, the distribution in the most constraining variable at 68\% c.l. (see Table~\ref{tab:sensitivity_ranking}). Only the shown Wilson coefficient is varied at each time, and the others are set to 0. 
  The sensitivity estimate for the OSWW+2j,  WZ+2j, ZZ+2j and ZV+2j channels includes contributions from the respective QCD induced processes. \label{fig:Combined_LL_Profiles_15}}  
\end{figure}

\begin{figure}[htbp]
  \centering 
  \vspace*{-1.7cm}
  \includegraphics[width=0.83\textwidth]{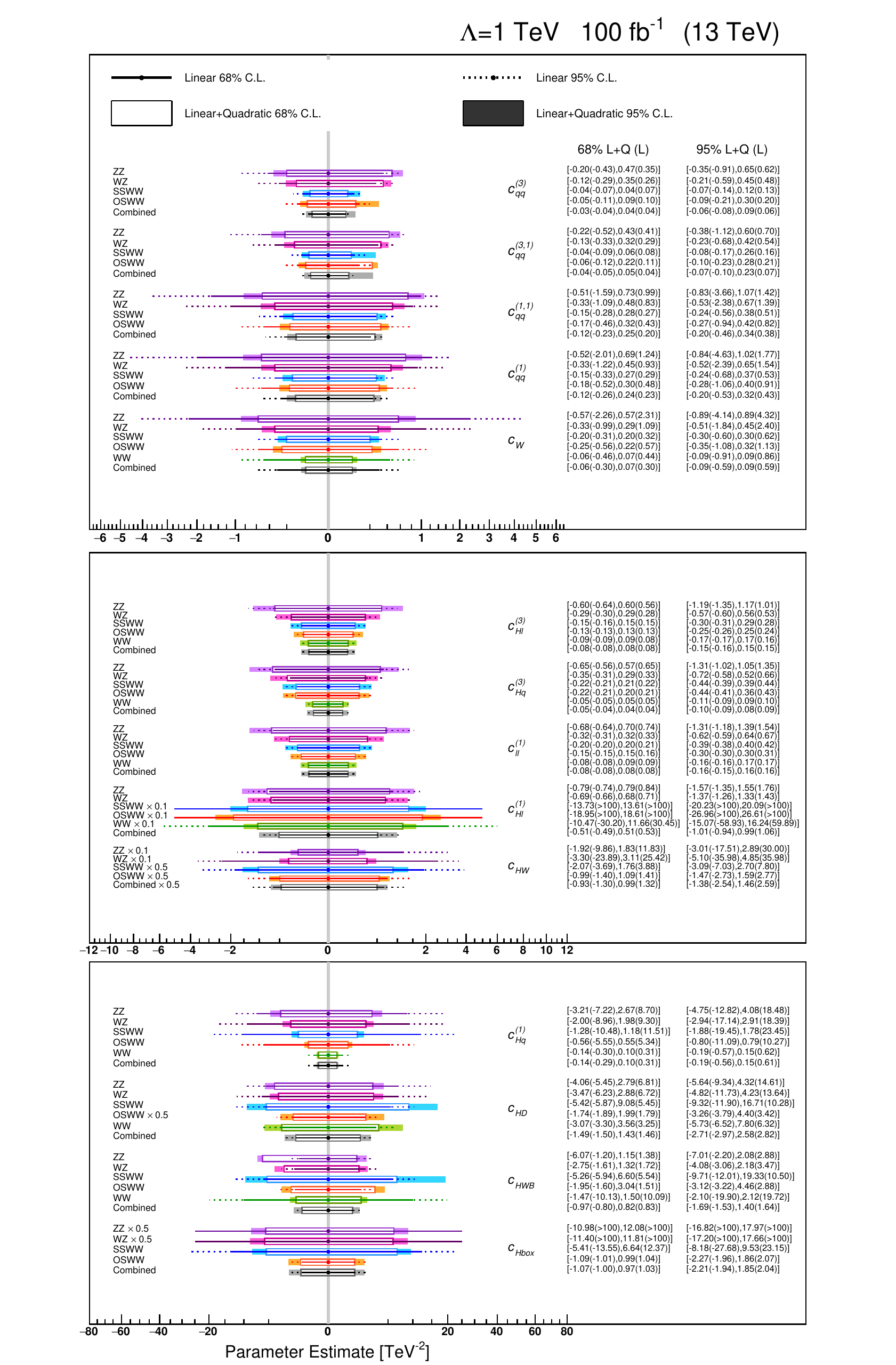}
  
  \caption{Individual expected constraints on Wilson coefficients from the leptonic VBS channels ZZ+2j, WZ+2j, SSWW+2j, OSWW+2j and diboson WW. The solid points represent the SM expectation. Solid (dashed) lines indicate the 68\%  (95\%) confidence intervals obtained including only terms linear in the Wilson coefficients in the signal predictions. Open (filled) boxes indicate 68\% (95\%) confidence intervals obtained including both linear and quadratic EFT components. 
  \label{fig:summary_2}}  
\end{figure}

\begin{figure}[htbp]
  \centering 
  \includegraphics[width=\textwidth]{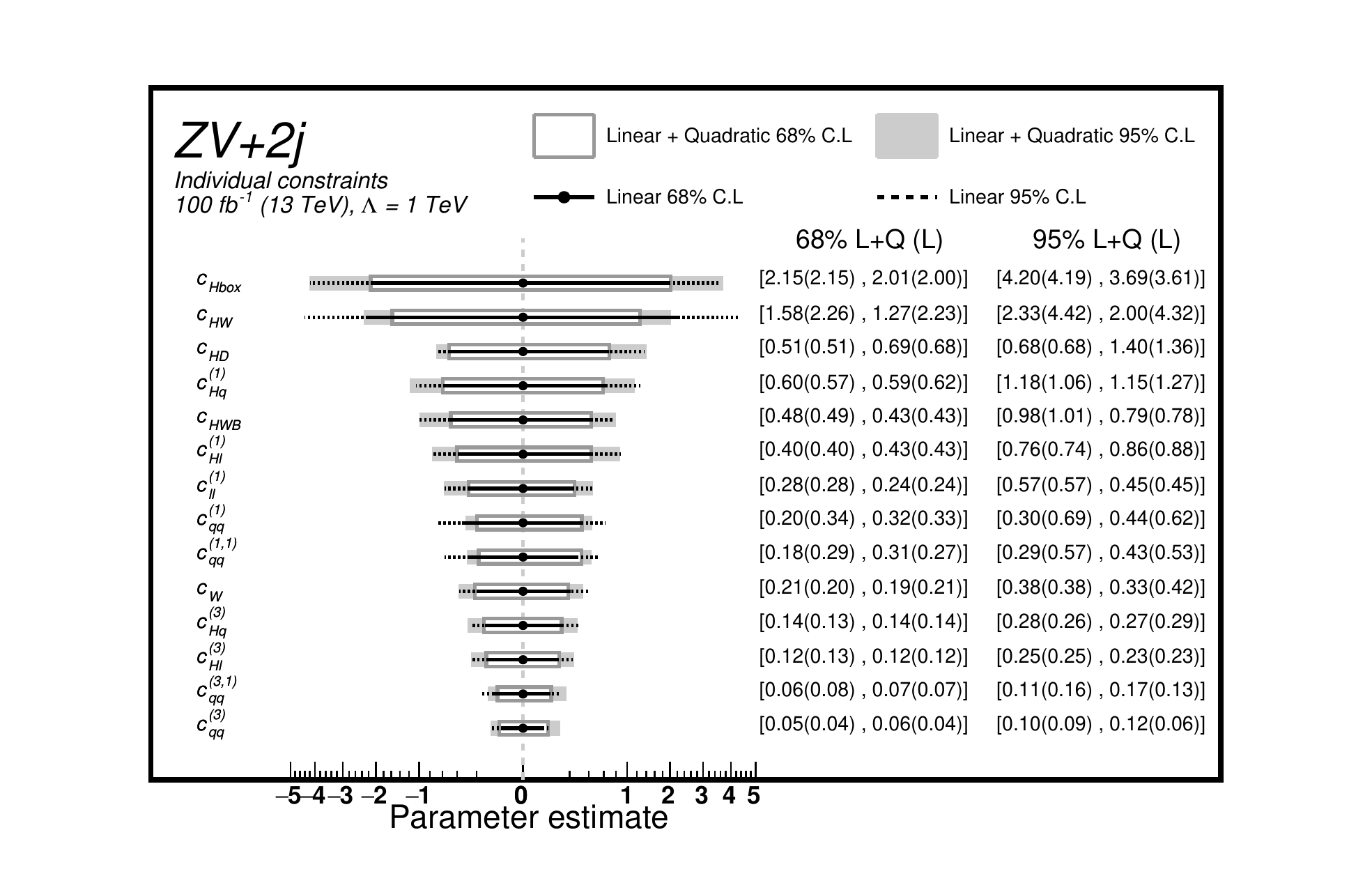}

  \caption{
  Individual expected constraints on Wilson coefficients from the VBS ZV+2j channel in the semileptonic final state. The solid points represent the SM expectation. Solid (dashed) lines indicate the 68\%  (95\%) confidence intervals obtained including only terms linear in the Wilson coefficients in the signal predictions. Open (filled) boxes indicate 68\% (95\%) confidence intervals obtained including both linear and quadratic EFT components. 
  The QCD induced EFT dependence was included when relevant. Note that the constraints shown in this figure neglect effects due to the main background for this final state, i.e. Z+jets Drell-Yan production. \label{fig:summary_ZV}}  
\end{figure}

Figure~\ref{fig:Combined_LL_Profiles_15} shows $-2\Delta \log \mathcal{L}$, profiled over the systematic nuisance parameter,
as a function of individual Wilson coefficients, for each final state and for their combination (excluding ZV+2j). 
The 68\% and 95\% c.l. intervals obtained are reported in Figure~\ref{fig:summary_2}
for the leptonic channels, 
and in Figure~\ref{fig:summary_ZV} for the semileptonic ZV+2j final state.
In the latter two figures,
the bands show the limits obtained with the baseline analysis,
while the thin lines correspond to those obtained
when neglecting all quadratic terms in  Eq.~\eqref{eq:2p2N}.
Note that the horizontal scale is not linear,
in order to better visualise very different ranges in the image.

\paragraph{Optimal observables.}
As explained in~\ref{sec:likelihood}, the likelihoods are constructed differently for each Wilson coefficient, by choosing optimal differential observables. For the 1D fits, the list of most sensitive observables is reported in Table~\ref{tab:sensitivity_ranking}.

Focusing on quadratic fits, for the four-quark operators,
the most sensitive variables are found to be
the transverse momentum of the first and second leading quarks $p_{T,j^1},\, p_{T,j^2}$. This is expected, as these operators intuitively only affect the kinematics of the tagging jets.
On the other hand, for the bosonic parameters $c_{HW}, c_{H\square},c_W$, that mainly enter the partonic scattering process,
the variables that play the most important role are related to the kinematic properties of the leptonic final state.
Transverse momentum variables related to the hadronic decay of the vector boson are the most sensitive to bosonic operators in the ZV+2j channel.

\paragraph{Impact of individual processes.}
As expected, the inclusive WW 
gives the strongest constraints for most operators. Nevertheless,
the VBS topology is found to be generally competitive. In the cases of four-quark and Higgs ($c_{HW},\,c_{H\square}$) operators, VBS is by construction the only constraining channel. In addition, as discussed in Sec.~\ref{sec:analyses}, OSWW+2j, ZZ+2j and WZ+2j are particularly competitive for $c_{Hl}^{(1)},c_{HD}$ and $c_{HWB}$.
Globally, among the leptonic VBS processes, the most sensitive to EFT effects are SSWW+2j and OSWW+2j. This is mainly due to their cross-sections being larger than for WZ+2j and ZZ+2j, that are suppressed by the $Z\to\ell\ell$ branching ratio.  
Constraints obtained from the ZV+2j final state
are also very competitive with the inclusive diboson ones,
justifying the interest in this channel for a EFT analysis,
and will deserve a more detailed study that includes
also the backgrounds due to the production of a single vector boson plus jets.

\paragraph{Results by operator.}
In the combined analysis, the most constrained coefficients are the four-quark interactions $c_{qq}^{(3)}, c_{qq}^{(3,1)}, c_{qq}^{(1)}, c_{qq}^{(1,1)}$, the parameter $c_W$, that modifies TGC and QGC in a highly momentum-enhanced way, and the coefficients $c_{Hq}^{(1)}$, $c_{Hq}^{(3)}$, $c_{Hl}^{(3)}$, $c_{ll}^{(1)}$. All these are bound to be below 0.15 at 68\% c.l. for $\Lambda=\unit[1]{TeV}$. This result is roughly consistent with the findings of Ref.~\cite{Gomez-Ambrosio:2018pnl} for the ZZ+2j case. 

The four-quark operators differ among them by the SU(2) and flavour structures. In particular $c_{qq}^{(3)}, c_{qq}^{(3,1)}$ enter charged-current interactions such as $(\bar u d)(\bar s c)$, which enhances their impact compared to $c_{qq}^{(1)}, c_{qq}^{(1,1)}$. This difference is particularly marked for the SSWW+2j and OSWW+2j channels, where the charged component is dominant, see Fig~\ref{fig:Combined_LL_Profiles_15}. On the other hand, the flavour structure has a subdominant impact on the constraints: a difference is only visible between $c_{qq}^{(3)}$ and $c_{qq}^{(3,1)}$, and again dominated by SSWW+2j.
 
The coefficients $c_{W}, c_{Hq}^{(1)}, c_{Hq}^{(3)}, c_{Hl}^{(3)}$ induce large effects to all the VBS channels and also to WW, with significant variations in the distributions shapes. They are all dominantly constrained by diboson, where  $c_{Hq}^{(1),(3)}$ give particularly momentum-enhanced signals in the longitudinally-polarized component and $c_W$ in the transverse one~\cite{Falkowski:2016cxu}.  

$c_{ll}^{(1)}$ only enters via corrections to the EW input quantities. Its dominant effect is a rescaling of the overall cross sections, with nearly no shape modification in the distributions. The strongest bounds on this operator come again from diboson.  $c_{HD}, c_{HWB}$ also dominantly enter via input corrections and lead to qualitatively similar effects, but their impact in the inclusive WW and SSWW+2j is smaller compared to $c_{ll}^{(1)}$, resulting in weaker constraints. 

The coefficients $c_{H\square}$ and $c_{HW}$ only affect HVV couplings and, among the processes considered here, they are dominantly constrained by OSWW+2j, which is the one with the largest cross-section with the Higgs entering in the $s$-channel.  These constraints are of course weaker than those imposed by available Higgs production and decay measurements.

Finally, the coefficient $c_{Hl}^{(1)}$ is significantly constrained in the ZZ+2j and WZ+2j VBS, which only give mild bounds. 

\paragraph{Linear vs quadratic EFT parameterization.}
As we consider kinematic distributions extending to high energies, the validity of the EFT expansion cannot be guaranteed \emph{a priori} over the entire parameter and phase spaces explored. 
Rather than implementing unitarisation or clipping procedures on the simulated events, that could introduce a dependence of the results on the specifics of these techniques, we provide a simple, qualitative assessment of the EFT validity by performing a comparison of the limits obtained retaining vs. neglecting quadratic SMEFT contributions (and omitting propagator corrections in both cases). Although exceptions are possible, a dominance of quadratic terms in the fit typically points to a breakdown of the EFT expansion, and indicates a potential sensitivity to neglected higher-dimensional operators.
In Figs.~\ref{fig:summary_2},~\ref{fig:summary_ZV}, one observes that quadratic terms significantly impact the combined results for less than a half of the operators, while, for the others, their inclusion has little consequence. 
Among the latter, $c_{Hq}^{(3)}$, $c_{Hl}^{(3)}$, $c_{ll}^{(1)}$ are dominantly constrained through their linear contributions in all the VBS processes as well as in the inclusive WW. This is consistent, for instance, with the corresponding plots in Figs.~\ref{fig:Distributions_inWW},~\ref{fig:Distributions_OSWW},~\ref{fig:Distributions_ZZ},~\ref{fig:Distributions_VZ}, which indicate that the linear contribution dominates in most bins  already for $c_\a/\Lambda^2=\unit[1]{TeV^{-2}}$.

In the case of $c_{Hl}^{(1)}$, $c_{HD}$, $c_{H\square}$, the sensitivity to quadratic terms varies between processes. However, it is very limited for the semileptonic channel and for the leptonic channels that dominate the combined constraints, i.e. WZ+2j for $c_{Hl}^{(1)}$ (see Fig.~\ref{fig:Distributions_WZ}) and OSWW+2j for  $c_{HD}$ and $c_{H\square}$ (see Fig.~\ref{fig:Distributions_OSWW}).
The case of $c_{HWB}$ is slightly different as several channels concur in constraining this operator. At quadratic level, the dominant constraint is WW. At linear level, this particular limit weakens significantly, but the effect is compensated by the constraints from ZZ+2j, WZ+2j and OSWW+2j, which become dominant and leave the final result nearly unchanged.

The four-quark Wilson coefficients $c_{qq}^{(3)}$ and $c_{qq}^{(3,1)}$ also show limited dependence on the inclusion of quadratic terms, while $c_{qq}^{(1)}, c_{qq}^{(1,1)}$ show the opposite behavior. All these operators are dominantly constrained by SSWW+2j, where they are measured in the $p_{T,j^i}$ distributions. As shown in Fig.~\ref{fig:Distributions_SSWW}, for $c_\a/\Lambda^2=\unit[1]{TeV^{-2}}$, the interference contributions for all four-quark operators are negative and close in size to the quadratic terms. 
Intuitively, for Wilson coefficients sufficiently small the interference becomes largest and drives the fit. This threshold lies within the ballpark of the 95\%~c.l. sensitivity, and the fit happens to be such that the upper bounds on $c_{qq}^{(3)}, c_{qq}^{(3,1)}$ are already within the linear dominance region, while those on $c_{qq}^{(1)}, c_{qq}^{(1,1)}$ are not.
This effect is also responsible for the asymmetry of the constraints on these operators, seen in Figs.~\ref{fig:Combined_LL_Profiles_15},~\ref{fig:summary_2}: the bound for $c_\a>0$ is weaker because it lies within a region where large cancellations take place between interference and quadratic terms. Indeed, neglecting the quadratic contribution spoils the cancellation and results in an improvement of this constraint, that is most visible for $c_{qq}^{(3,1)}$. 

Finally, the coefficients that show the largest differences in the constraints derived with and without the inclusion of quadratic terms are $c_{qq}^{(1)},c_{qq}^{(1,1)},c_W,c_{HW}$ and $c_{Hq}^{(1)}$. 
For all these, the difference is generally due to the quadratic contribution being significantly enhanced compared to the interference in the most relevant processes. This is can indeed be observed in the distribution plots shown in Appendix~\ref{app:distributions} (all the distributions can be found in \href{https://github.com/MultibosonEFTStudies/D6EFTPaperPlots/tree/master/Distributions}{\faGithub}).

\begin{figure}[tbp]
    \centering
    \includegraphics[width=0.72\textwidth]{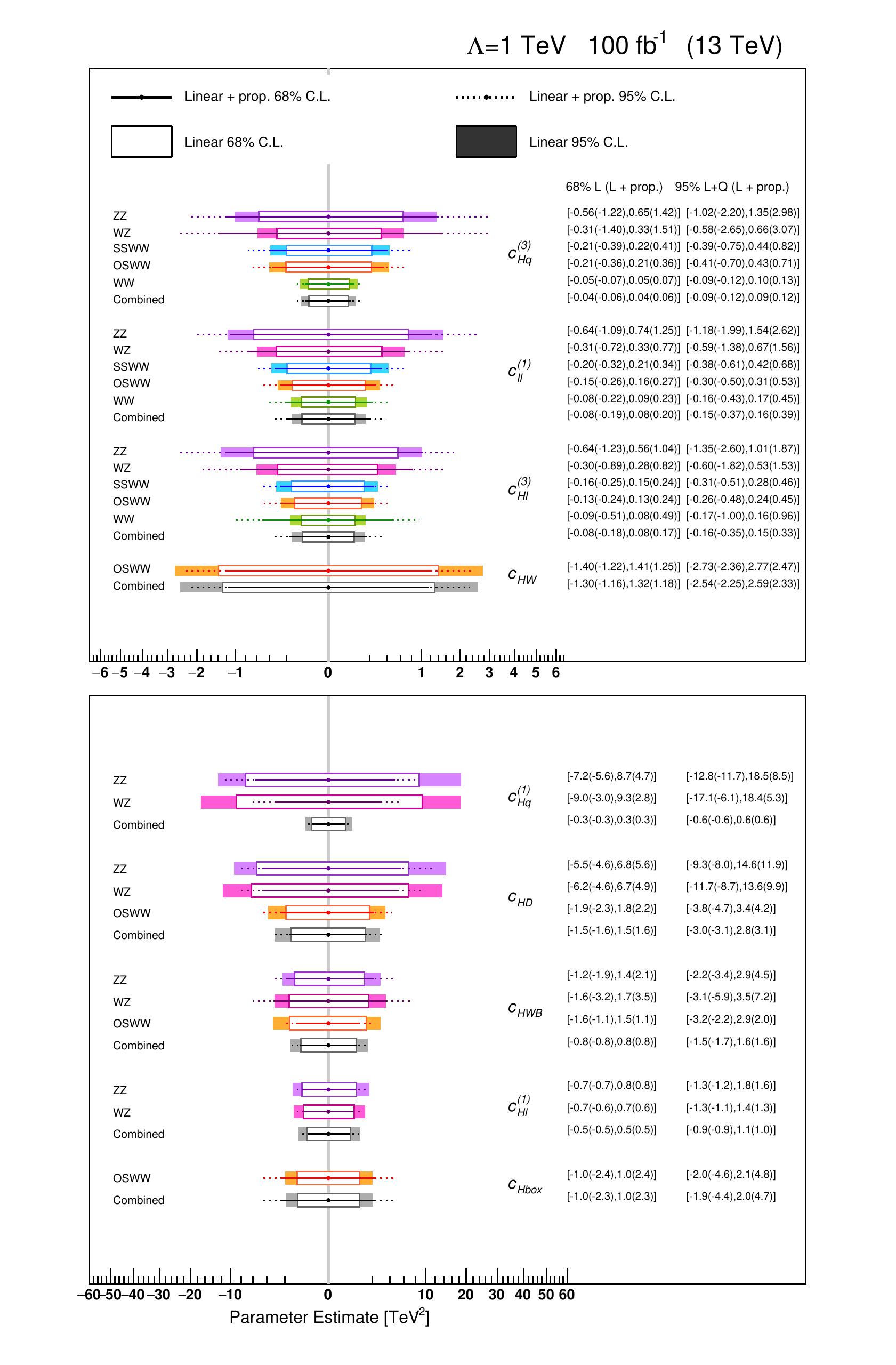}
    \caption{Impact of linear EFT contributions entering via corrections to the W, Z and Higgs propagators. Solid (dashed) lines indicate the 68\% (95\%) confidence level intervals obtained in individual fits, including corrections from both vertex and propagator insertions. Open (filled) boxes indicate 68\% (95\%) confidence level intervals obtained including only vertex contributions. The figure only shows Wilson coefficients and channels for which the propagator contribution is non-negligible. In all other cases, the results coincide with the linear-only limits shown in Fig.~\ref{fig:summary_2}. The combined constraints are computed with all channels, including those not shown. }
    \label{fig:propagators}
\end{figure}

\paragraph{Impact of SMEFT corrections in propagators.}\label{propagators}
As discussed in Sec.~\ref{sec.EFT}, SMEFT operators can contribute via both vertex and propagator corrections, due to the presence of EFT contributions to the total widths of the W, Z and Higgs bosons. 
Propagator corrections were omitted in the baseline results discussed so far, because currently available simulation tools only allow their consistent estimate at linear order in the Wilson coefficients. Therefore, we examine their impact within a purely linear setup: Figure~\ref{fig:propagators} shows a comparison of the results obtained in individual fits retaining vs. neglecting propagator contributions.
For consistency, the latter were added to both the EW and QCD-induced processes, and the two fits are performed using the same optimal variables in both cases.
In fact we have verified that the inclusion of propagator effects does not alter the variables sensitivity ranking.

The width corrections are only sizeable in the phase space region where the intermediate boson is approximately on-shell. For this reason, contributions from $\d\Gamma_H$ are very suppressed in all channels except OSWW+2j, and, in general, we observe that in each channel the most relevant propagator corrections are those associated to the reconstructed bosons.

The constraints that are most impacted by the inclusion of propagator corrections are those on $c_{Hq}^{(3)}, c_{ll}^{(1)}$, and $c_{Hl}^{(3)}$.  This is expected because these Wilson coefficients enter the total widths of both W and Z, and therefore  enter all diagrams. In addition they give the largest contributions to both $\delta\Gamma_W$ and $\delta\Gamma_Z$, see Eq.~\eqref{eq.dW_expressions}.\footnote{It is interesting to note that these 3 operators give nearly identical contributions to $\d\Gamma_W$ and $\d\Gamma_Z$, therefore they affect all the VBS channels considered in very similar ways. The net impact is typically more marked for ZZ+2j and WZ+2j compared to WW channels, simply because the $m_{ll}$ cut ensures a stronger dominance of the on-shell phase space in the signal region, which enhances the relative impact of propagator corrections.} 
These corrections always partially cancel against the vertex contributions, leading to a large worsening of the constraints for all channels, that reaches a factor 2 in most cases and even a factor 7.5 for $c_{Hl}^{(3)}$ in inclusive WW. It is important to underline that for such large cancellations to take place consistently across the entire fitted distribution, the vertex and propagator corrections need to have similar shapes. In the NWA, the latter amounts to an overall rescaling of the SM prediction in the resonant region. Therefore, cancellations can only happen for operators whose vertex corrections also "factor" into a SM rescaling when the intermediate boson is on-shell. This is indeed the case for $c_{Hq}^{(3)}, c_{ll}^{(1)}$, and $c_{Hl}^{(3)}$ in all processes under study.

In OSWW, the coefficients $c_{HW},\,c_{H\square},\,c_{HD}$ and $c_{HWB}$ enter dominantly through $\d\Gamma_H$. Among these, $c_{H\square}$ is the only one affected by the introduction of propagator corrections, as the bound on this coefficient worsens by about a factor 2. The reason can again be traced back to a large cancellation between vertex and propagator insertions, which does not take place for the other operators, since their vertex correction have very different shapes compared to the SM Higgs-mediated process: $c_{HW}$ induces a kinematic enhancement compared to the SM, and $c_{HD}$, $c_{HWB}$ enter several non-Higgs diagrams.

In ZZ+2j and WZ+2j, the operators $c_{Hq}^{(1)},\, c_{Hl}^{(1)},c_{HD}$ and $c_{HWB}$ enter exclusively through $\d\Gamma_Z$. The constraints on $c_{Hq}^{(1)}$ and $c_{HWB}$ change by over a factor of 2 in both processes, although in opposite directions. The constraints on $c_{HD}$ and $c_{Hl}^{(1)}$ vary much less significantly, because they only give smaller contributions to $\d\Gamma_Z$, see Eq.~\eqref{eq.dW_expressions}.

Among the combined results, only those on $c_{Hl}^{(3)}, c_{ll}^{(1)}$ and $C_{H\square}$ are affected significantly by the introduction of propagator corrections, as most variations observed in individual channels either compensate each other, or are over-ruled by constraints from processes that are insensitive to propagator corrections (e.g. the constraint on $c_{Hq}^{(1)}$ is still dominated by inclusive WW).

\paragraph{Impact of the QCD-induced sample.}\label{qcd_effect}

The QCD-induced processes are typically considered as a background to EW VBS analyses. Nevertheless, they are also generally modified in the presence of EFT operators. An interesting question is whether the dependence of the QCD background on the Wilson coefficients can have any significant impact on the constraints extracted from the statistical analysis. 
Here we compare the one-dimensional baseline results reported in Sec.~\ref{Individual_constraints} with those obtained fixing the QCD component to its SM shape.

\begin{figure}[tbp]
\begin{minipage}{0.73\textwidth}
\centering
\includegraphics[width=0.49\textwidth]{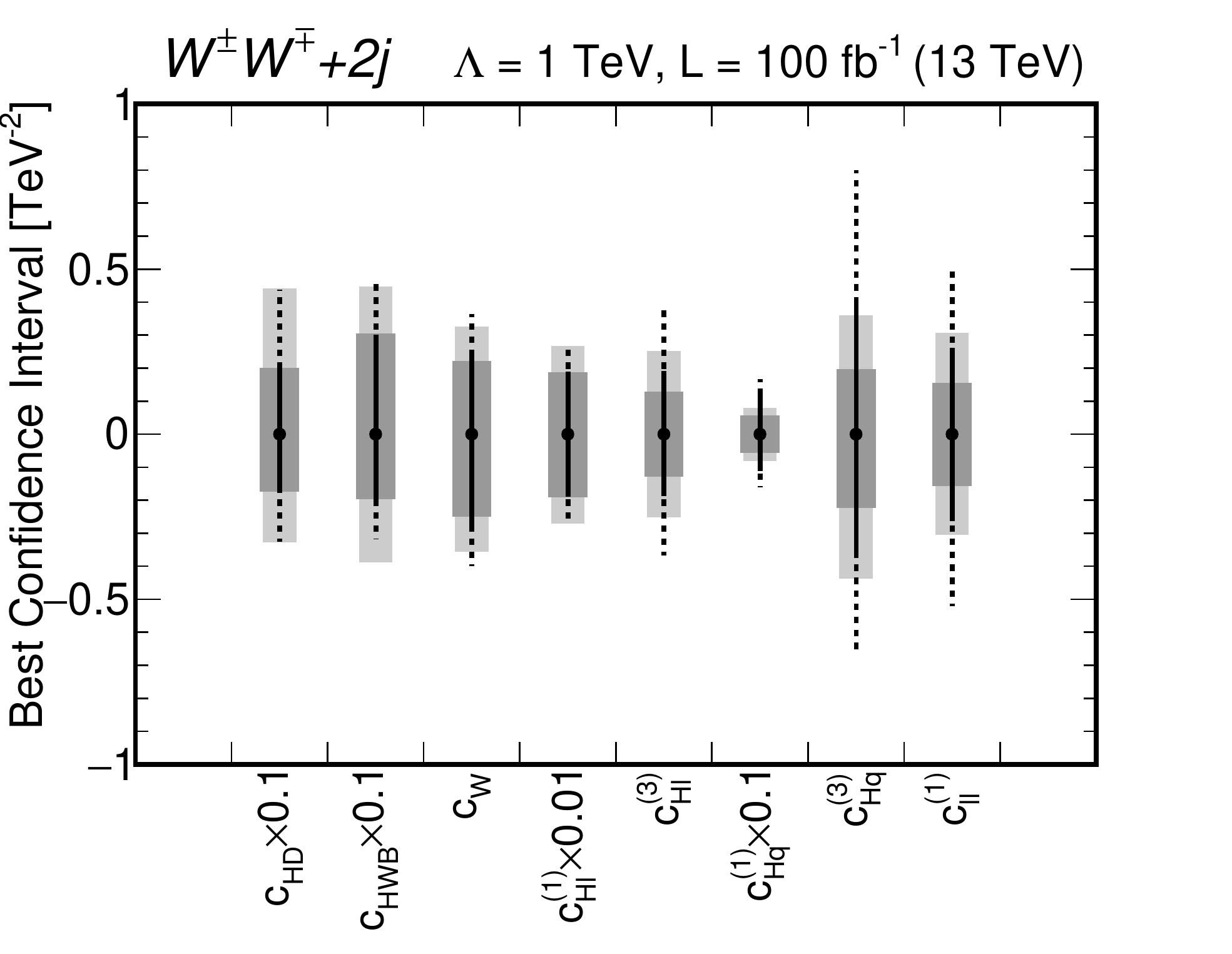}
\hfill
\includegraphics[width=0.49\textwidth]{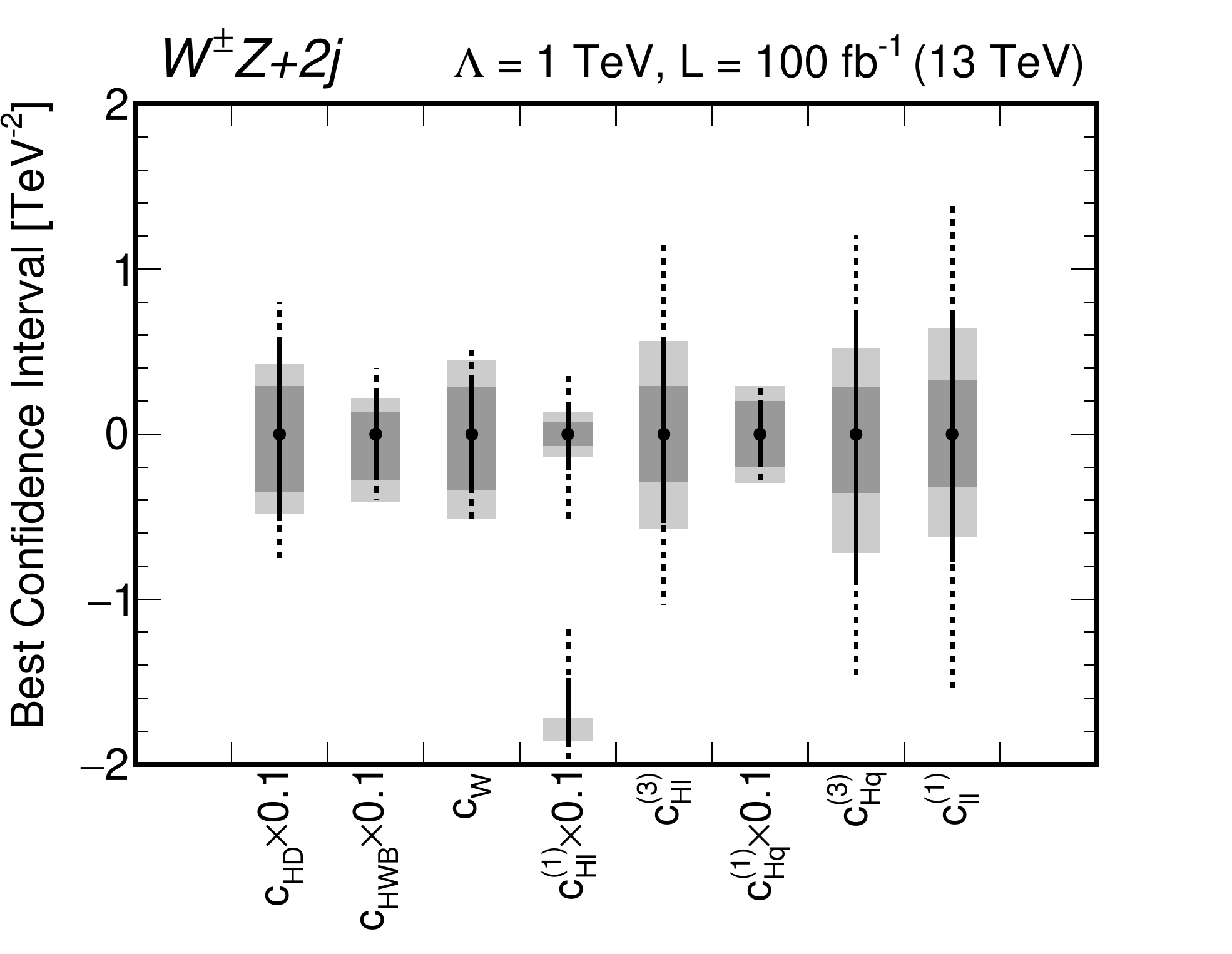}\\
\includegraphics[width=0.49\textwidth]{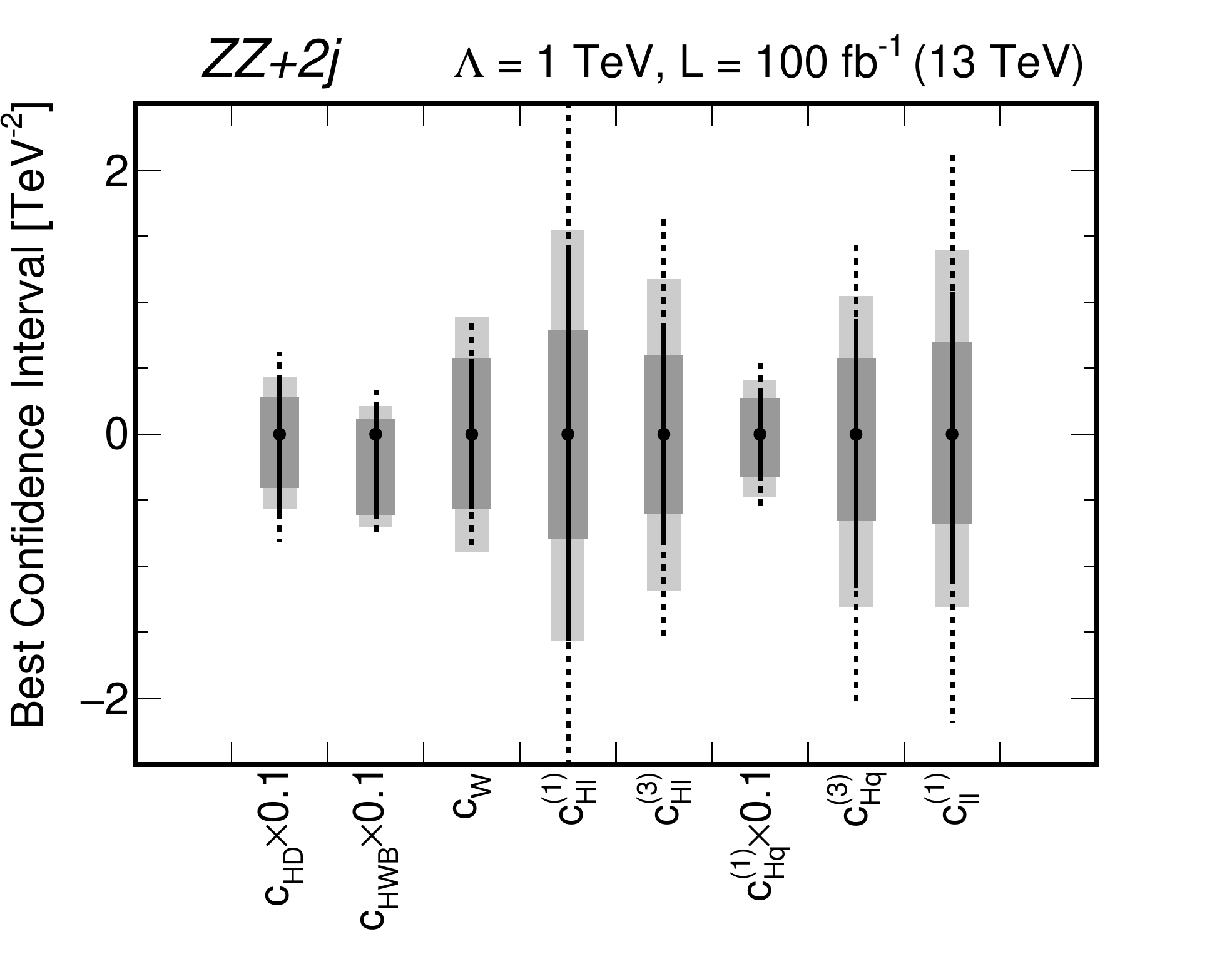}
\hfill
\includegraphics[width=0.49\textwidth]{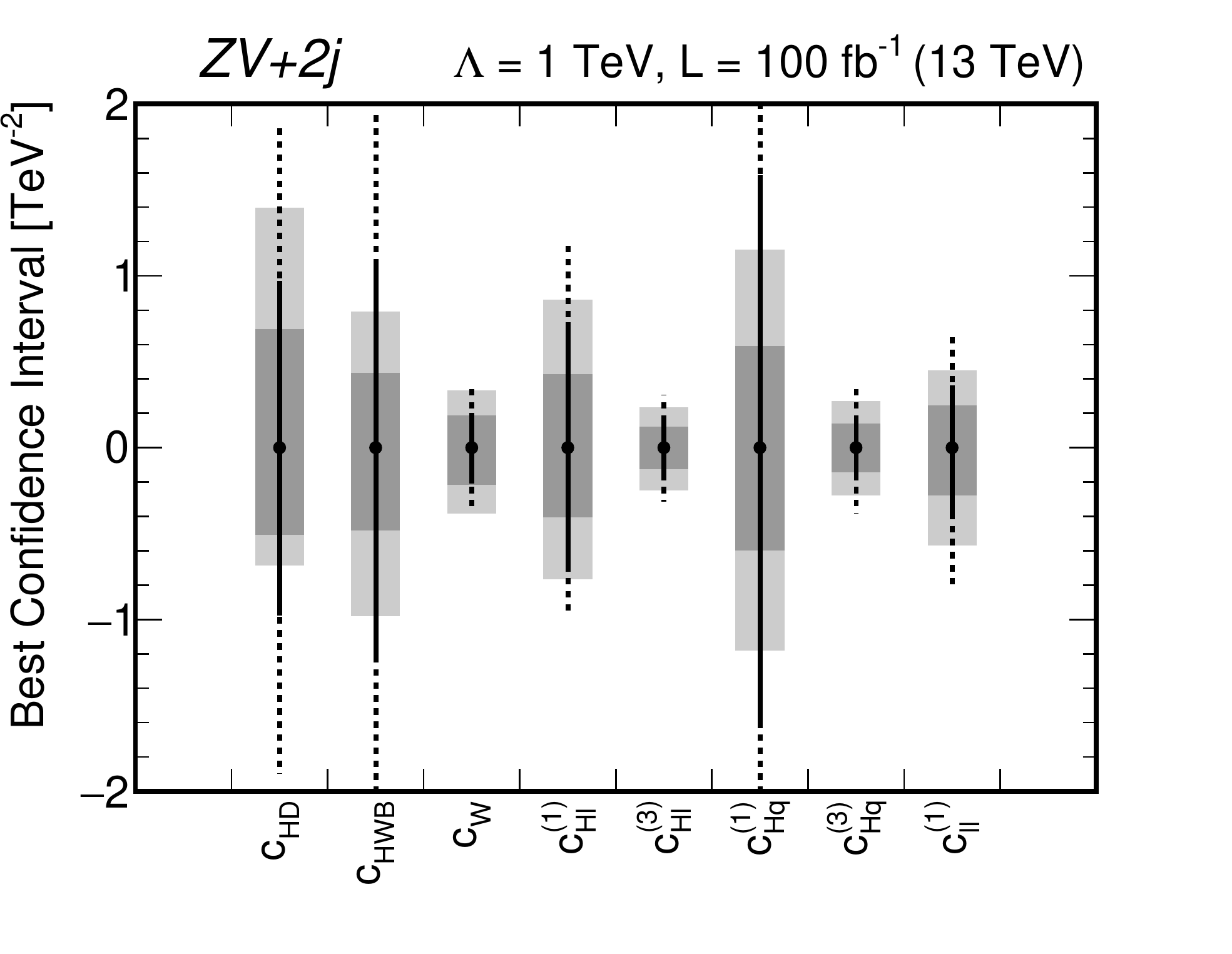}
\end{minipage}
\hfill
\begin{minipage}{0.26\textwidth}
\centering
\includegraphics[width=\textwidth]{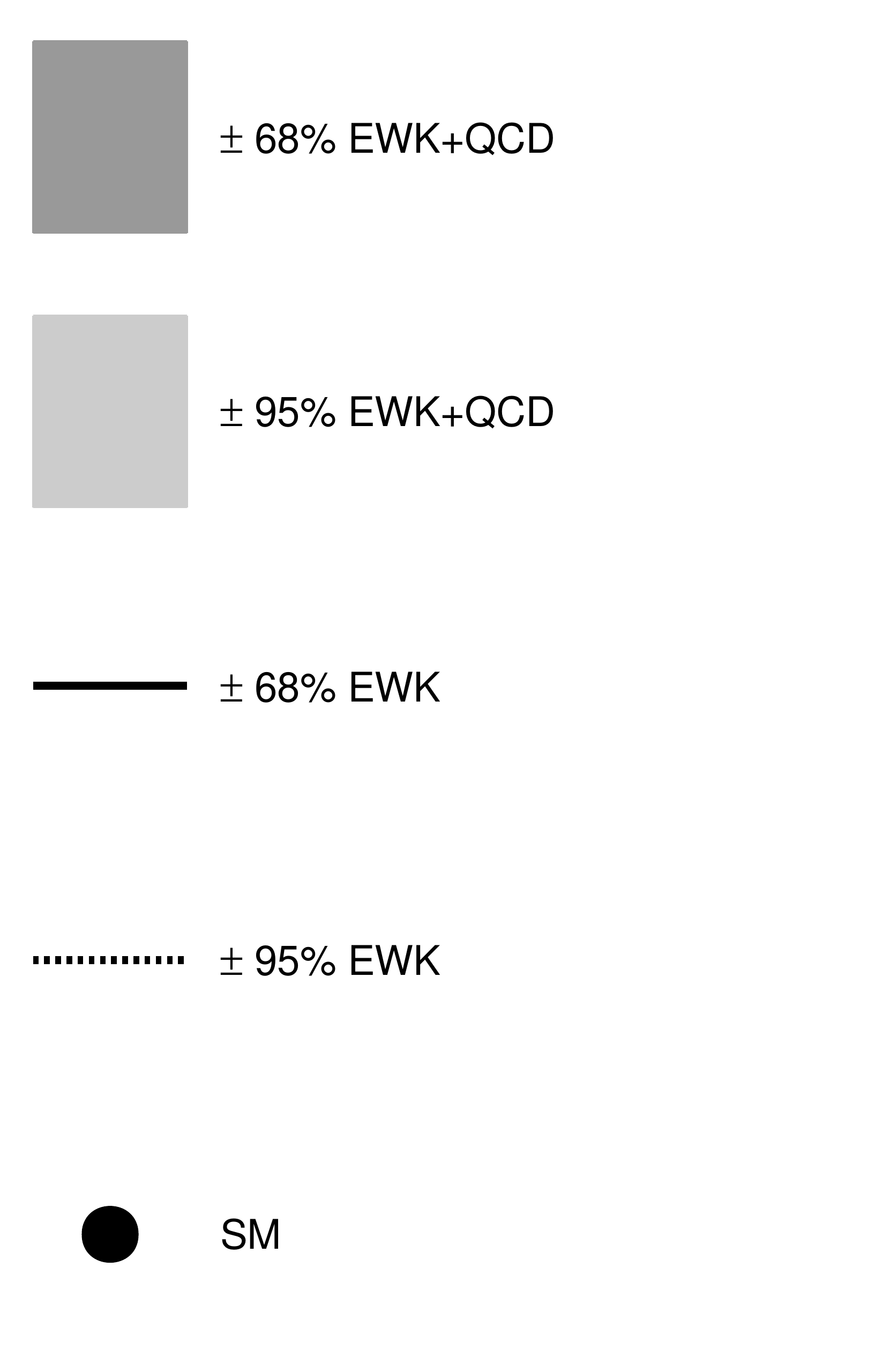}
\end{minipage}
\caption{ Impact of EFT corrections to the QCD backgrounds on the fit sensitivity, for the channels OSWW+2j (upper left), WZ+2j (upper right), ZZ+2j (lower left) and ZV+2j (lower right). The black solid (dashed) lines represent the 68\%~c.l. (95\%~c.l.) expected limits obtained neglecting QCD EFT contributions. The dark (light) grey bands represent the 68\%~c.l. (95\%~c.l.) limits including the QCD EFT dependence in the fits. The black points correspond to the SM expectation. \label{fig:qcdeffect}}  
\end{figure}

Figure~\ref{fig:qcdeffect} shows that for all operators and in all channels, the inclusion of the QCD EFT dependence never weakens the sensitivity: in some cases its impact is negligible, but in many others it leads to an improvement of the constraints by up to a factor of 2.
In first approximation, this behavior can be understood considering that a worsening of the constraints could only occur if (partial) cancellations between EW and QCD SMEFT contributions took place in all bins of a distribution. However, for the vast majority of the operators and observables considered, this is not the case.  

As a general rule, constraints for which the EW+QCD and EW results are very close, are numerically dominated by the EW-induced process alone. Among the constraints that show a significant improvement with the introduction of QCD, those on $c_{Hq}^{(3)}$ and $c_{Hq}^{(1)}$ in $W^\pm W^\mp+2j$ are strongly dominated by the pure QCD-induced contribution, while the others results from a non-trivial interplay between the EW and QCD bounds.
These conclusions were checked against fits to the QCD-induced components only.

\subsection{Individual vs. profiled bounds}\label{individual_vs_profiled}

When interpreting the EFT formalism as a low energy manifestation of a UV complete theory,
it has been shown that its matching to the SMEFT basis typically introduces more than one non-zero Wilson coefficient~\cite{de_Blas_2018,Jiang:2016czg}. The precise mapping of each subset of SMEFT operators to a UV complete model strictly depends on the model tested. In this context, upper bounds to the sensitivity can be estimated in the worst case scenario where all the operators under study are present simultaneously with non-zero Wilson coefficients.

We compare the results in Sec.~\ref{Individual_constraints} to those obtained allowing all 14 Wilson coefficients to float in the likelihood maximisation,
and profiling over all of them except the one of interest.
For each operator, 
the template analysis of the profiled constraint is carried out using, for each process, 
the same best variable obtained from individual constraints 
described in Section~\ref{Individual_constraints}.

\begin{figure}[t]
  \centering
  \includegraphics[width=\textwidth]{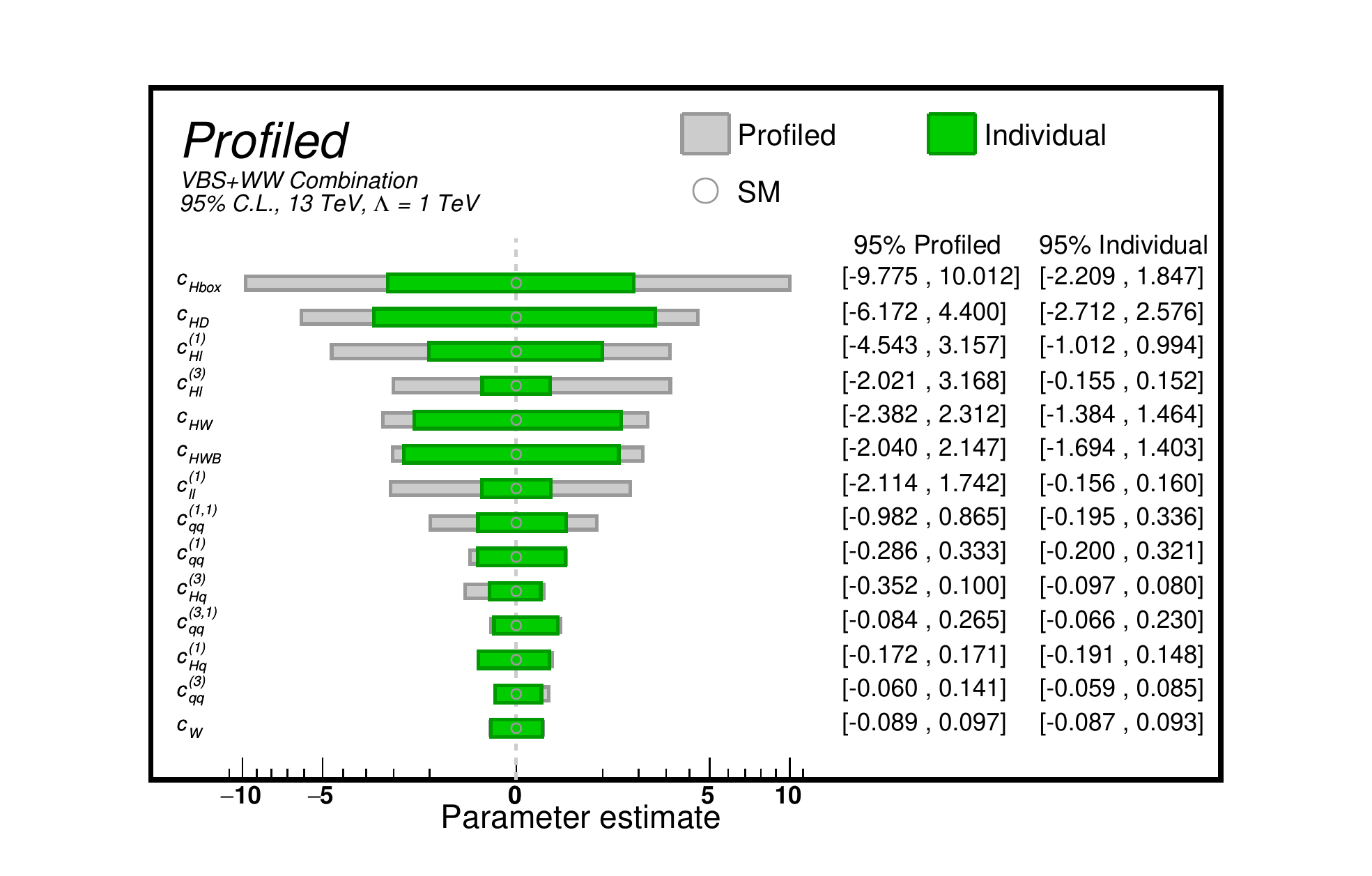}
  
  \caption{Sensitivity of the combined analysis of VBS SSWW+2j, OSWW+2j, WZ+2j, ZZ+2j and diboson WW to the dimension-six operators considered, when the remaining Wilson coefficients are set to zero (green) or profiled away (grey).   The QCD-induced EFT dependence was included where relevant. \label{fig:profiled}}  
\end{figure}

The comparison is shown in Figure~\ref{fig:profiled}. By definition, the profiled constraints are always equal to or worse than the individual ones. The differences observed between the two are quite heterogeneous, and they vary between a factor 1 (i.e. no difference) and 20. In particular, the constraints on $c_{W}$, $c_{Hq}^{(1)}$, $c_{qq}^{(3)}$, $c_{qq}^{(3,1)}$ are nearly unaffected by the introduction of extra degrees of freedom, suggesting that these directions are well-resolved already within the individual processes that dominate the bounds, i.e. inclusive WW and SSWW+2j for the first and last two respectively. 
The largest deterioration in the constraints is observed for $c_{ll}^{(1)}$ and $c_{Hl}^{(3)}$. This can be easily traced back to a combination of the corresponding Wilson coefficients, approximately close to $(c_{Hl}^{(3)}-c_{ll}^{(1)})$, remaining nearly unconstrained in the fit. This is clearly visible in Figs.~\ref{fig:2DCombined} and~\ref{fig:Global_2D_constraints} and further discussed below.

\subsection{Two-dimensional constraints}
\begin{figure}[t]
  \centering 
  \vspace*{-1.3cm}
  \includegraphics[width=.9\textwidth,trim=2.6cm 4mm 2.6cm 4mm,clip]{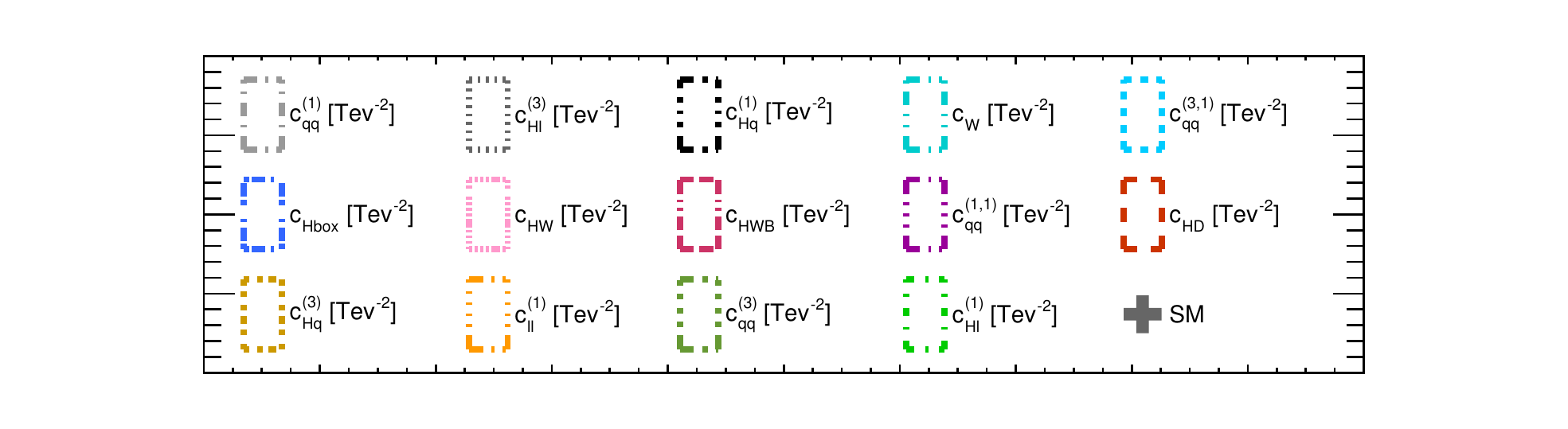}
  
  \includegraphics[width=.45\textwidth,trim=2mm 6mm 2cm 1.5cm,clip]{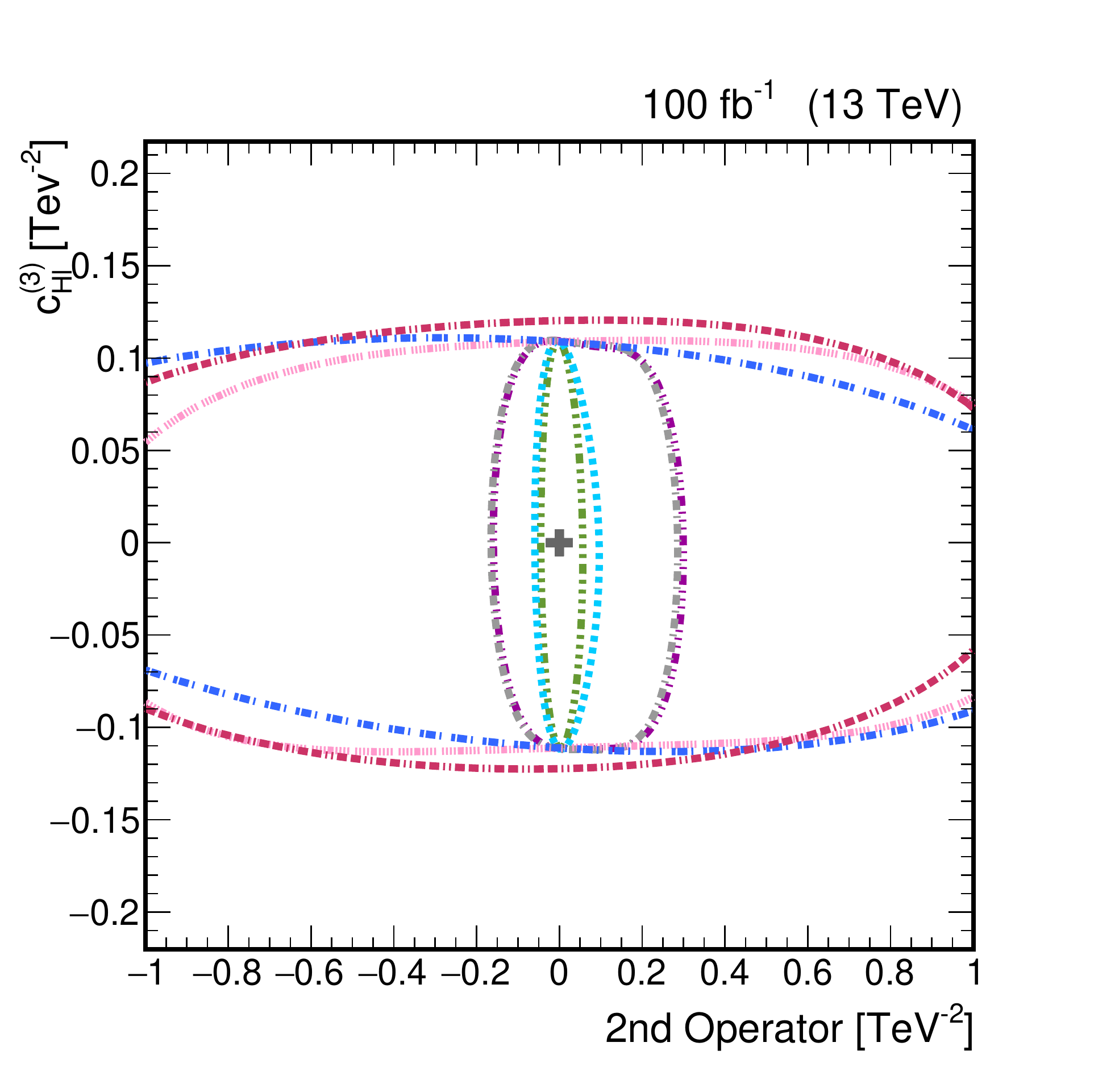}
  \includegraphics[width=.45\textwidth,trim=2mm 6mm 2cm 1.5cm,clip]{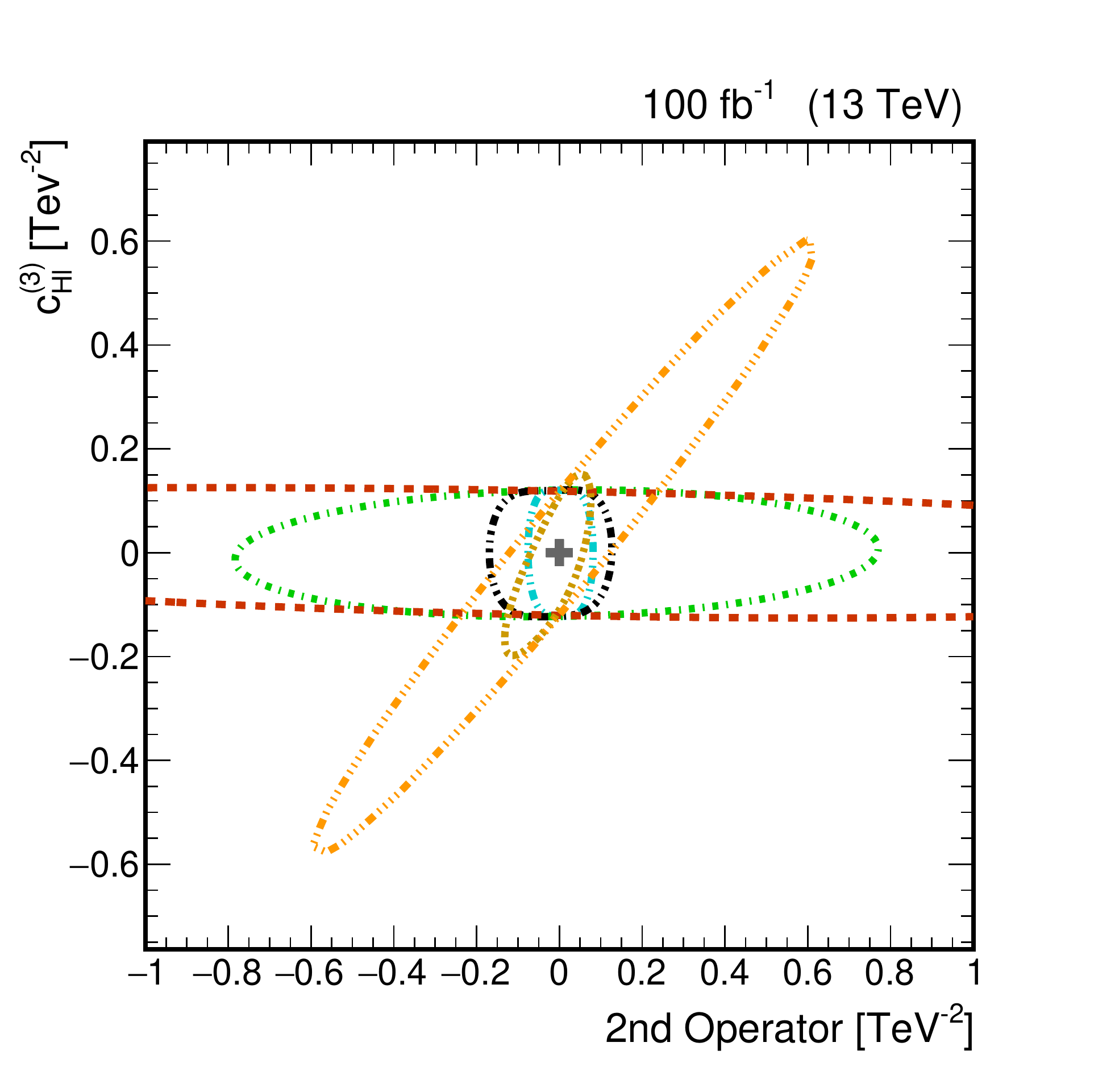}
  
  \includegraphics[width=.45\textwidth,trim=2mm 6mm 2cm 1.5cm,clip]{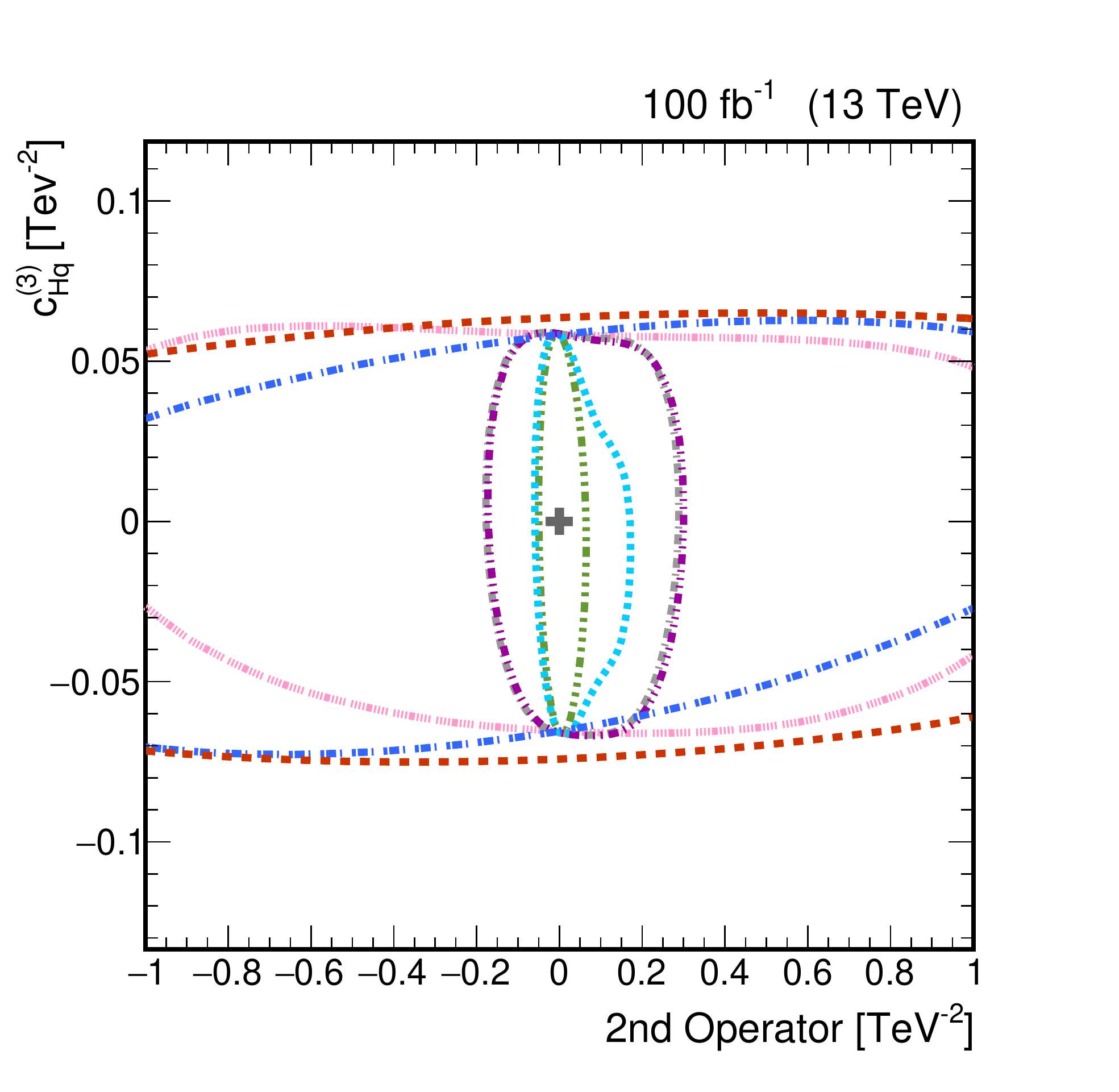}
  \includegraphics[width=.45\textwidth,trim=2mm 6mm 2cm 1.5cm,clip]{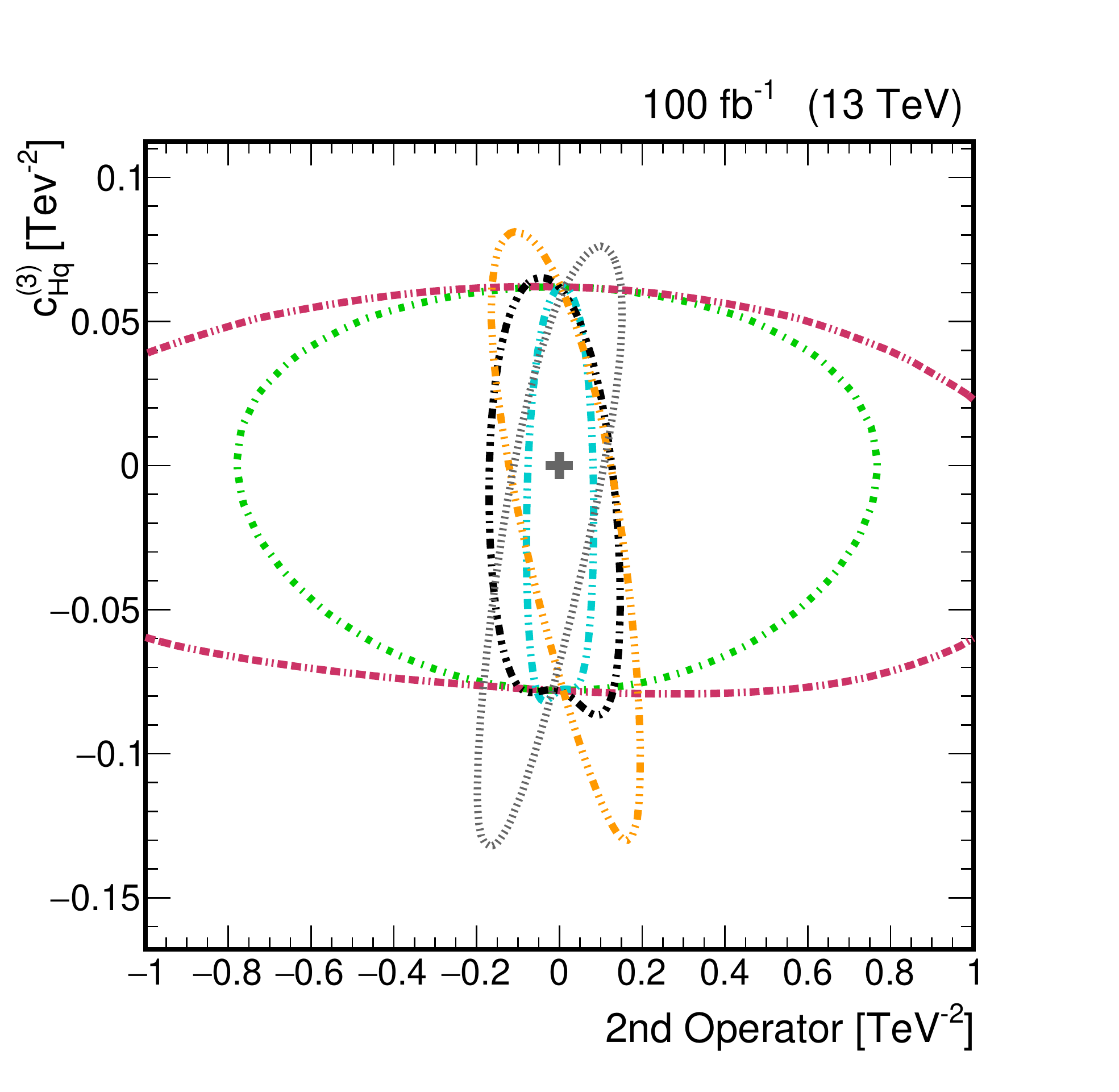}
  
  \caption{Bi-dimensional likelihood contours for $-2\Delta \log \mathcal{L} = 2.3$  for the combination of diboson and leptonic VBS channels, for all coefficient pairs involving $c_{Hq}^{(3)}$ or $c_{Hl}^{(3)}$. The VBS $W^{+}W^{-}+2j$, $W^{\pm}Z+2j$, $ZZ+2j$ channels include the respective QCD-induced processes. Only two Wilson coefficients are varied at a time, while the others are fixed to zero. Quadratic EFT contributions are included in all cases. In each panel, the operator on the $y$ axis is  fixed, while the one on the $x$ axis is different for each curve (see legend).    }  
  \label{fig:Global_2D_constraints}
\end{figure}

In this section we discuss constraints obtained allowing two operators to vary at the same time, fixing the remaining ones to zero. The analysis follows a strategy analogous to the one employed for the individual studies, as described in Sections~\ref{sec:analysis}~and~\ref{Individual_constraints}.
As above, EFT contributions to QCD-induced components of the VBS processes are accounted for in the fit, whenever pertinent. The list of optimal observables employed for each channel and operator pair is provided in Tables~\ref{tab.sensitivity_2d_ssWW}--\ref{tab.sensitivity_2d_inWW} in Appendix~\ref{app:ranking}.
Figures~\ref{fig:2DCombined} and~\ref{fig:Global_2D_constraints} show a subset of the likelihood scans obtained. The first figure illustrates the interplay between different processes for fixed operator pairs, while the second compares the combined sensitivity among all coefficient pairs containing $c_{Hq}^{(3)}$ and $c_{Hl}^{(3)}$.
The complete set of scans is available at the GitHub repository \href{https://github.com/MultibosonEFTStudies/D6EFTPaperPlots}{\faGithub}.

\begin{figure}[t]
  \centering 
  \vspace*{-1.3cm}
  \includegraphics[width=.40\textwidth,trim=3mm 6mm 2cm 1.5cm,clip]{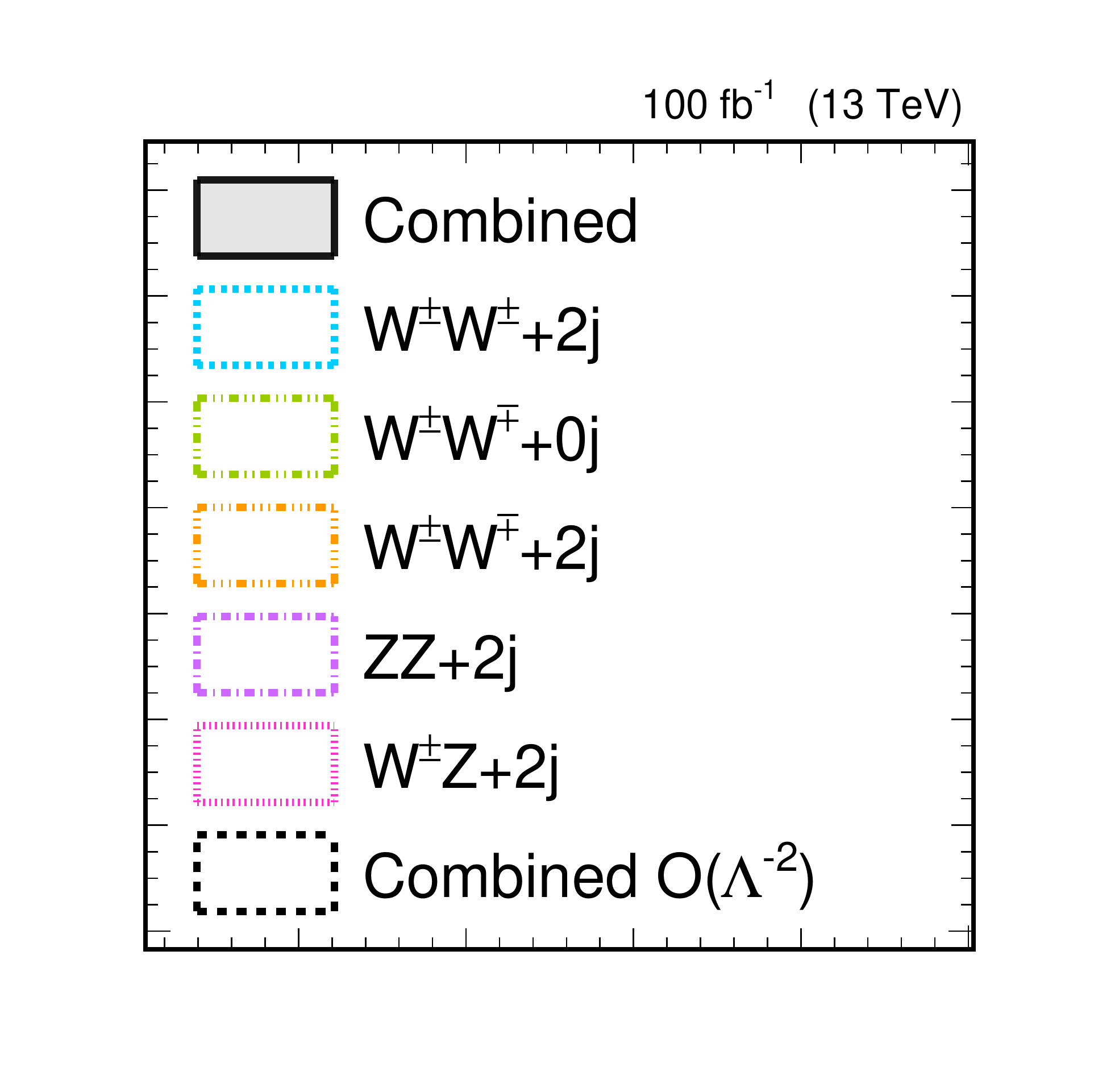}
  \includegraphics[width=.40\textwidth,trim=3mm 6mm 2cm 1.5cm,clip]{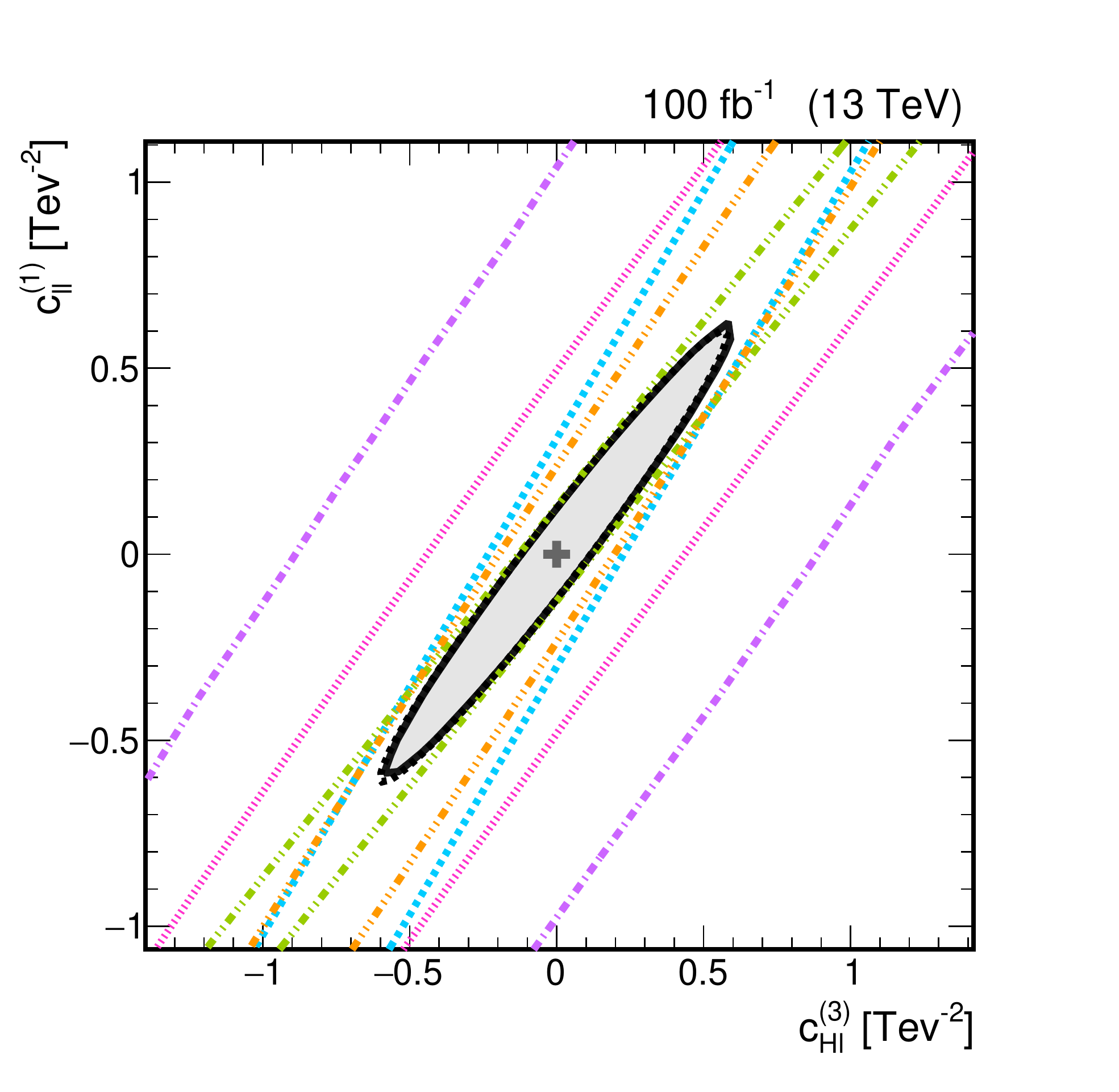}
 
  \includegraphics[width=.40\textwidth,trim=3mm 6mm 2cm 1.5cm,clip]{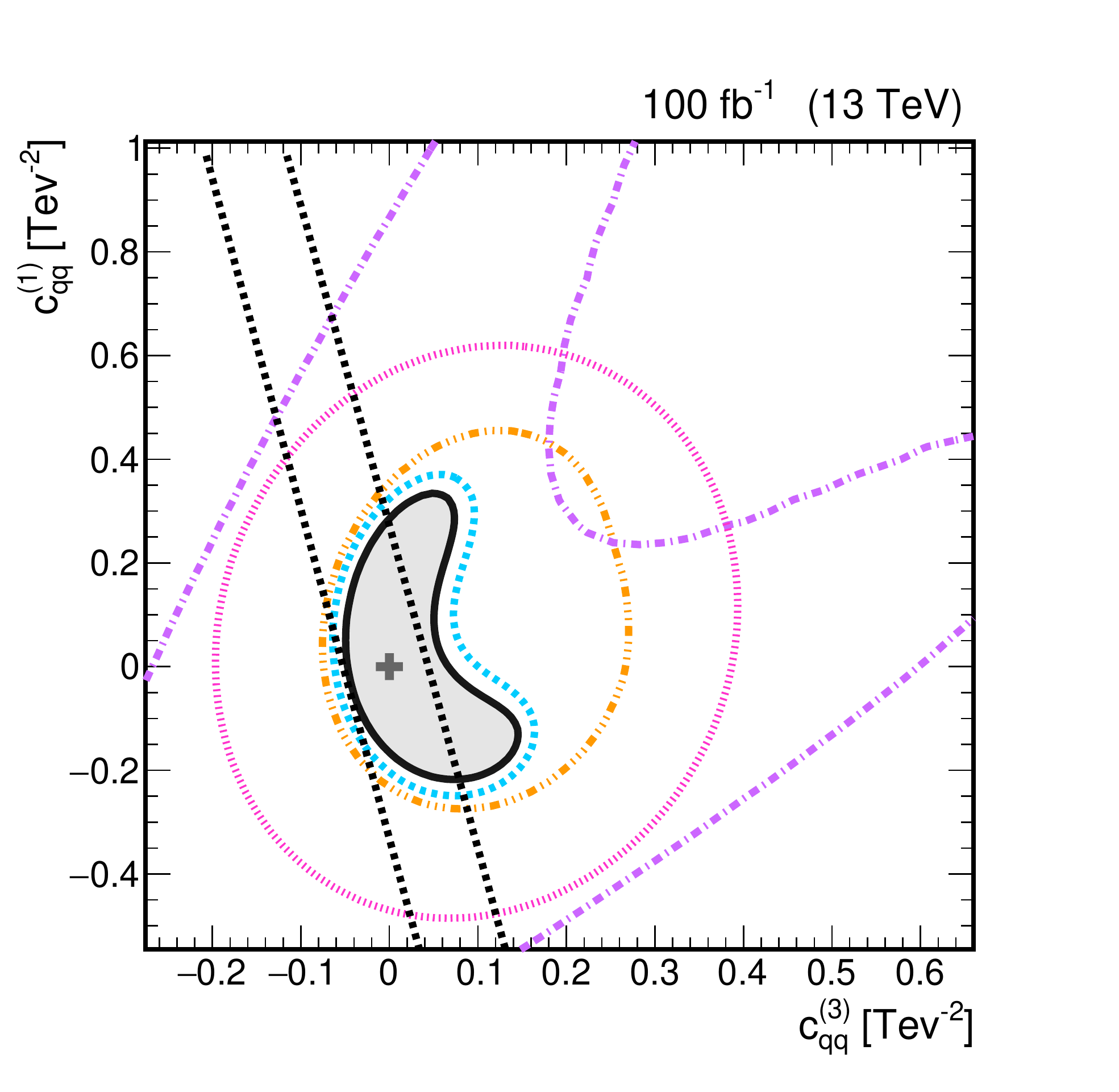}
  \includegraphics[width=.40\textwidth,trim=3mm 6mm 2cm 1.5cm,clip]{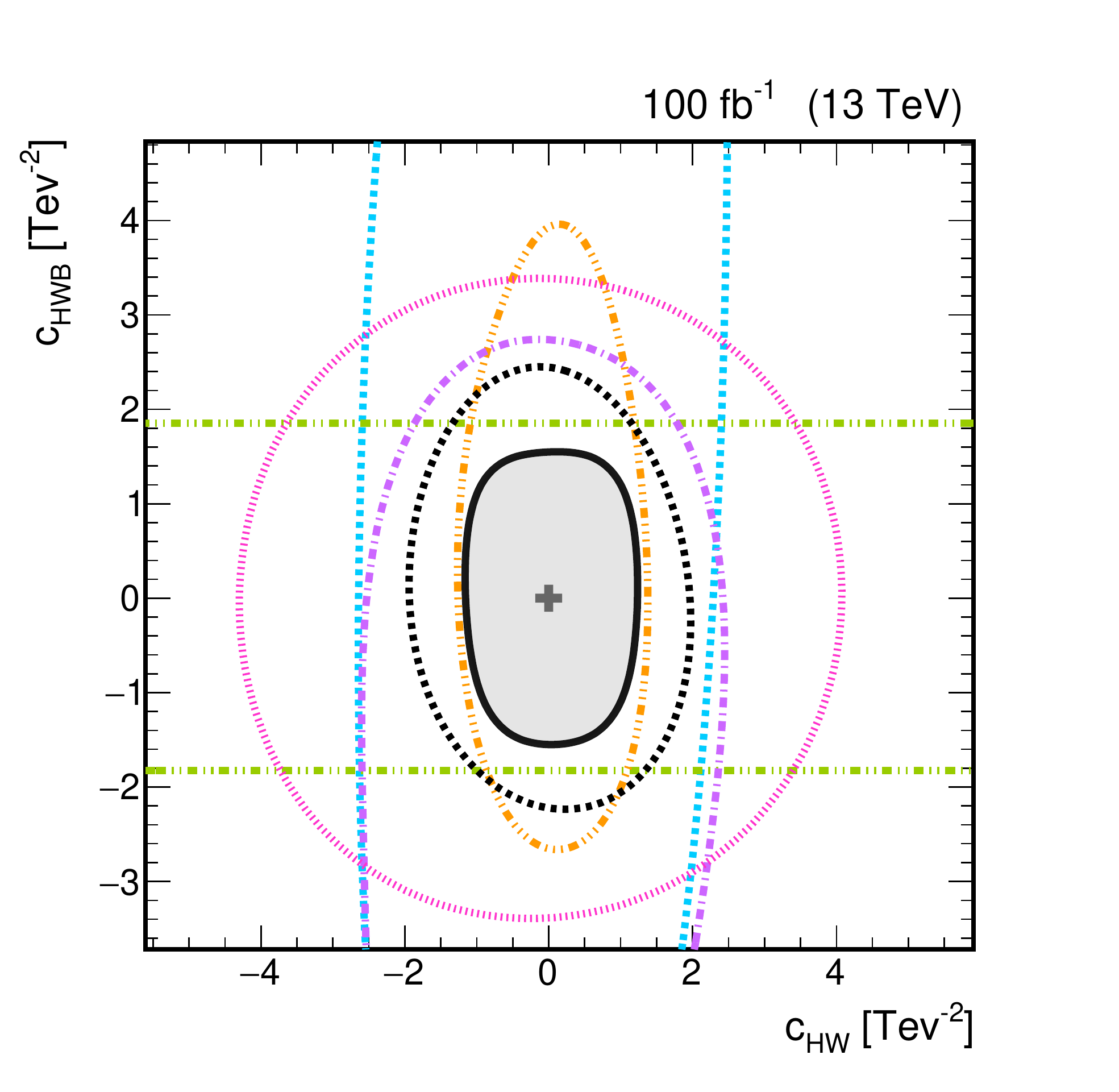}
  
  \includegraphics[width=.40\textwidth,trim=3mm 6mm 2cm 1.5cm,clip]{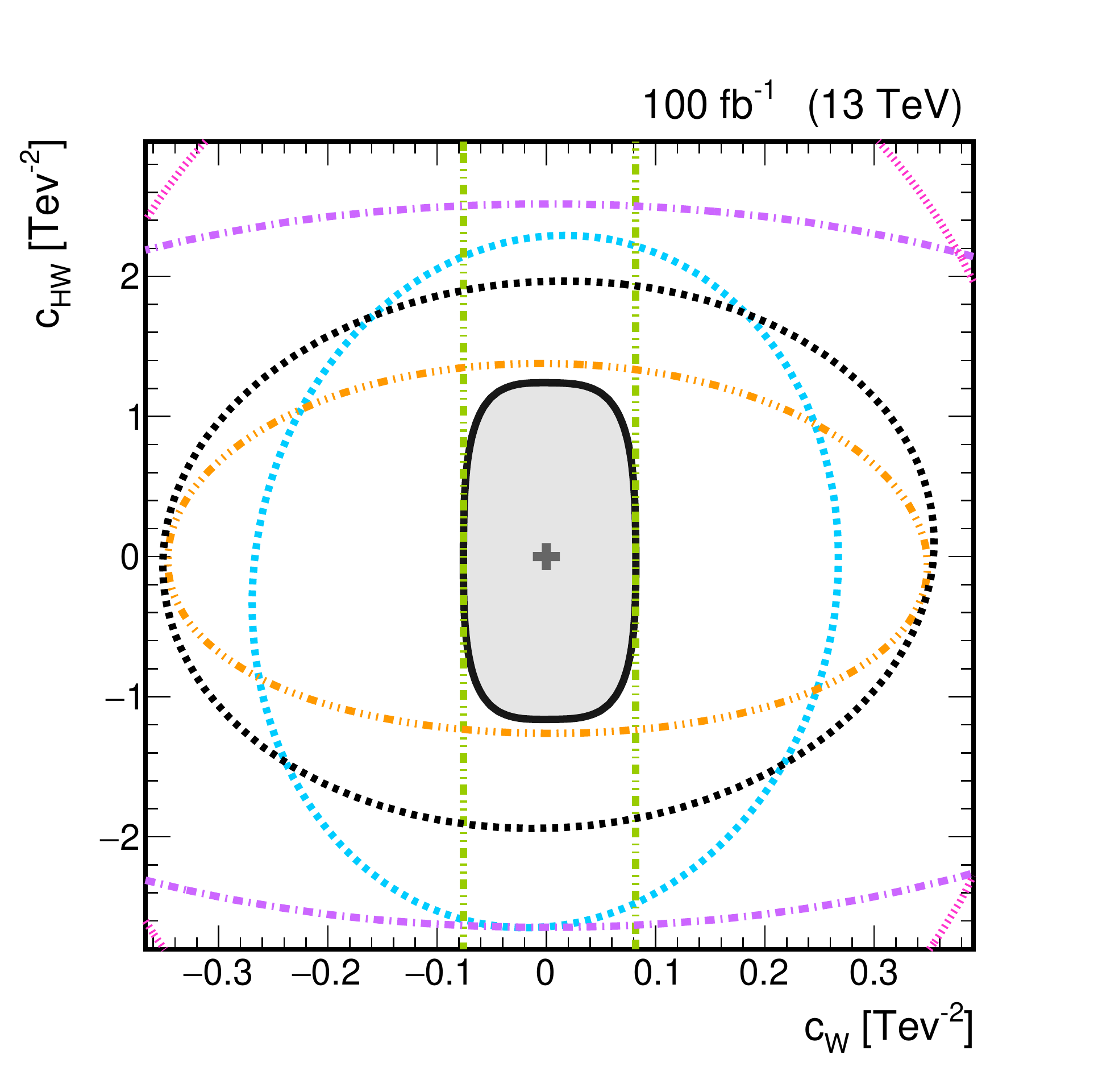}
  \includegraphics[width=.40\textwidth,trim=3mm 6mm 2cm 1.5cm,clip]{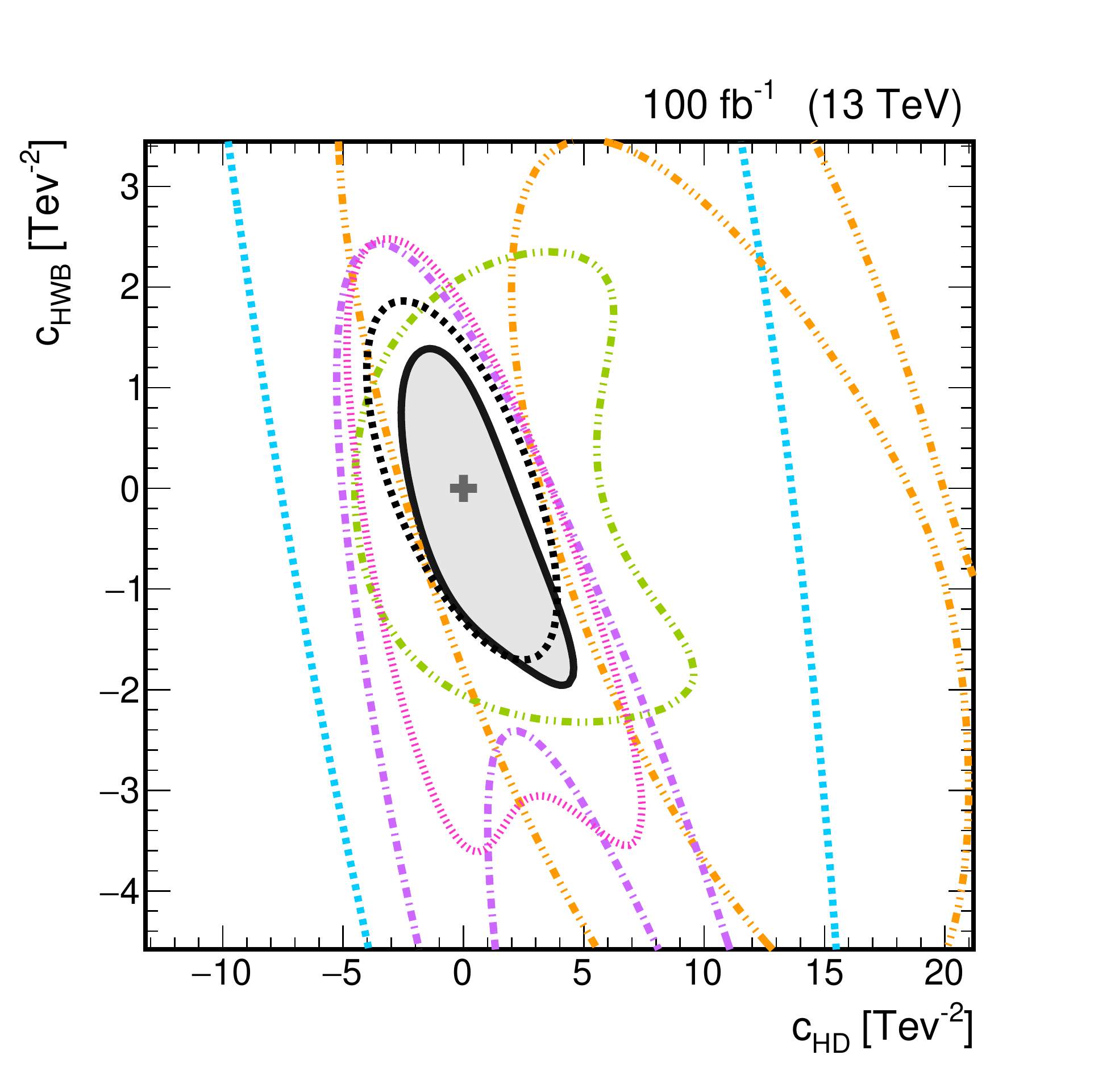}
  
  \caption{Bi-dimensional likelihood contours for $-2\Delta \log \mathcal{L} = 2.3$ (68\%c.l.),  for individual channels (in color) and for their combination (black). The VBS $W^{+}W^{-}+2j$, $W^{\pm}Z+2j$, $ZZ+2j$ channels include the respective QCD-induced processes. Only two Wilson coefficients are varied at a time, while the others are fixed to zero. Quadratic EFT contributions are included in all baseline cases. For comparison, the black dashed line shows the contour for the combined likelihood obtained truncating the EFT parametrization at the linear level. } \label{fig:2DCombined} 
\end{figure}

\paragraph{Interplay between measurements.}
For most operators, the profiled bounds are always dominated by a single process, that coincides with the leading constraint in the 1-dimensional fit. This is the case, for instance, of $c_{HW},c_{H\square}$ (OSWW+2j), $c_W,c_{Hq}^{(1)},c_{Hq}^{(3)}, c_{Hl}^{(3)}, c_{ll}^{(1)}$ (WW)\footnote{The constraints on $c_{Hl}^{(3)}$ and $c_{ll}^{(1)}$ are always dominated by diboson, except when the pair formed by these two operators is considered, as in this case an unconstrained direction emerges. See below. } and the 4-quark operators (SSWW+2j). Examples are shown in the $(c_W,c_{HW})$ and $(c_{HW},c_{HWB})$ panels in Fig.~\ref{fig:2DCombined}, where the diboson and VBS constraints are nearly orthogonal to each other.
The interplay between different channels is found to be most relevant for $c_{HD}$ and $c_{HWB}$, see the corresponding panel in Fig.~\ref{fig:2DCombined}. All channels play a role in constraining these operators. The dominant bounds vary depending on the operator pair considered, although in most cases the constraints are led by WW and OSWW+2j. 

\paragraph{Resolving degeneracies.}
Studying bi-dimensional likelihoods can provide insights about how potential degeneracies between operators are resolved in the fit. These mainly arise in two cases: among four-quark operators in VBS and between $c_{Hl}^{(3)}$ and $c_{ll}^{(1)}$ in all processes.  

All 4-quark operator pairs exhibit an unconstrained direction in the linear case, which is removed by the introduction of quadratic terms, see e.g. the $(c_{qq}^{(3)},c_{qq}^{(1)})$ panel in Fig~\ref{fig:2DCombined}. Moreover, for all pairs, the combined constraint is dominated by the SSWW+2j channel alone, consistently with the discussion in Sec.~\ref{individual_vs_profiled}. 
The degeneracy of singlet vs triplet $SU(2)$ contractions (i.e. $c_{qq}^{(3),(3,1)}$ vs $c_{qq}^{(1),(1,1)}$) is generally better resolved compared to the degeneracy between different flavor contractions, and leads to stronger projected constraints. This is again consistent with the 1-dimensional result (Sec.~\ref{Individual_constraints}) and with the 2-dimensional fits shown in Fig.~\ref{fig:Global_2D_constraints} (left panels), where the curves for $c_{qq}^{(1)}$ and $c_{qq}^{(1,1)}$ are fully overlapping.

As shown in the corresponding panel of Fig.~\ref{fig:2DCombined}, all leptonic
processes exhibit a near-degeneracy between $c_{Hl}^{(3)}$ and $c_{ll}^{(1)}$ that leaves a direction close to the diagonal $c_{ll}^{(1)}\simeq c_{Hl}^{(3)}$ essentially unconstrained within the fit range. In the combined fit, this is not resolved by the inclusion of quadratic terms, but rather thanks to the unconstrained directions of the various processes having slightly different slopes.
Specifically, the combined constraint is dominated by the interplay of inclusive WW and SSWW+2j, whose
scattering amplitudes scale with  $(c_{Hl}^{(3)}-c_{ll}^{(1)})$  and  $(4c_{Hl}^{(3)}-3c_{ll}^{(1)})$  respectively. 
This dependence follows from the interplay between corrections to vertices entering the $WW$ production processes, that scale with $\Delta G_F=2c_{Hl}^{(3)}-c_{ll}^{(1)}$, and corrections to $Wl\nu$ vertices, that scale with $c_{ll}^{(1)}$ only.

\paragraph{Linear vs quadratic EFT parameterization.}

Analogously to the individual limits above, 
we compare the combined results obtained with a linear or quadratic EFT parameterisation (neglecting all propagator contributions). 
In Fig.~\ref{fig:2DCombined}, these are indicated respectively by dashed and solid black lines.
We find that the impact of quadratic terms on our fit is generally sizeable for all the processes considered. 

Examining the bounds obtained profiling over one of the two parameters in each 2D fit, the behavior observed in the 2D fits is qualitatively consistent with the one observed in the individual case.
The Wilson coefficients whose constraints vary the most depending on whether or not the quadratics are retained are $c_W$ and $c_{Hq}^{(1)}$. The 4-quark operators are most sensitive to the quadratic terms in 2D fits where they are paired with one another, due to the presence of the unconstrained direction in the linear case. When paired with other coefficients, only $c_{qq}^{(3)}, c_{qq}^{(3,1)}$ show significant differences between the linear and quadratic fits. In this particular case, the upper bound \emph{worsens} when quadratics are introduced, because of the cancellation between linear and quadratic contribution discussed in Sec.~\ref{Individual_constraints}.
The coefficients $c_{Hl}^{(1)},c_{Hl}^{(3)},c_{Hq}^{(3)},c_{ll}^{(1)}$ are the least sensitive to the introduction of quadratics, as their projected bounds remain essentially unchanged in all 2D planes.  
For the remaining parameters ($c_{HW},c_{HD},c_{HWB},c_{H\square}$) the bounds obtained with and without quadratics generally differ, but the size of the variation depends on the operator pair considered.

At the bi-dimensional level, the introduction of quadratic terms generally leads to a deformation of the likelihood contours, such that in many cases one can identify either parameter space regions (sufficiently away from the SM point) that are 68\%~c.l. allowed in the quadratic fit but not in the linear one, and regions for which the opposite is true. Two examples are shown in the $(c_{qq}^{(3)},c_{qq}^{(1)})$ and $(c_{HD},c_{HWB})$ panels in Fig.~\ref{fig:2DCombined}.

\subsection{Projected constraints for LHC Run III and HL-LHC}\label{Projection}

The LHC Run III is expected to deliver roughly $\unit[200]{fb^{-1}}$ after 3 years of activity. 
Combined with the Run I and Run II statistics, 
the data set will then amount to more than $\unit[300]{fb^{-1}}$.
The Run III will be followed by a long shut-down of the accelerator 
in order to prepare the machine for its high luminosity phase (HL-LHC). 
The instantaneous luminosity will be increased up to about $\unit[5-7.5 \times 10^{34}]{cm^{-2}s^{-1}}$,
allowing the LHC to deliver approximately $\unit[3000]{fb^{-1}}$ in 10 years of data taking.
This section presents a sensitivity projection to these two scenarios 
(keeping the centre-of-mass energy of the proton collisions at $\unit[13]{TeV}$ in both cases).

\begin{figure}[t]
  \centering 
  \includegraphics[width=\textwidth,trim={1cm 1mm 1cm 1cm},clip]{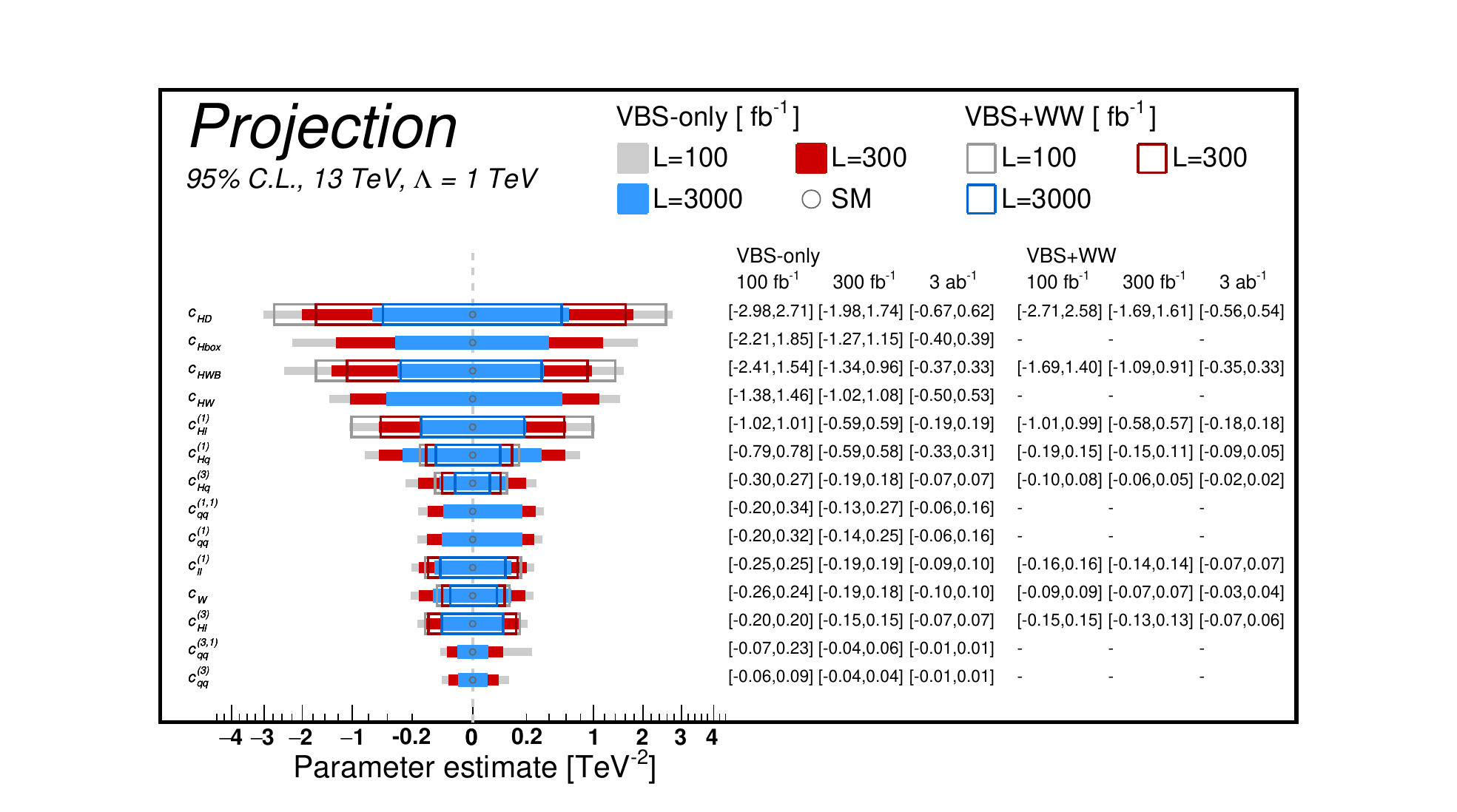}

  \caption{Expected $95\%$~c.l. constraints on individual SMEFT operators for an integrated luminosity of $\unit[100]{fb^{-1}}$ (grey), $\unit[300]{fb^{-1}}$ (red) and $\unit[3]{ab^{-1}}$ (blue). Constraints expected from the combination of the VBS channels  SSWW+2j, OSWW+2j, WZ+2j, ZZ+2j are depicted as filled colored boxes while the combination of VBS and WW channels as hollow boxes with a colored border line.
  EFT effects in the  QCD-induced processes are included whenever relevant. The result for VBS+WW is not indicated for operators that do not enter WW. In this case the limits from the VBS+WW combination would coincide with those from VBS only. }{\label{fig:projection}}  
\end{figure}

The projections are obtained simply by scaling the expected number of events computed for the individual constraints in Sec.~\ref{Individual_constraints} by a factor that accounts for the increase in luminosity.
No scaling of the constraint on the 2\% luminosity uncertainty is applied. 
Figure~\ref{fig:projection} shows the individual exclusion ranges at 95\%~c.l. 
expected for the VBS-only and for the VBS and inclusive WW combinations 
at  $\unit[100, 300, 3000]{fb^{-1}}$, 
including the relevant QCD-induced contributions and quadratic terms in the EFT predictions.

The projection study highlights that the VBS combination should be able to constrain at parton-level all the operators to less than $[-1,1]$ at $95\%$~c.l. by the end of the HL-LHC, 
while the inclusion of the inclusive WW channel further improves these bounds down to the $[-0.5,0.5]$ level, reaching a few percent for the most constrained operators.

\section{Summary and conclusions}\label{sec:conclusions}

We have presented a comprehensive study of the sensitivity of VBS and diboson measurements to dimension-six SMEFT operators.  The full set of plots for the differential distributions considered and likelihood scans performed are available at the dedicated GitHub repository~\href{https://github.com/MultibosonEFTStudies/D6EFTPaperPlots}{\tt github.com/MultibosonEFTStudies/D6EFTPaperPlots}.

We estimated the expected limits on 14 dimension-six operators, that are representative of all classes of SMEFT corrections entering VBS and diboson. The baseline constraints are estimated for LHC measurement with $\sqrt s=\unit[13]{TeV}$ and with an integrated luminosity of $\unit[100]{fb^{-1}}$, working at parton- and tree-level, and including terms that are quadratic in the Wilson coefficients.
We considered VBS processes as full $2\to 6$ scatterings, including non-resonant diagrams in the event generation. This allowed us to perform a consistent template analysis of a significant number of differential distributions, for variables typically exploited in realistic LHC analyses.
This study indicates that four-quark operators are mostly probed by the kinematics of the two tagging jets, while leptonic variables provide the best constraints on bosonic operators, see App.~\ref{app:ranking}. Interestingly, angular observables are often among the most sensitive ones, indicating that the phenomenology of SMEFT effects in VBS processes is quite rich, and it goes beyond enhancements in transverse momenta or invariant mass distributions.

Overall, we find that VBS processes are particularly sensitive to four-quark operators, that, individually, can be constrained even more strongly than operators affecting TGC and QGC. Operators correcting fermion-gauge interactions also give large effects to the VBS distributions, often with marked shape distortions. On the other hand, the sensitivity to operators modifying the Higgs-gauge couplings is poorer, due to Higgs contributions being suppressed by the selection cuts. 

The constraints were derived for one and for two Wilson coefficient at a time, and also for each coefficient after profiling over the other 14. The comparison between individual and profiled constraints has shown degradation in sensitivity up to  an order of magnitude. 
The results obtained in the 2D-scans highlight the complementarity between the different channels considered, as well as between VBS and diboson WW measurements. In most cases, combining all channels increases significantly the sensitivity to BSM effects. %
The impact of quadratic EFT contributions in the fit was also assessed, by comparing individual constraints obtained with and without retaining them in the EFT parameterization. We found that the inclusion of quadratic terms enhances considerably the constraining power of both the VBS-only and VBS + WW combinations for the dimension-six operators that induce the most energy-enhanced effects. On the other hand, for about a half of the operators considered, the linear contributions are dominant. 

Limited to the linear analysis, we also investigated the impact of SMEFT corrections stemming from operator insertions in the propagators of intermediate W, Z and H states. We found that, although these contributions can have a strong impact for individual channels, in the combined analysis their introduction leads to a mild worsening of the constraints for three operators only, namely $c_{ll}^{(1)}, c_{Hl}^{(3)}$ and $c_{H\square}$. The inclusion of propagator corrections in quadratic fits is not possible at present, due to ambiguities in the distinction between conventional quadratic terms and double insertions.

Where relevant, our analysis kept into account EFT corrections to the QCD-induced background processes at order $\alpha^{4}_{EW}\alpha^{2}_{QCD}$.
Interestingly, we find that the inclusion of EFT effects in the QCD-induced processes never weakens the constraints on individual Wilson coefficients. On the contrary, in several cases is enhances the sensitivity by up to a factor 2.

The sensitivity of the ZV+2j process with a semileptonic final state was studied separately, due to the lack of simulated Z+jets background events. The sensitivity of this process to BSM physics is largely unexplored in the literature, but was found to be competitive with that of inclusive diboson WW, at least at the level of individual constraints. Although this result should be taken as optimistic, due to the incomplete background treatment, it is certainly promising and worth further investigations in the future.

Finally, we derived projections for the individual constraints at the integrated luminosities expected for the end of the LHC Run III ($\unit[300]{fb^{-1}}$) and of the HL-LHC ($\unit[3]{ab^{-1}}$). In the latter scenario the VBS combination is expected to constrain all the operators in the $(-1,1)$ range at the 95\%~c.l. while the addition of the diboson WW channel improves the constraints to the $(-0.5,0.5)$ range.

Our study is a first step in the exploration of the potential of VBS measurements to constrain EFT effects. It can be improved and extended in several ways, for instance by including a larger set of dimension-six operators in the statistical analysis, or by fully accounting for the hadronisation and detector effects. A more refined analysis could also consider further (reducible) background sources, particularly for the semileptonic ZV+2j channel, where the Z+jets background was neglected altogether. One interesting question, that we leave for a future work, is whether the kinematic richness of VBS processes could help discriminating between different operators, for instance by exploiting angular information or by getting access to the polarisations of the weak bosons in final state.

\acknowledgments

The authors acknowledge the support of the CA16108 VBSCan COST Action.\\
A. Vagnerini is supported by the "Fondazione CRT" grant n. 2020.453.

\appendix
\section{Optimal observables}
\label{app:ranking}
The Tables in this Appendix indicate the optimal observable employed in the statistical analysis for each Wilson coefficient and VBS process. Table~\ref{tab:sensitivity_ranking} reports optimal observables for individual and profiled 1D fits, Tables~\ref{tab.sensitivity_2d_ssWW}--\ref{tab.sensitivity_2d_inWW} report those employed for 2D fits. 

\renewcommand{\arraystretch}{1.5} 
\begin{sidewaystable}[!htbp]
\begin{tabular}{l|*5{cc@{\hspace*{8mm}}}cc} \toprule
    \multirow{2}{*}[-0.5\dimexpr \aboverulesep + \belowrulesep + \cmidrulewidth]{Op.} & \multicolumn{2}{c}{{SSWW+2j}} &
    \multicolumn{2}{c}{{OSWW+2j}} & 
    \multicolumn{2}{c}{{WZ+2j}} & 
    \multicolumn{2}{c}{{ZZ+2j}} & 
    \multicolumn{2}{c}{{ZV+2j}} & 
    \multicolumn{2}{c}{{WW}}\\
    \cmidrule(lr){2-3} \cmidrule(lr){4-5} \cmidrule(lr){6-7} \cmidrule(lr){8-9} \cmidrule(lr){10-11} \cmidrule(lr){12-13}
    & L & L+Q & L & L+Q & L & L+Q & L & L+Q & L & L+Q & L & L+Q\\
    \midrule
   
    $c_{Hl}^{(1)}$  & - &  $m_{ll}$ & - & MET & $m_{ee}{}^{\dagger}$ & $m_{WZ}$ & $p_{T,e^{-}\mu^{-}}{}^{\dagger}$ & $p_{T,e^{-}\mu^{-}}{}^{\dagger}$ & $p_{T,j_1}^{V}$ & $p_{T,j_1}^{V}$ & $p_{T,l^{1}}$ & MET\\ 
   
    $c_{Hl}^{(3)}$  & $\Delta\eta_{jj}{}^{\dagger}$ & $\Delta\eta_{jj}{}^{\dagger}$ &  $m_{jj}{}^{\dagger}$ & $m_{jj}{}^{\dagger}$ & $m_{jj}{}^{\dagger}$ & $m_{jj}$   & $m_{jj}{}^{\dagger}$ & $m_{jj}{}^{\dagger}$ & $\Delta\eta_{jj}^{V}$ & $\Delta\eta_{jj}^{V}$ & $m_{ll}{}^{\dagger}$ & $m_{ll}{}^{\dagger}$\\ 

    $c_{Hq}^{(1)}$  &  $p_{T,j^{1}}$ & $p_{T,j^{1}}$ & $m_{jj}$ & $m_{ll}$ & $m_{jj}$ & $p_{T,j^{1}}$ & $m_{jj}$ & $p_{T,j^{1}}$ & $m_{jj}^{VBS}$ & $m_{jj}^{VBS}$ & MET & MET\\ 
    
    $c_{Hq}^{(3)}$  &  $\Delta\phi_{jj}$ & $\Delta\phi_{jj}$ & $m_{ll}$ & $m_{ll}$ & $\Delta\phi_{jj}{}^{\dagger}$ & $p_{T,l^{1}}$ &  $\Delta\phi_{jj}{}^{\dagger}$  & $p_{T,l^{4}}$ & $p_{T,j_2}^{VBS}$ & $p_{T,j_2}^{VBS}$ & $p_{T,l^{1}}$ & $p_{T,l^{1}}$\\

     $c_{qq}^{(3)}$ & $m_{ll}{}^{\dagger}$ & $p_{T,j^{2}}$ & $m_{jj}$ & $p_{T,j^{2}}$ & $m_{jj}$ & $p_{T,j^{2}}$ & $m_{jj}$  & $p_{T,j^{1}}$ & $p_{T,l^{1}}{}^{\dagger}$ & $\Delta\phi_{jj}^{VBS}$ & - & -\\
     
     $c_{qq}^{(3,1)}$  & $\Delta\phi_{jj}$ & $p_{T,j^{2}}$ & $m_{jj}$ & $p_{T,j^{2}}$ & $m_{jj}$ & $p_{T,j^{2}}$  & $m_{jj}$ & $p_{T,j^{1}}$ & $\Delta\eta_{jj}^{V}{}^{\dagger}$ & $\Delta\phi_{jj}^{VBS}$ & - & -\\
     
    $c_{qq}^{(1,1)}$  & $\Delta\phi_{jj}$ & $p_{T,j^{1}}$ & $p_{T,j^{2}}$ & $p_{T,j^{2}}$ & $p_{T,j^{2}}$ & $p_{T,j^{1}}$ & $p_{T,j^{2}}$ & $p_{T,j^{2}}$ & $\Delta\phi_{jj}^{VBS}$ & $p_{T,j_1}^{VBS}$ & - & -\\
    
    $c_{qq}^{(1)}$  & $p_{T,j^{1}}$ & $p_{T,j^{1}}$ & $p_{T,j^{2}}$ & $p_{T,j^{2}}$ & $p_{T,j^{2}}$ & $p_{T,j^{2}}$ & $p_{T,j^{2}}$ & $p_{T,j^{2}}$ & $\Delta\phi_{jj}^{VBS}$ & $p_{T,j_1}^{VBS}$ & - & -\\ 
   
    $c_{HD}$  &  $p_{T,j^{1}}$ & $m_{ll}$ & $\Delta\eta_{jj}$ & $\Delta\eta_{jj}$ & $m_{ee}$ & $\Delta\eta_{jj}{}^{\dagger}$ & $p_{T,e^{+}\mu^{+}}$ & $p_{T,e^{+}\mu^{+}}{}^{\dagger}$ & $p_{T,l^{2}}$ & $p_{T,l^{2}}$ & $p_{T,l^{1}}$ & $p_{T,l^{1}}$\\ 
    
    $c_{H \square}$  & $p_{T,j^{1}}$ & $m_{ll}$ & $m_{ll}$ & $m_{ll}$ & - & $m_{WZ}$  & - & $\Delta\eta_{jj}$ & $p_{T,j_2}^{V}$ & $p_{T,j_2}^{V}$ & - & -\\
     
    $c_{HW}$  &  $\Delta\phi_{jj}$ & $m_{ll}$ & $\Delta\phi_{jj}$ & $m_{ll}$ & $\eta_{l^{3}}{}^{\dagger}$ & $m_{WZ}$ & $m_{jj}$ & $m_{4l}$ & $p_{T,j_1}^{VBS}$ & $p_{T,j_2}^{V}$ & - & -\\ 
 
    $c_{HWB}$  & $p_{T,j^{1}}$  & $p_{T,j^{1}}$ & $\Delta\eta_{jj}$ & $m_{ll}$ & $m_{ee}$ & $m_{WZ}$ & $m_{\mu\mu}{}^{\dagger}$ & $\Delta\eta_{jj}$ & $\Delta\eta_{jj}^{V}$ & $\Delta\eta_{jj}^{V}$ & $p_{T,l^{1}}$ & MET\\
    
    $c_{W}$  & $\Delta\phi_{jj}$ & $p_{T,ll}$ & $\Delta\phi_{jj}$ & $m_{ll}$ & $p_{T,l^{1}}$ & $m_{WZ}$  & $\Delta\phi_{jj}$ & $p_{T,l^{4}}$ & $\Delta\phi_{jj}^{VBS}{}^{\dagger}$ & $\Delta\phi_{jj}^{VBS}{}^{\dagger}$ & MET & MET\\ 
    
    $c_{ll}^{(1)}$  &  $m_{jj}{}^{\dagger}$ & $m_{jj}{}^{\dagger}$ & $m_{jj}{}^{\dagger}$ & $m_{jj}{}^{\dagger}$ & $m_{jj}{}^{\dagger}$ & $m_{jj}$  & $m_{jj}{}^{\dagger}$ & $m_{jj}{}^{\dagger}$ &  $\Delta\eta_{jj}^{V}{}^{\dagger}$ & $\Delta\eta_{jj}^{V}{}^{\dagger}$ & $p_{T,ll}{}^{\dagger}$ & $p_{T,l^{2}}$\\ 
    
    \bottomrule
\end{tabular}
\caption{Most sensitive observable for each VBS process and  SMEFT Wilson coefficient, inferred from individual fits and employed 
in the final statistical combination. The columns marked with L and L+Q indicate the results for fits including respectively linear and linear+quadratic terms in the Wilson coefficients. A $-$ indicates no sensitivity to an operator in a given process. A ${}^{\dagger}$ indicates that the preference for the variable indicated is very mild,  as most observables give similar sensitivity.\\ EFT corrections to QCD backgrounds were retained for OSWW+2j, WZ+2j, ZZ+2j and ZV+2j.}\label{tab:sensitivity_ranking}
\end{sidewaystable}

\begin{sidewaystable}[!htbp]
\setlength\tabcolsep{0pt}
\begin{tabular*}{\textwidth}{@{\extracolsep{\fill}} l *{15}{S[table-format=-1.3]} }
\toprule
&  $c_{qq}^{(3,1)}$ & $c_{qq}^{(1,1)}$ & $c_{qq}^{(1)}$ & $c_{W}$ & $c_{Hl}^{(3)}$ & $c_{Hq}^{(3)}$ & $c_{ll}^{(1)}$ & $c_{Hl}^{(1)}$ & $c_{HW}$ & $c_{Hq}^{(1)}$ & $c_{HD}$ & $c_{HWB}$ & $c_{H\square}$  \\ 
\midrule
 $c_{qq}^{(3)}$ & $p_{T,j^2}$ & $p_{T,j^1}$ & $p_{T,j^1}$ & $\Delta\phi_{jj}$ & $p_{T,j^2}$ & $p_{T,j^2}$ & $p_{T,j^2}$ & $p_{T,j^2}$ & $\Delta\phi_{jj}$ & $p_{T,j^1}$ & $p_{T,j^2}$ & $p_{T,j^1}$ & $p_{T,j^2}$   \\ 
 $c_{qq}^{(3,1)}$  & \si{-} & $p_{T,j^2}$ & $p_{T,j^1}$ & $\Delta\phi_{jj}$ & $\Delta\phi_{jj}$ & $p_{T,j^1}$ & $\Delta\phi_{jj}$ & $\Delta\phi_{jj}$ & $\Delta\phi_{jj}$ & $p_{T,j^1}$ & $p_{T,j^1}$ & $p_{T,j^1}$ & $\Delta\phi_{jj}$   \\ 
 $c_{qq}^{(1,1)}$  & \si{-} & \si{-} & $p_{T,j^1}$ & $p_{T,j^1}$ & $p_{T,j^1}$ & $p_{T,j^1}$ & $p_{T,j^1}$ & \si{MET} & $p_{T,l^1}$ & $p_{T,j^1}$ & $p_{T,j^1}$ & $p_{T,j^1}$ & $p_{T,j^1}$  \\ 
 $c_{qq}^{(1)}$  & \si{-} & \si{-} & \si{-} & $p_{T,j^1}$ & $p_{T,j^1}$ & $p_{T,j^1}$ & $p_{T,j^1}$ & \si{MET} & $p_{T,l^1}$ & $p_{T,j^1}$ & $p_{T,j^1}$ & $p_{T,j^1}$ & $p_{T,j^1}$ \\ 
 $c_{W}$  & \si{-} & \si{-} & \si{-} & \si{-} & $p_{T,l^1}$ & $p_{T,l^1}$ & $p_{T,l^1}$ & $p_{T,l^1}$ & $m_{ll}$ & $p_{T,l^1}$ & $p_{T,l^1}$ & $p_{T,j^1}$ & $m_{ll}$   \\ 
 $c_{Hl}^{(3)}$  & \si{-} & \si{-} & \si{-} & \si{-} & \si{-} & $p_{T,j^1}$ & $\eta_{l^2}$ & $m_{ll}$ & $m_{ll}$ & $p_{T,j^1}$ & $m_{ll}$ & $p_{T,j^1}$ & $m_{ll}$   \\ 
 $c_{Hq}^{(3)}$  & \si{-} & \si{-} & \si{-} & \si{-} & \si{-} & \si{-} & $p_{T,j^1}$ & $p_{T,l^1}$ & $m_{ll}$ & $p_{T,j^1}$ & $p_{T,l^1}$ & $p_{T,j^1}$ & $m_{ll}$  \\ 
 $c_{ll}^{(1)}$  & \si{-} & \si{-} & \si{-} & \si{-} & \si{-} & \si{-} & \si{-} & $m_{ll}$ & $m_{ll}$ & $p_{T,j^1}$ & $m_{ll}$ & $p_{T,j^1}$ & $m_{ll}$   \\ 
 $c_{Hl}^{(1)}$  & \si{-} & \si{-} & \si{-} & \si{-} & \si{-} & \si{-} & \si{-} & \si{-} & $m_{ll}$ & $p_{T,l^1}$ & $m_{ll}$ & $p_{T,l^1}$ & $m_{ll}$  \\ 
 $c_{HW}$  & \si{-} & \si{-} & \si{-} & \si{-} & \si{-} & \si{-} & \si{-} & \si{-} & \si{-} & $p_{T,l^1}$ & $m_{ll}$ & $m_{ll}$ & $m_{ll}$  \\ 
 $c_{Hq}^{(1)}$  & \si{-} & \si{-} & \si{-} & \si{-} & \si{-} & \si{-} & \si{-} & \si{-} & \si{-} & \si{-} & $p_{T,l^1}$ & $p_{T,j^1}$ & $p_{T,l^1}$  \\ 
 $c_{HD}$  & \si{-} & \si{-} & \si{-} & \si{-} & \si{-} & \si{-} & \si{-} & \si{-} & \si{-} & \si{-} & \si{-} & $m_{ll}$ & $m_{ll}$  \\ 
 $c_{HWB}$  & \si{-} & \si{-} & \si{-} & \si{-} & \si{-} & \si{-} & \si{-} & \si{-} & \si{-} & \si{-} & \si{-} & \si{-} & $m_{ll}$   \\ 

 \bottomrule 
 \end{tabular*}
\caption{Most sensitive observable for each pair of SMEFT Wilson coefficients for 2D fits to SSWW+2j-EW and including both linear and quadratic terms in the Wilson coefficients.} \label{tab.sensitivity_2d_ssWW}
\end{sidewaystable}

\begin{sidewaystable}
\setlength\tabcolsep{0pt}
\begin{tabular*}{\textwidth}{@{\extracolsep{\fill}} l *{14}{S[table-format=-1.3]} }
\toprule
&  $c_{qq}^{(3,1)}$ & $c_{qq}^{(1,1)}$ & $c_{qq}^{(1)}$ & $c_{W}$ & $c_{Hl}^{(3)}$ & $c_{Hq}^{(3)}$ & $c_{ll}^{(1)}$ & $c_{Hl}^{(1)}$ & $c_{HW}$ & $c_{Hq}^{(1)}$ & $c_{HD}$ & $c_{HWB}$ & $c_{H\square}$   \\ 
\midrule
 $c_{qq}^{(3)}$  & $p_{T,j^2}$ & $p_{T,j^2}$ & $p_{T,j^2}$ & $p_{T,j^1}$ & $p_{T,j^2}$ & $p_{T,j^1}$ & $p_{T,j^2}$ & \si{MET} & $p_{T,j^1}$ & $p_{T,j^1}$ & $\Delta\eta_{jj}$ & $p_{T,j^1}$ & $m_{ll}$  \\ 
 $c_{qq}^{(3,1)}$  & \si{-} & $p_{T,j^2}$ & $p_{T,j^2}$ & $p_{T,j^1}$ & $p_{T,j^2}$ & $p_{T,j^1}$ & $p_{T,j^2}$ & \si{MET} & $p_{T,j^1}$ & $p_{T,j^1}$ & $\Delta\eta_{jj}$ & $p_{T,j^1}$ & $m_{ll}$  \\ 
 $c_{qq}^{(1,1)}$  & \si{-} & \si{-} & $p_{T,j^2}$ & $p_{T,j^1}$ & $p_{T,j^2}$ & $p_{T,j^1}$ & $p_{T,j^2}$ & \si{MET} & $p_{T,j^1}$ & $p_{T,j^1}$ & $\Delta\eta_{jj}$ & $p_{T,j^1}$ & $m_{ll}$  \\ 
 $c_{qq}^{(1)}$  & \si{-} & \si{-} & \si{-} & $p_{T,j^1}$ & $p_{T,j^2}$ & $p_{T,j^1}$ & $p_{T,j^2}$ & \si{MET} & $p_{T,j^1}$ & $p_{T,j^1}$ & $\Delta\eta_{jj}$ & $p_{T,j^1}$ & $m_{ll}$  \\ 
 $c_{W}$  & \si{-} & \si{-} & \si{-} & \si{-} & $m_{ll}$ & $m_{ll}$ & $m_{ll}$ & $m_{ll}$ & $m_{ll}$ & $m_{ll}$ & $\Delta\eta_{jj}$ & $m_{ll}$ & $m_{ll}$ \\ 
 $c_{Hl}^{(3)}$  & \si{-} & \si{-} & \si{-} & \si{-} & \si{-} & $m_{ll}$ & $m_{jj}$ & \si{MET} & $m_{ll}$ & $m_{ll}$ & $\Delta\eta_{jj}$ & $m_{ll}$ & $m_{ll}$  \\ 
 $c_{Hq}^{(3)}$  & \si{-} & \si{-} & \si{-} & \si{-} & \si{-} & \si{-} & $m_{ll}$ & $m_{ll}$ & $m_{ll}$ & $m_{ll}$ & $\Delta\eta_{jj}$ & $m_{ll}$ & $m_{ll}$  \\ 
 $c_{ll}^{(1)}$  & \si{-} & \si{-} & \si{-} & \si{-} & \si{-} & \si{-} & \si{-} & \si{MET} & $m_{ll}$ & $m_{ll}$ & $\Delta\eta_{jj}$ & $m_{ll}$ & $m_{ll}$   \\ 
 $c_{Hl}^{(1)}$  & \si{-} & \si{-} & \si{-} & \si{-} & \si{-} & \si{-} & \si{-} & \si{-} & $m_{ll}$ & $m_{ll}$ & $\Delta\eta_{jj}$ & $m_{ll}$ & $m_{ll}$  \\ 
 $c_{HW}$  & \si{-} & \si{-} & \si{-} & \si{-} & \si{-} & \si{-} & \si{-} & \si{-} & \si{-} & $p_{T,j^1}$ & $\Delta\eta_{jj}$ & $m_{ll}$ & $m_{ll}$  \\ 
 $c_{Hq}^{(1)}$  & \si{-} & \si{-} & \si{-} & \si{-} & \si{-} & \si{-} & \si{-} & \si{-} & \si{-} & \si{-} & $\Delta\eta_{jj}$ & $m_{ll}$ & $m_{ll}$  \\ 
 $c_{HD}$ & \si{-} & \si{-} & \si{-} & \si{-} & \si{-} & \si{-} & \si{-} & \si{-} & \si{-} & \si{-} & \si{-} & $\Delta\eta_{jj}$ & $m_{ll}$  \\ 
 $c_{HWB}$  & \si{-} & \si{-} & \si{-} & \si{-} & \si{-} & \si{-} & \si{-} & \si{-} & \si{-} & \si{-} & \si{-} & \si{-} & $m_{ll}$  \\
 \bottomrule 
 \end{tabular*}
\caption{Most sensitive observable for each pair of SMEFT Wilson coefficients for 2D fits to OSWW+2j (EW and QCD) and including both linear and quadratic terms in the Wilson coefficients.} \label{tab.sensitivity_2d_osWW}
\end{sidewaystable}

\begin{sidewaystable}[!htbp]
\setlength\tabcolsep{0pt}
\begin{tabular*}{\textwidth}{@{\extracolsep{\fill}} l *{14}{S[table-format=-1.3]} }
\toprule
&  $c_{qq}^{(3,1)}$ & $c_{qq}^{(1,1)}$ & $c_{qq}^{(1)}$ & $c_{W}$ & $c_{Hl}^{(3)}$ & $c_{Hq}^{(3)}$ & $c_{ll}^{(1)}$ & $c_{Hl}^{(1)}$ & $c_{HW}$ & $c_{Hq}^{(1)}$ & $c_{HD}$ & $c_{HWB}$ & $c_{H\square}$   \\ 
\midrule
 $c_{qq}^{(3)}$ & $p_{T,j^2}$ & $p_{T,j^2}$ & $p_{T,j^1}$ & $p_{T,j^1}$ & $p_{T,j^2}$ & $p_{T,j^1}$ & $p_{T,j^2}$ & $p_{T,j^2}$ & $m_{WZ}$ & $p_{T,j^1}$ & $p_{T,j^2}$ & $m_{WZ}$ & $m_{WZ}$  \\ 
 $c_{qq}^{(3,1)}$  & \si{-} & $p_{T,j^2}$ & $p_{T,j^2}$ & $p_{T,j^1}$ & $p_{T,j^2}$ & $p_{T,j^1}$ & $p_{T,j^2}$ & $p_{T,j^2}$ & $m_{WZ}$ & $p_{T,j^1}$ & $p_{T,j^2}$ & $m_{WZ}$ & $m_{WZ}$  \\ 
 $c_{qq}^{(1,1)}$  & \si{-} & \si{-} & $p_{T,j^2}$ & $p_{T,j^1}$ & $p_{T,j^2}$ & $p_{T,j^1}$ & $p_{T,j^2}$ & $p_{T,j^2}$ & $m_{WZ}$ & $p_{T,j^1}$ & $p_{T,j^2}$ & $m_{WZ}$ & $m_{WZ}$ \\ 
 $c_{qq}^{(1)}$  & \si{-} & \si{-} & \si{-} & $p_{T,j^1}$ & $p_{T,j^2}$ & $p_{T,j^1}$ & $p_{T,j^2}$ & $p_{T,j^2}$ & $m_{WZ}$ & $p_{T,j^1}$ & $p_{T,j^2}$ & $p_{T,j^1}$ & $m_{WZ}$  \\ 
 $c_{W}$ & \si{-} & \si{-} & \si{-} & \si{-} & $m_{WZ}$ & $m_{WZ}$ & $m_{WZ}$ & $m_{WZ}$ & $m_{WZ}$ & $m_{WZ}$ & $m_{WZ}$ & $m_{WZ}$ & $m_{WZ}$ \\ 
 $c_{Hl}^{(3)}$  & \si{-} & \si{-} & \si{-} & \si{-} & \si{-} & $m_{WZ}$ & $m_{jj}$ & $m_{WZ}$ & $m_{WZ}$ & $m_{WZ}$ & $m_{ee}$ & $m_{WZ}$ & $m_{WZ}$  \\ 
 $c_{Hq}^{(3)}$  & \si{-} & \si{-} & \si{-} & \si{-} & \si{-} & \si{-} & $m_{WZ}$ & $m_{WZ}$ & $m_{WZ}$ & $p_{T,j^1}$ & $p_{T,l^2}$ & $m_{WZ}$ & $m_{WZ}$  \\ 
 $c_{ll}^{(1)}$  & \si{-} & \si{-} & \si{-} & \si{-} & \si{-} & \si{-} & \si{-} & $m_{WZ}$ & $m_{WZ}$ & $m_{WZ}$ & $m_{ee}$ & $m_{WZ}$ & $m_{WZ}$  \\ 
 $c_{Hl}^{(1)}$  & \si{-} & \si{-} & \si{-} & \si{-} & \si{-} & \si{-} & \si{-} & \si{-} & $m_{WZ}$ & $m_{WZ}$ & $m_{ee}$ & $m_{WZ}$ & $m_{WZ}$ \\ 
 $c_{HW}$ &  \si{-} & \si{-} & \si{-} & \si{-} & \si{-} & \si{-} & \si{-} & \si{-} & \si{-} & $m_{WZ}$ & $m_{WZ}$ & $m_{WZ}$ & $m_{WZ}$  \\ 
 $c_{Hq}^{(1)}$  & \si{-} & \si{-} & \si{-} & \si{-} & \si{-} & \si{-} & \si{-} & \si{-} & \si{-} & \si{-} & $p_{T,j^1}$ & $m_{WZ}$ & $m_{WZ}$ \\ 
 $c_{HD}$  & \si{-} & \si{-} & \si{-} & \si{-} & \si{-} & \si{-} & \si{-} & \si{-} & \si{-} & \si{-} & \si{-} & $m_{WZ}$ & $m_{WZ}$ \\ 
 $c_{HWB}$  & \si{-} & \si{-} & \si{-} & \si{-} & \si{-} & \si{-} & \si{-} & \si{-} & \si{-} & \si{-} & \si{-} & \si{-} & $m_{WZ}$  \\ 
 
 \bottomrule 
 \end{tabular*}
\caption{Most sensitive observable for each pair of SMEFT Wilson coefficients for 2D fits to WZ+2j (EW and QCD)  and including both linear and quadratic terms in the Wilson coefficients.} \label{tab.sensitivity_2d_WZ}
\end{sidewaystable}

\begin{sidewaystable}
\setlength\tabcolsep{0pt}
\begin{tabular*}{\textwidth}{@{\extracolsep{\fill}} l *{14}{S[table-format=-1.3]} }
\toprule
 & $c_{qq}^{(3,1)}$ & $c_{qq}^{(1,1)}$ & $c_{qq}^{(1)}$ & $c_{W}$ & $c_{Hl}^{(3)}$ & $c_{Hq}^{(3)}$ & $c_{ll}^{(1)}$ & $c_{Hl}^{(1)}$ & $c_{HW}$ & $c_{Hq}^{(1)}$ & $c_{HD}$ & $c_{HWB}$ & $c_{H\square}$  \\ 
\midrule
 $c_{qq}^{(3)}$ & $p_{T,j^1}$ & $p_{T,j^1}$ & $p_{T,j^2}$ & $p_{T,l^1}$ & $p_{T,j^2}$ & $p_{T,j^1}$ & $p_{T,j^2}$ & $p_{T,j^2}$ & $p_{T,l^1}$ & $p_{T,j^1}$ & $p_{T,j^1}$ & $p_{T,j^1}$ & $p_{T,j^1}$  \\ 
 $c_{qq}^{(3,1)}$  & \si{-} & $p_{T,j^1}$ & $p_{T,j^2}$ & $p_{T,l^1}$ & $p_{T,j^2}$ & $p_{T,j^1}$ & $p_{T,j^2}$ & $p_{T,j^2}$ & $p_{T,l^1}$ & $p_{T,j^1}$ & $p_{T,j^2}$ & $p_{T,j^1}$ & $p_{T,j^1}$  \\ 
 $c_{qq}^{(1,1)}$ & \si{-} & \si{-} & $p_{T,j^2}$ & $p_{T,l^1}$ & $p_{T,j^2}$ & $p_{T,j^1}$ & $p_{T,j^2}$ & $p_{T,j^2}$ & $p_{T,l^1}$ & $p_{T,j^1}$ & $p_{T,j^2}$ & $p_{T,j^1}$ & $p_{T,j^1}$  \\ 
 $c_{qq}^{(1)}$ & \si{-} & \si{-} & \si{-} & $p_{T,l^1}$ & $p_{T,j^2}$ & $p_{T,j^1}$ & $p_{T,j^2}$ & $p_{T,j^2}$ & $p_{T,l^1}$ & $p_{T,j^1}$ & $p_{T,j^2}$ & $p_{T,j^1}$ & $p_{T,j^1}$  \\ 
 $c_{W}$ & \si{-} & \si{-} & \si{-} & \si{-} & $p_{T,l^1}$ & $p_{T,l^1}$ & $p_{T,l^1}$ & $p_{T,l^1}$ & $p_{T,l^1}$ & $p_{T,l^1}$ & $p_{T,l^1}$ & $p_{T,l^1}$ & $m_{4l}$  \\ 
 $c_{Hl}^{(3)}$ & \si{-} & \si{-} & \si{-} & \si{-} & \si{-} & $p_{T,j^1}$ & $m_{jj}$ & $m_{jj}$ & $p_{T,ee}$ & $p_{T,j^1}$ & $m_{jj}$ & $m_{jj}$ & $m_{4l}$  \\ 
 $c_{Hq}^{(3)}$ & \si{-} & \si{-} & \si{-} & \si{-} & \si{-} & \si{-} & $p_{T,j^1}$ & $p_{T,j^1}$ & $p_{T,l^1}$ & $p_{T,j^1}$ & $p_{T,l^1}$ & $p_{T,l^1}$ & $p_{T,l^2}$  \\ 
 $c_{ll}^{(1)}$ & \si{-} & \si{-} & \si{-} & \si{-} & \si{-} & \si{-} & \si{-} & $m_{jj}$ & $p_{T,Z}$ & $p_{T,j^1}$ & $p_{T,e^{-}\mu^{-}}$ & $p_{T,Z}$ & $p_{T,j^2}$  \\ 
 $c_{Hl}^{(1)}$  & \si{-} & \si{-} & \si{-} & \si{-} & \si{-} & \si{-} & \si{-} & \si{-} & $m_{4l}$ & $p_{T,j^1}$ & $p_{T,e^{+}\mu^{+}}$ & $p_{T,ee}$ & $\Delta\eta_{jj}$  \\
 $c_{HW}$ &  \si{-} & \si{-} & \si{-} & \si{-} & \si{-} & \si{-} & \si{-} & \si{-} & \si{-} & $p_{T,l^1}$ & $m_{4l}$ & $p_{T,Z}$ & $m_{4l}$  \\ 
 $c_{Hq}^{(1)}$ & \si{-} & \si{-} & \si{-} & \si{-} & \si{-} & \si{-} & \si{-} & \si{-} & \si{-} & \si{-} & $p_{T,e^{-}\mu^{-}}$ & $p_{T,e^{-}\mu^{-}}$ & $p_{T,l^2}$  \\ 
 $c_{HD}$ &  \si{-} & \si{-} & \si{-} & \si{-} & \si{-} & \si{-} & \si{-} & \si{-} & \si{-} & \si{-} & \si{-} & $m_{4l}$ & $\Delta\eta_{jj}$  \\ 
 $c_{HWB}$ &  \si{-} & \si{-} & \si{-} & \si{-} & \si{-} & \si{-} & \si{-} & \si{-} & \si{-} & \si{-} & \si{-} & \si{-} & $m_{4l}$  \\ 

 \bottomrule 
 \end{tabular*}
\caption{Most sensitive observable for each pair of SMEFT Wilson coefficients for 2D fits to ZZ+2j (EW and QCD)  and including both linear and quadratic terms in the Wilson coefficients. The variable $p_{T,Z}$ is defined as the transverse momentum of the same-flavour dilepton system with invariant mass closest to $m_Z$ } \label{tab.sensitivity_2d_ZZ}
\end{sidewaystable}

\begin{sidewaystable}[!htbp]
\setlength\tabcolsep{0pt}
\begin{tabular*}{\textwidth}{@{\extracolsep{\fill}} l *{15}{S[table-format=-1.3]} }
\toprule
&   $c_{Hl}^{(3)}$ & $c_{Hq}^{(3)}$ & $c_{ll}^{(1)}$ & $c_{Hl}^{(1)}$ & $c_{Hq}^{(1)}$ & $c_{HD}$ & $c_{HWB}$  \\ 
\midrule
 $c_{W}$  & \si{MET} & \si{MET} & \si{MET} & \si{MET} & \si{MET} & \si{MET} & \si{MET}  \\ 
 $c_{Hl}^{(3)}$ & \si{-} & \si{MET} & $\eta_{l^2}$ & \si{MET} & \si{MET} & $p_{T,l^1}$ & \si{MET}  \\ 
 $c_{Hq}^{(3)}$ & \si{-} & \si{-} & $p_{T,l^1}$ & \si{MET} & \si{MET} & $p_{T,l^1}$ & \si{MET}  \\ 
 $c_{ll}^{(1)}$ & \si{-} & \si{-} & \si{-} & \si{MET} & \si{MET} & $p_{T,l^1}$ & \si{MET}  \\ 
 $c_{Hl}^{(1)}$ & \si{-} & \si{-} & \si{-} & \si{-} & \si{MET} & \si{MET} & \si{MET}  \\ 
 $c_{Hq}^{(1)}$ & \si{-} & \si{-} & \si{-} & \si{-} & \si{-} & \si{MET} & \si{MET}  \\ 
 $c_{HD}$ & \si{-} & \si{-} & \si{-} & \si{-} & \si{-} & \si{-} & $p_{T,l^1}$  \\ 

 \bottomrule 
 \end{tabular*}
\caption{Most sensitive observable for each pair of SMEFT Wilson coefficients for 2D fits to diboson WW and including both linear and quadratic terms in the Wilson coefficients.} \label{tab.sensitivity_2d_inWW}
\end{sidewaystable}

\clearpage
\section{Representative kinematic distributions}
\label{app:distributions}

\begin{figure}[htbp]
  \centering 
  \includegraphics[width=.49\textwidth]{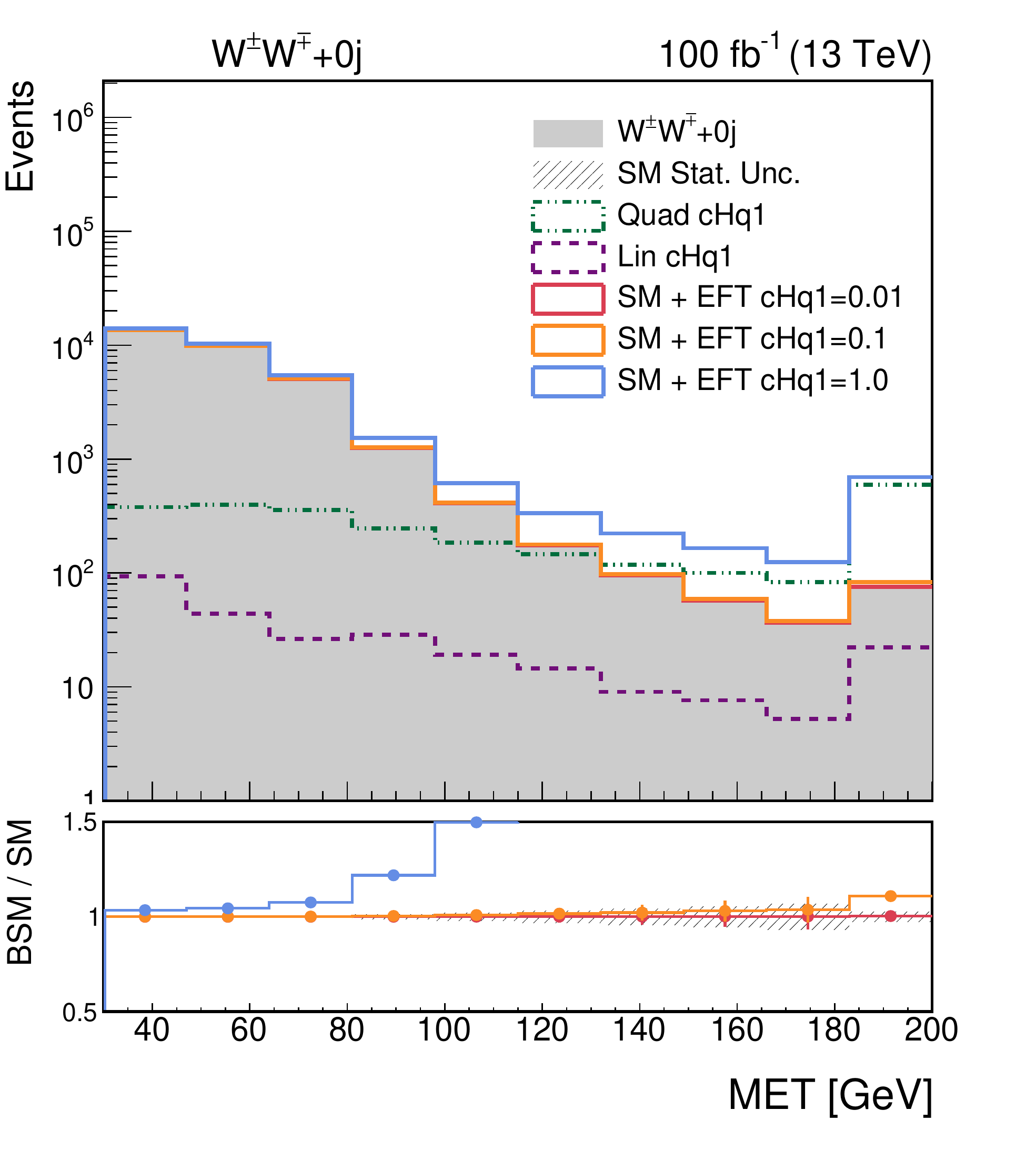}
  \includegraphics[width=.49\textwidth]{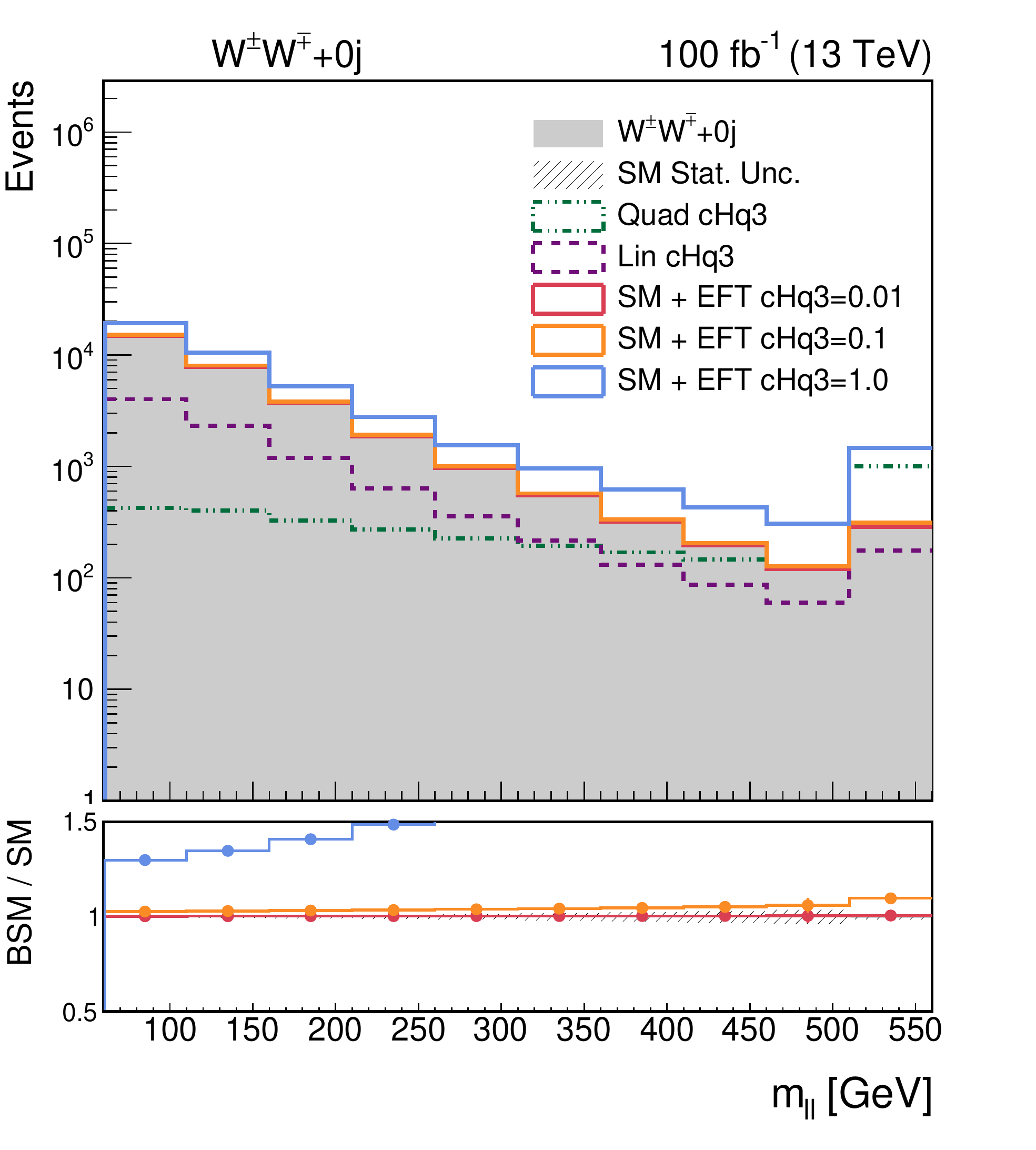}   \\
  \includegraphics[width=.49\textwidth]{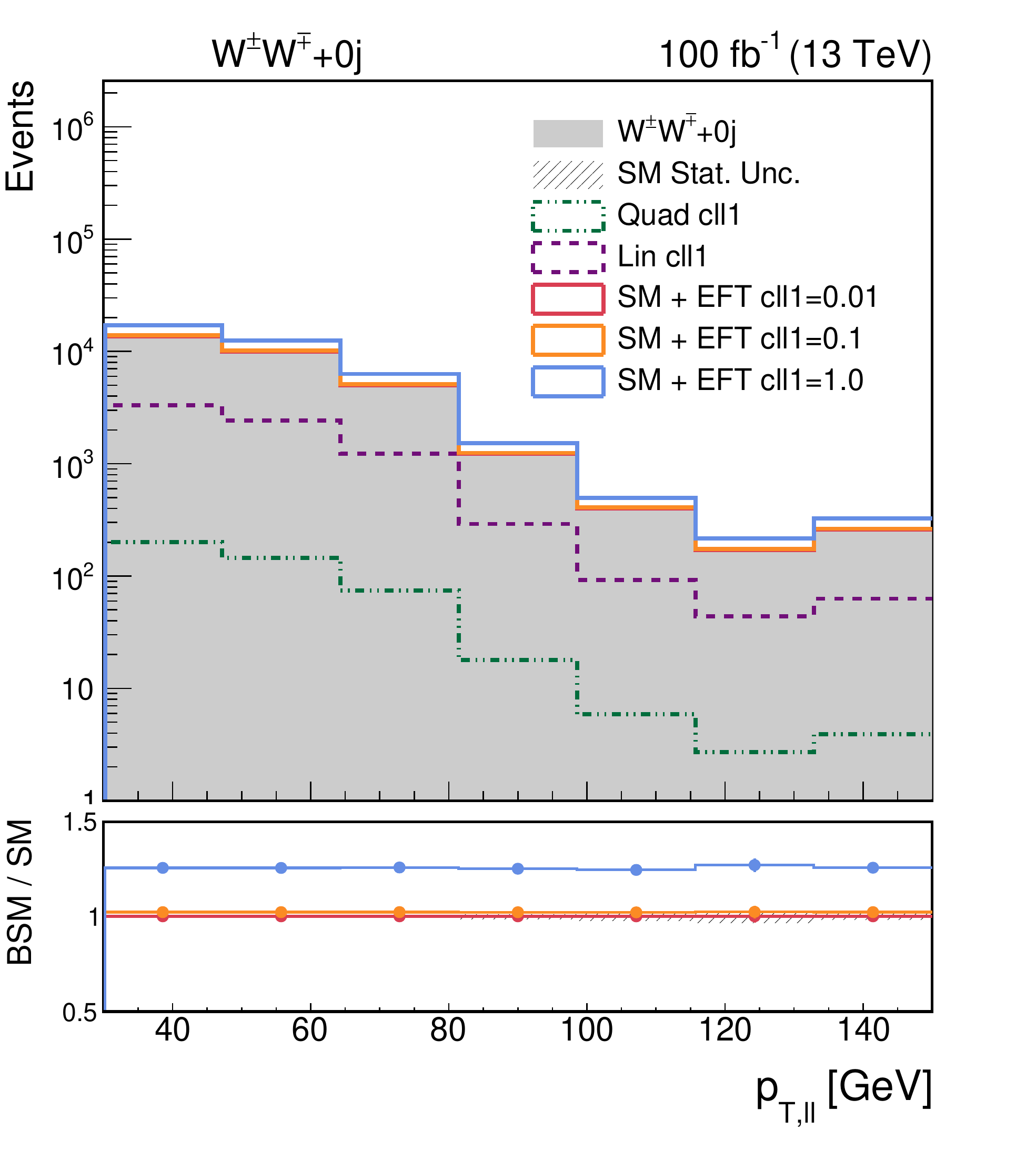}
  \includegraphics[width=.49\textwidth]{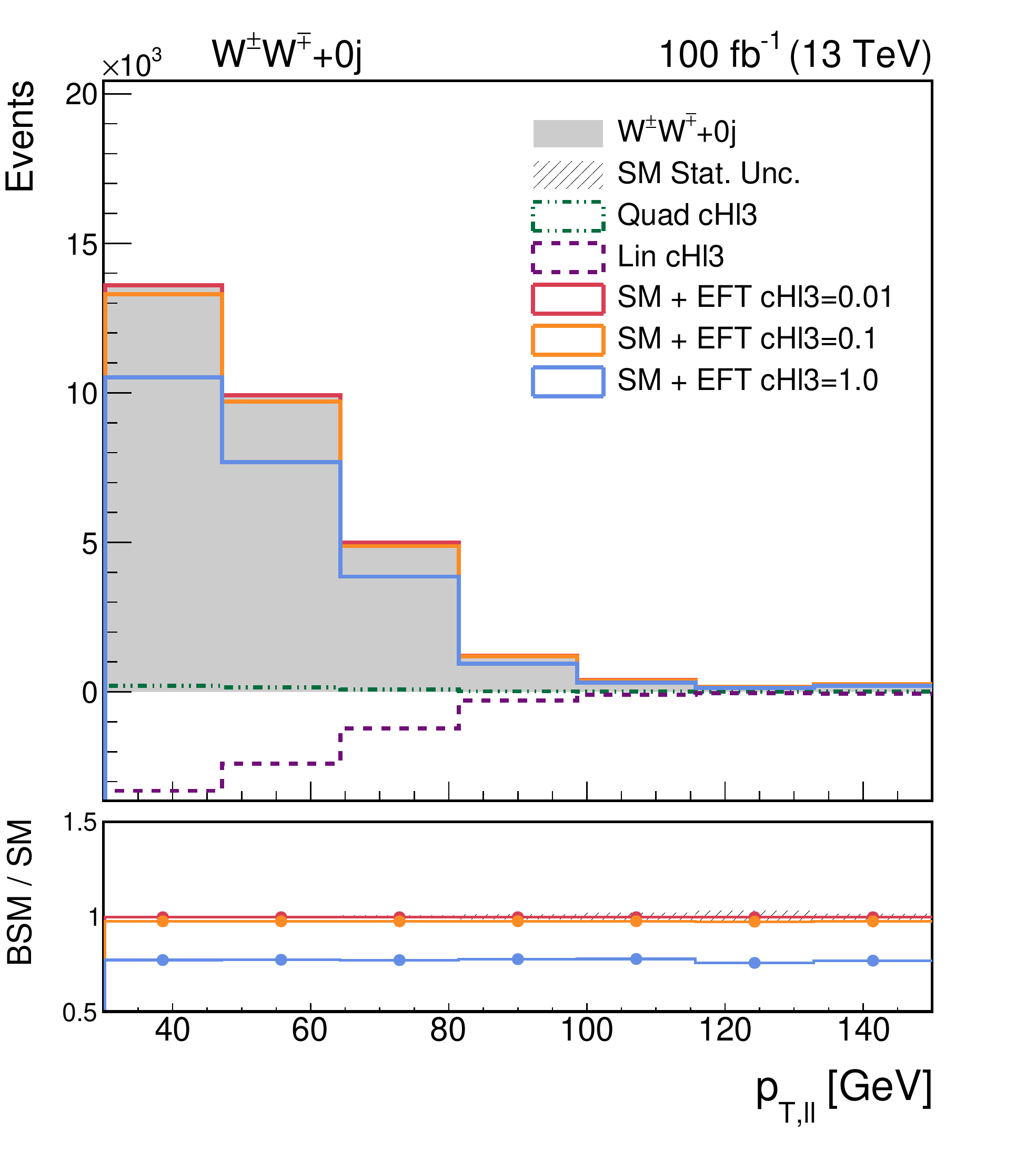}
  \caption{
  Comparison of SM (filled histograms) and BSM (lines)
      expected number of events 
      after the event selection in Table~\ref{tab:variables},
      for various observables and Wilson coefficients in the inclusive WW process, at an integrated luminosity of $\unit[100]{fb^{-1}}$.
      Solid lines show the total prediction for one Wilson coefficient at a time, with $c_\a/\Lambda^2 = 0.01$ (red), $0.1$ (orange) or $\unit[1]{TeV^{-2}}$ (blue).
      The pure interference (quadratic) EFT component, normalized to $c_\a/\Lambda^2=\unit[1]{TeV^{-2}}$, is indicated with a purple (green) dashed line.
      For all distributions, the last bin comprises all the overflow events. 
      \label{fig:Distributions_inWW}  }  
\end{figure}

\begin{figure}[htbp]
  \centering 
    \includegraphics[width=.49\textwidth]{Figures/Distributions_2/SSWW/cqq1_ptj1_stat.pdf}
    \includegraphics[width=.49\textwidth]{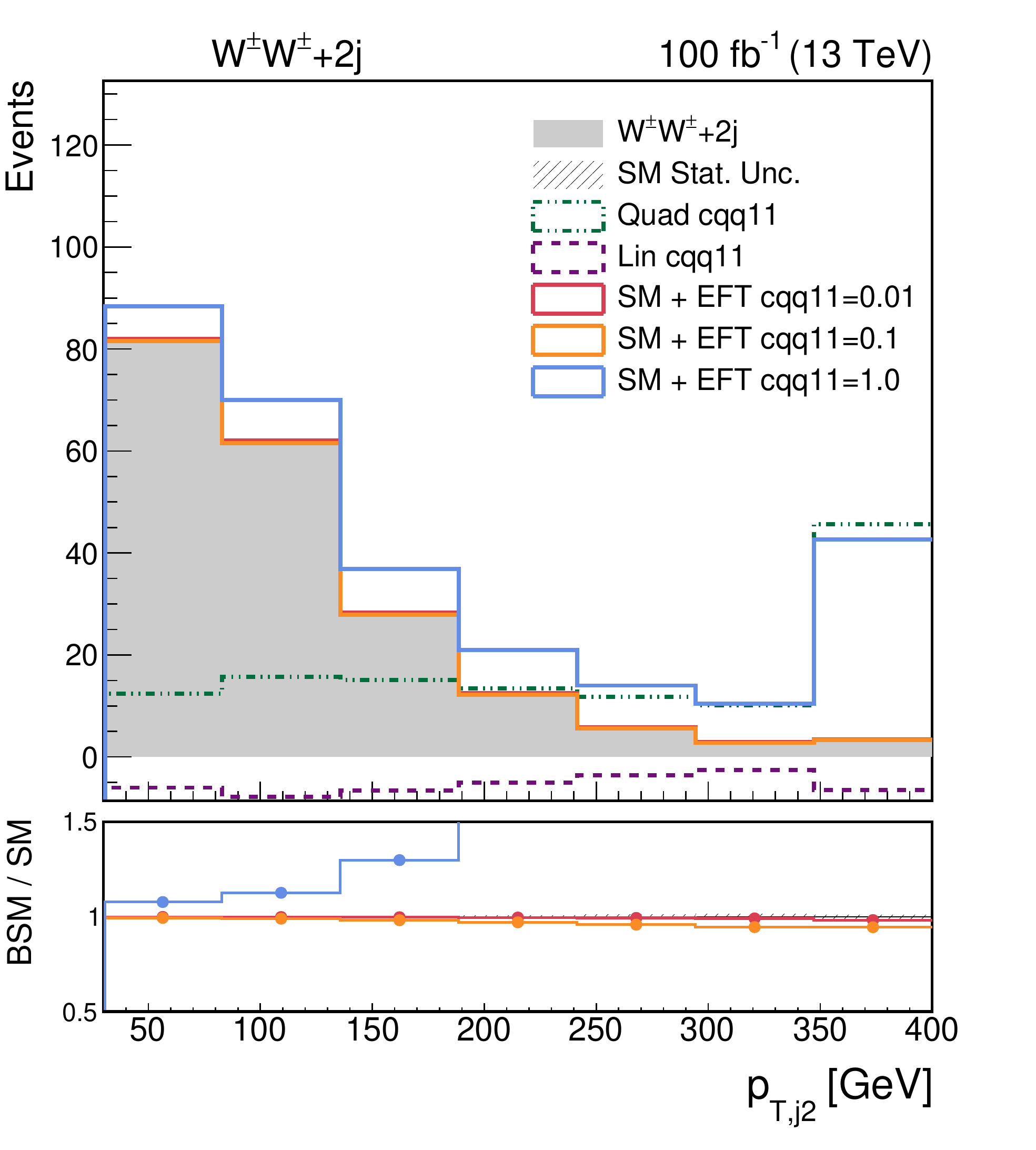}  \\
    \includegraphics[width=.49\textwidth]{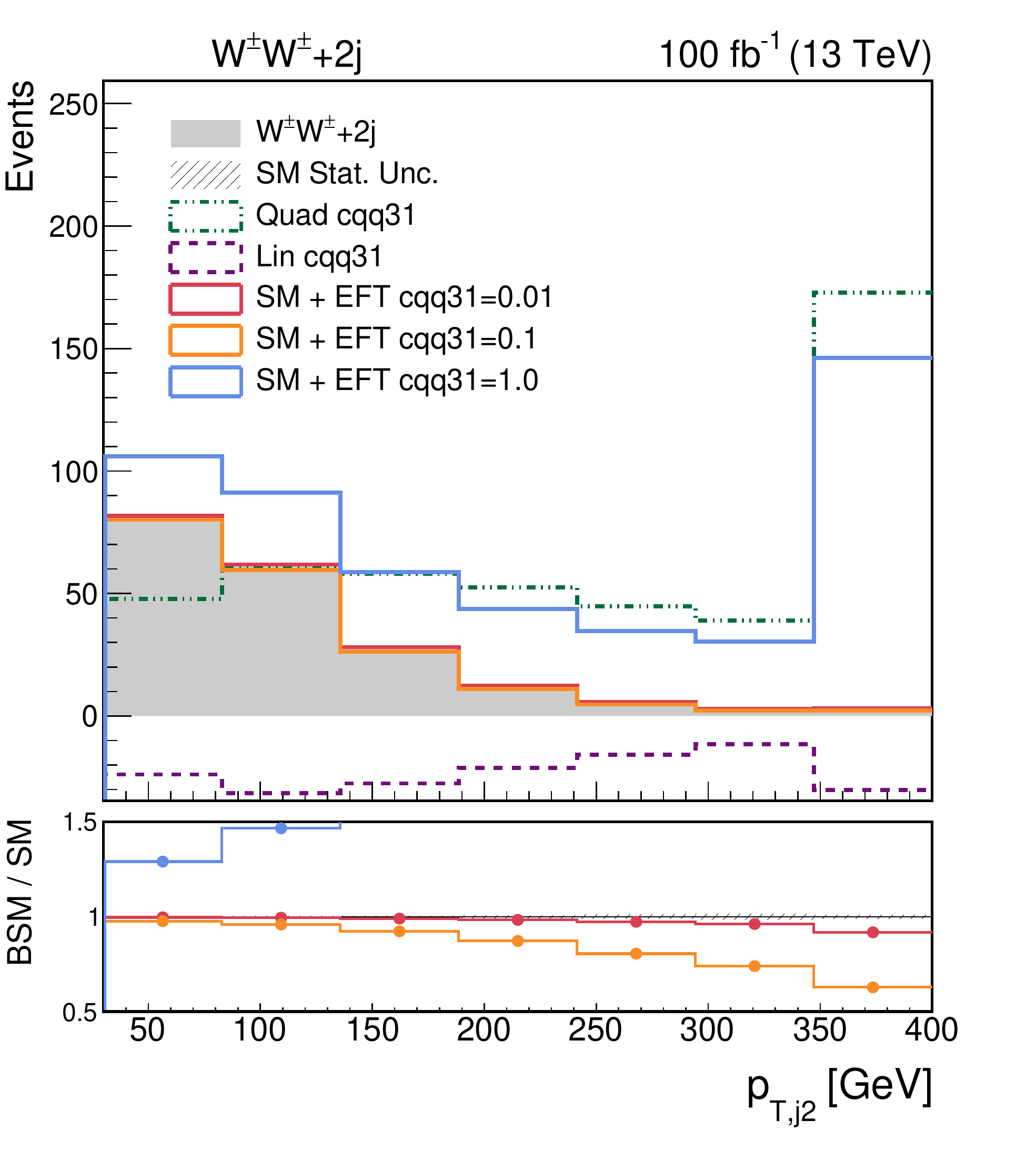}
    \includegraphics[width=.49\textwidth]{Figures/Distributions_2/SSWW/cW_mll_stat.pdf}
  \caption{Comparison of SM (filled histograms) and BSM (lines)
      expected number of events 
      after the event selection in Table~\ref{tab:variables},
      for various observables and Wilson coefficients in the SSWW+2j-EW process, at an integrated luminosity of $\unit[100]{fb^{-1}}$.
      Solid lines show the total prediction for one Wilson coefficient at a time, with $c_\a/\Lambda^2 = 0.01$ (red), $0.1$ (orange) or $\unit[1]{TeV^{-2}}$ (blue).
      The pure interference (quadratic) EFT component, normalized to $c_\a/\Lambda^2=\unit[1]{TeV^{-2}}$, is indicated with a purple (green) dashed line.
      For all distributions, the last bin comprises all the overflow events.\label{fig:Distributions_SSWW} }  
\end{figure}

\begin{figure}[htbp]
  \centering \vspace*{-5mm}
  \includegraphics[width=.49\textwidth]{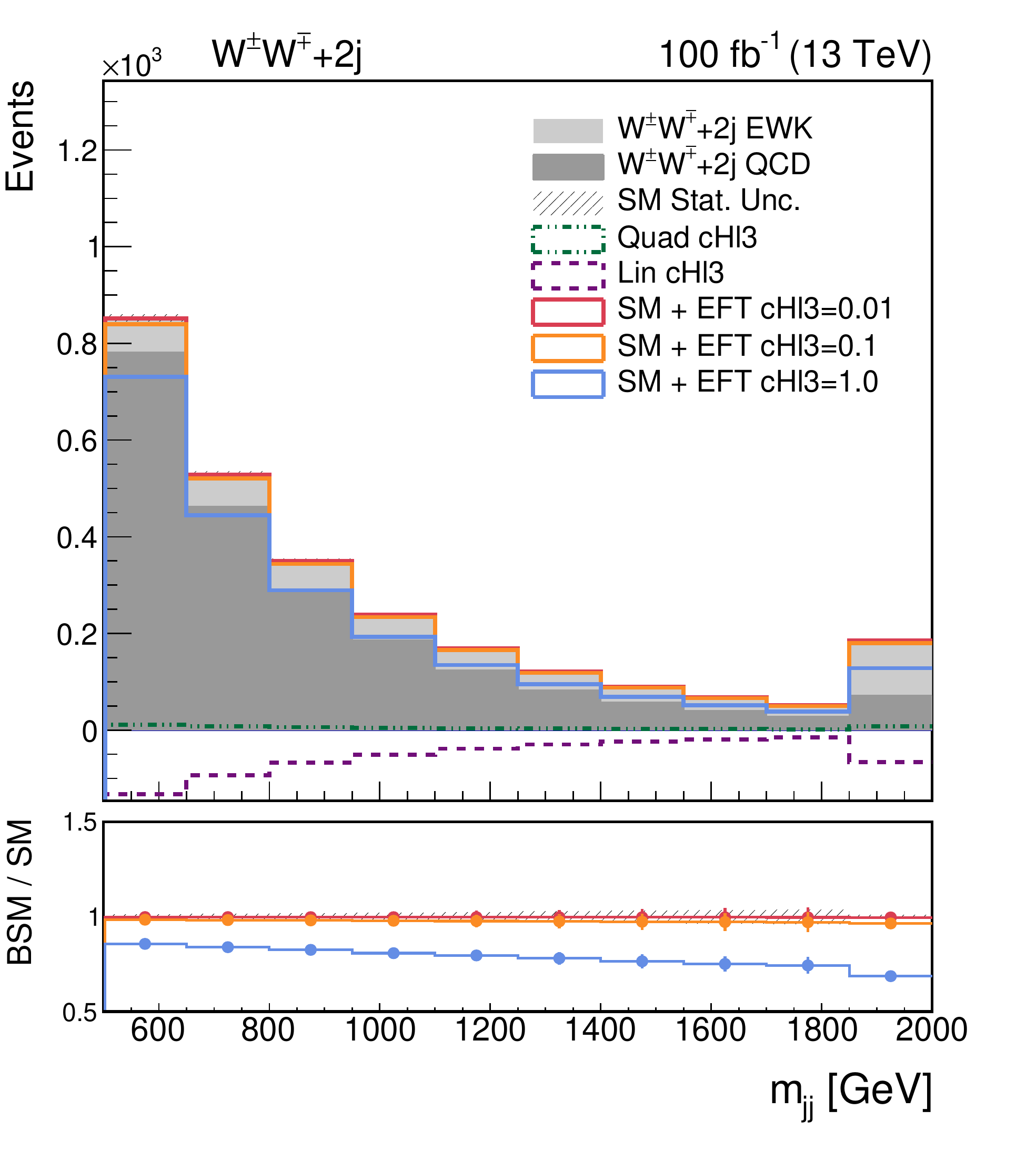}    
  \includegraphics[width=.49\textwidth]{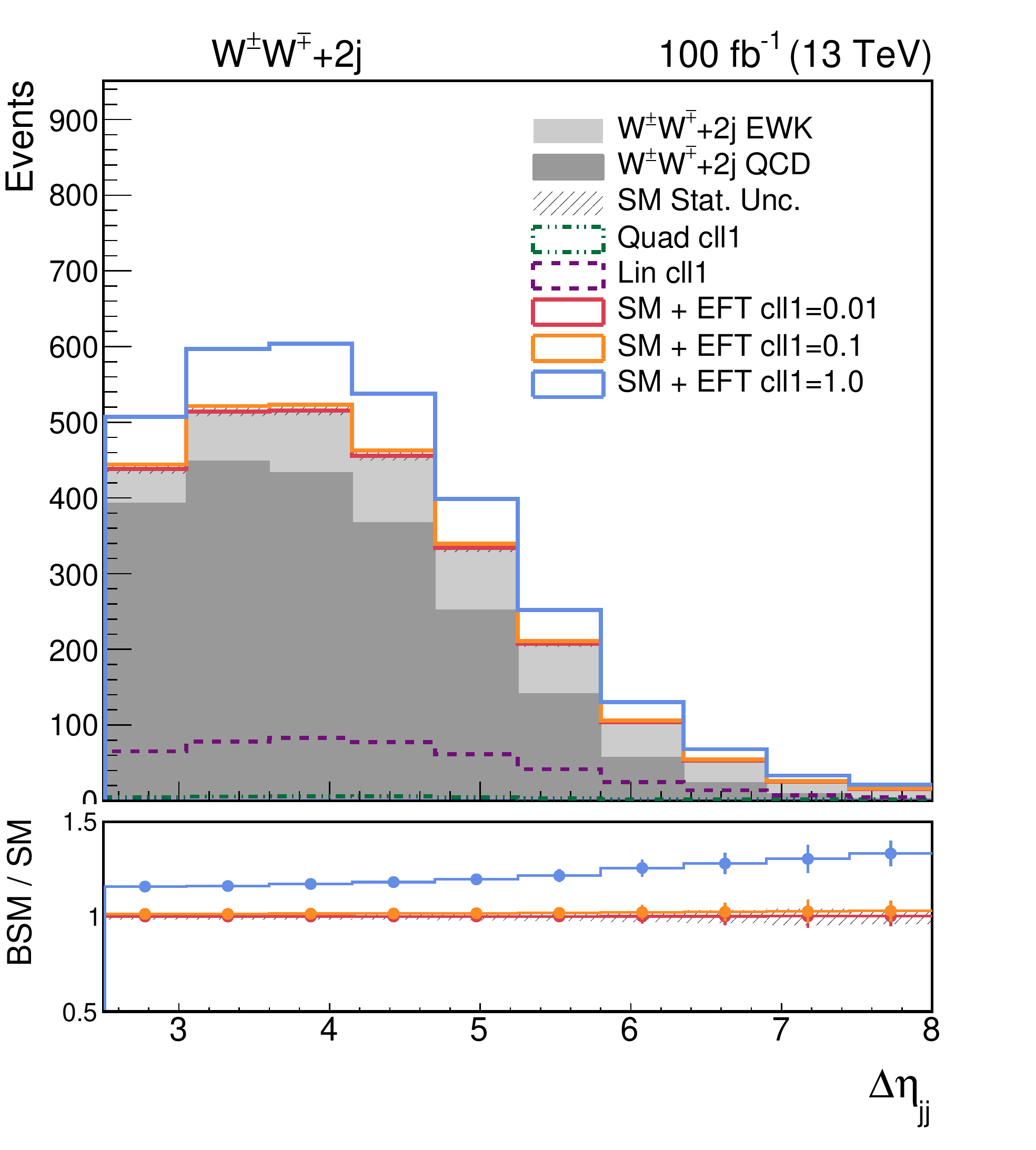}
  \includegraphics[width=.49\textwidth]{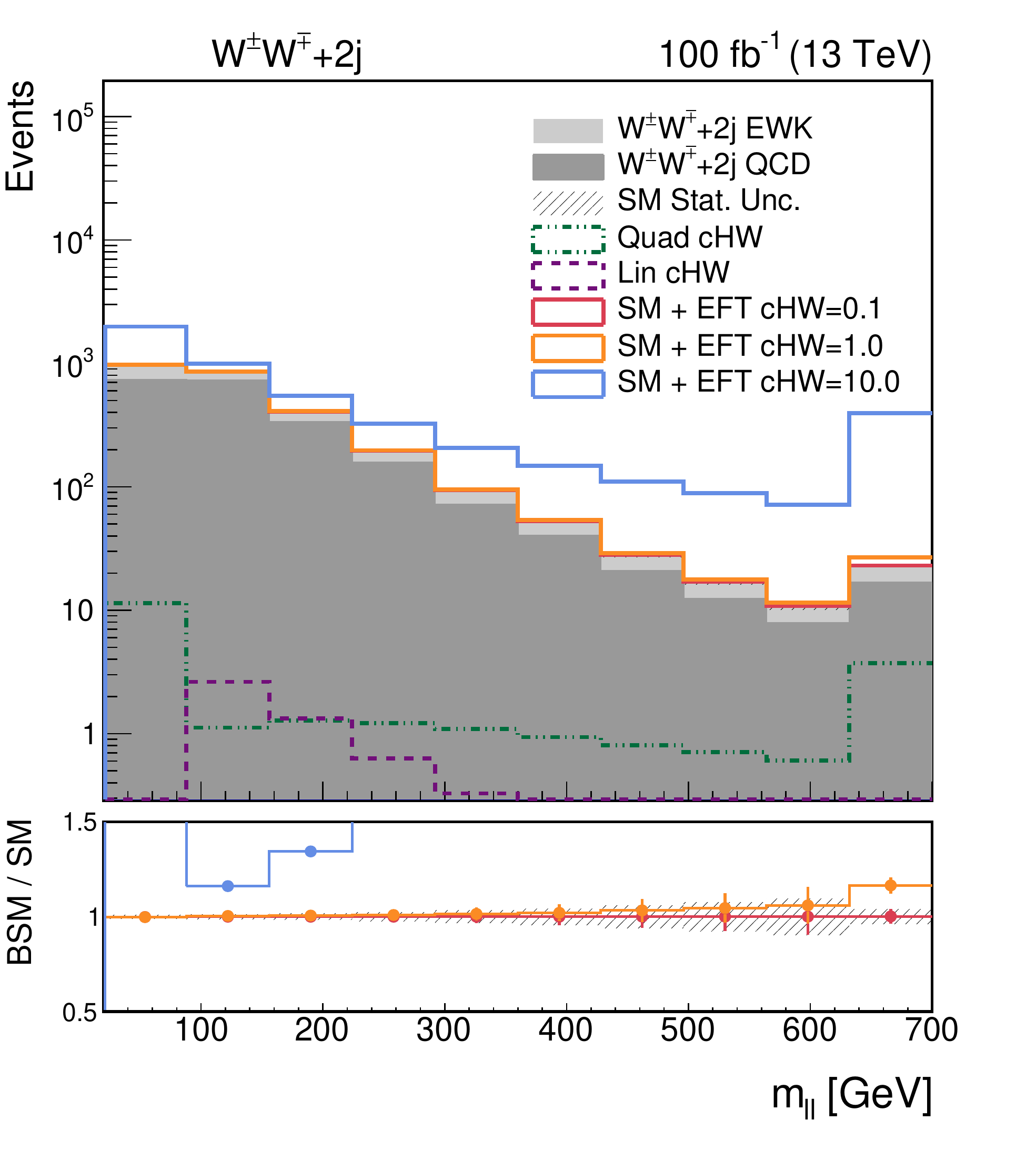}
  \includegraphics[width=.49\textwidth]{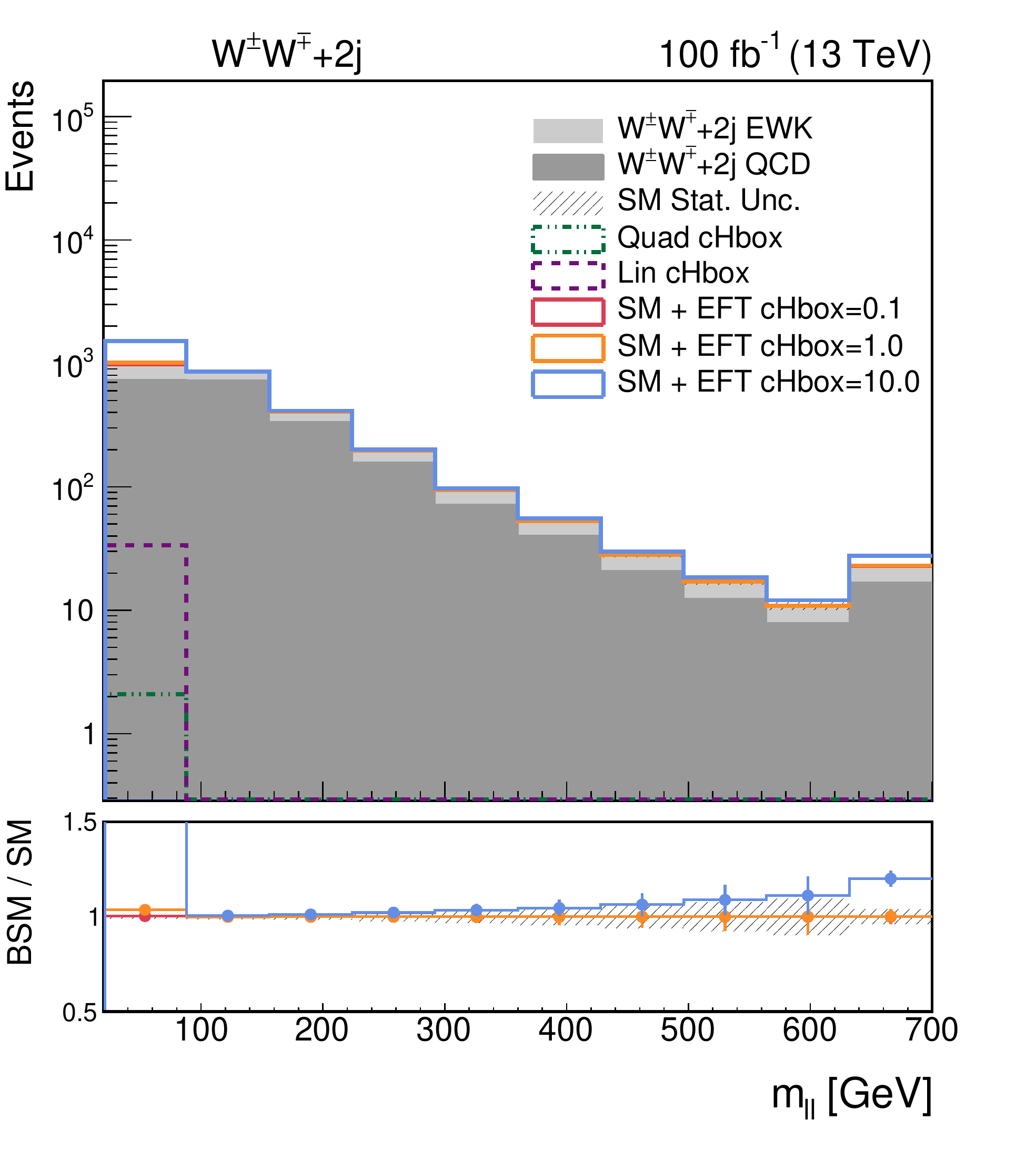}
  \caption{ Comparison of SM (filled histograms) and BSM (lines)
      expected number of events 
      after the event selection in Table~\ref{tab:variables},
      for various observables and Wilson coefficients in the OSWW+2j-EW and OSWW+2j-QCD processes, at an integrated luminosity of $\unit[100]{fb^{-1}}$.
      The SM distribution is represented as a stacked histogram, summing EW (light grey) and QCD (dark grey) components.
      Solid lines show the total prediction for one Wilson coefficient at a time, with $c_\a/\Lambda^2 = 0.01$ (red), $0.1$ (orange) or $\unit[1]{TeV^{-2}}$ (blue).
      The pure interference (quadratic) EFT component, normalized to $c_\a/\Lambda^2=\unit[1]{TeV^{-2}}$, is indicated with a purple (green) dashed line.
      For all distributions, the last bin comprises all the overflow events.
      \label{fig:Distributions_OSWW} }  
\end{figure}

\begin{figure}[hptb]
  \centering
  \vspace*{-5mm}
  \includegraphics[width=.49\textwidth]{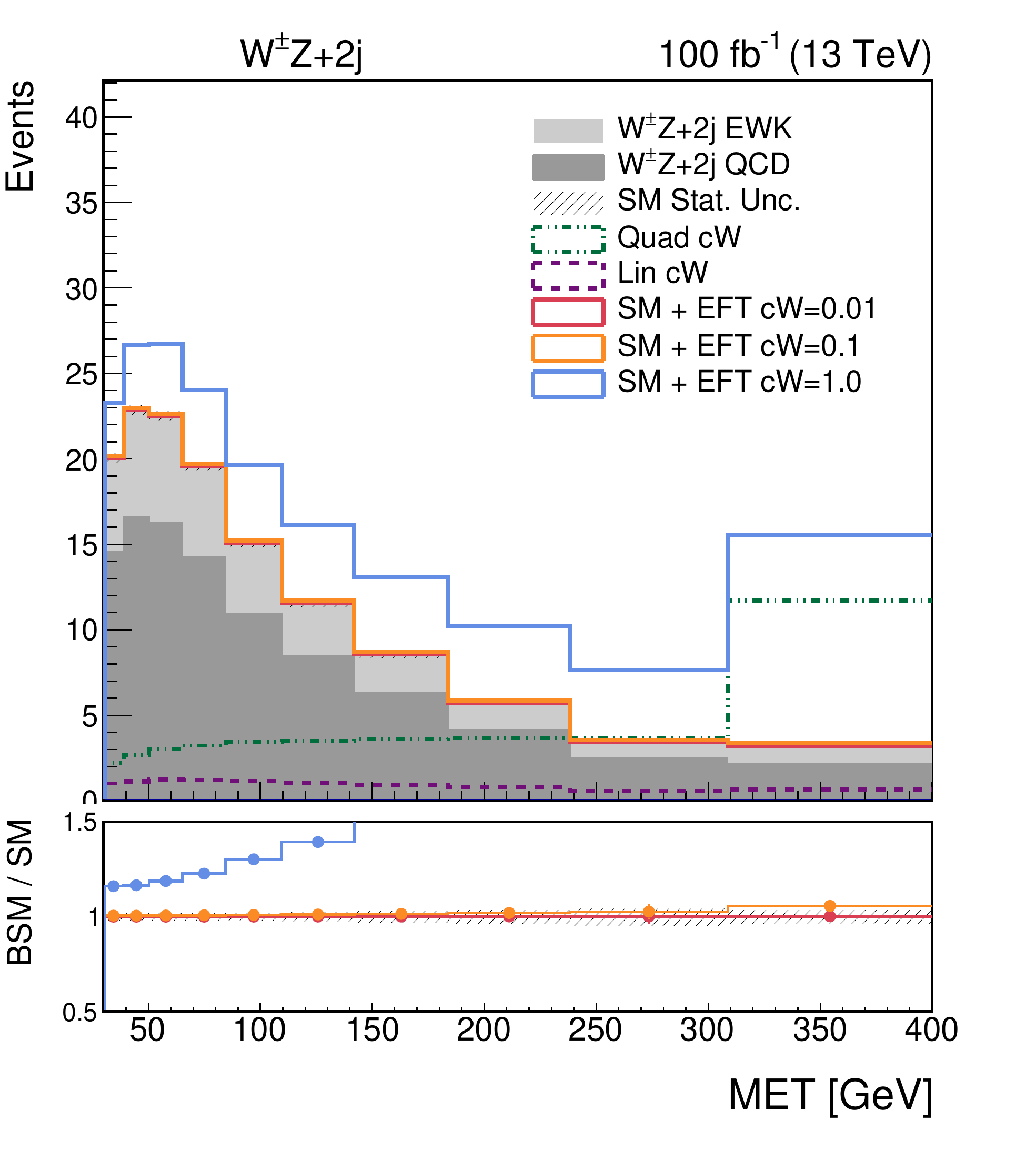}
  \includegraphics[width=.49\textwidth]{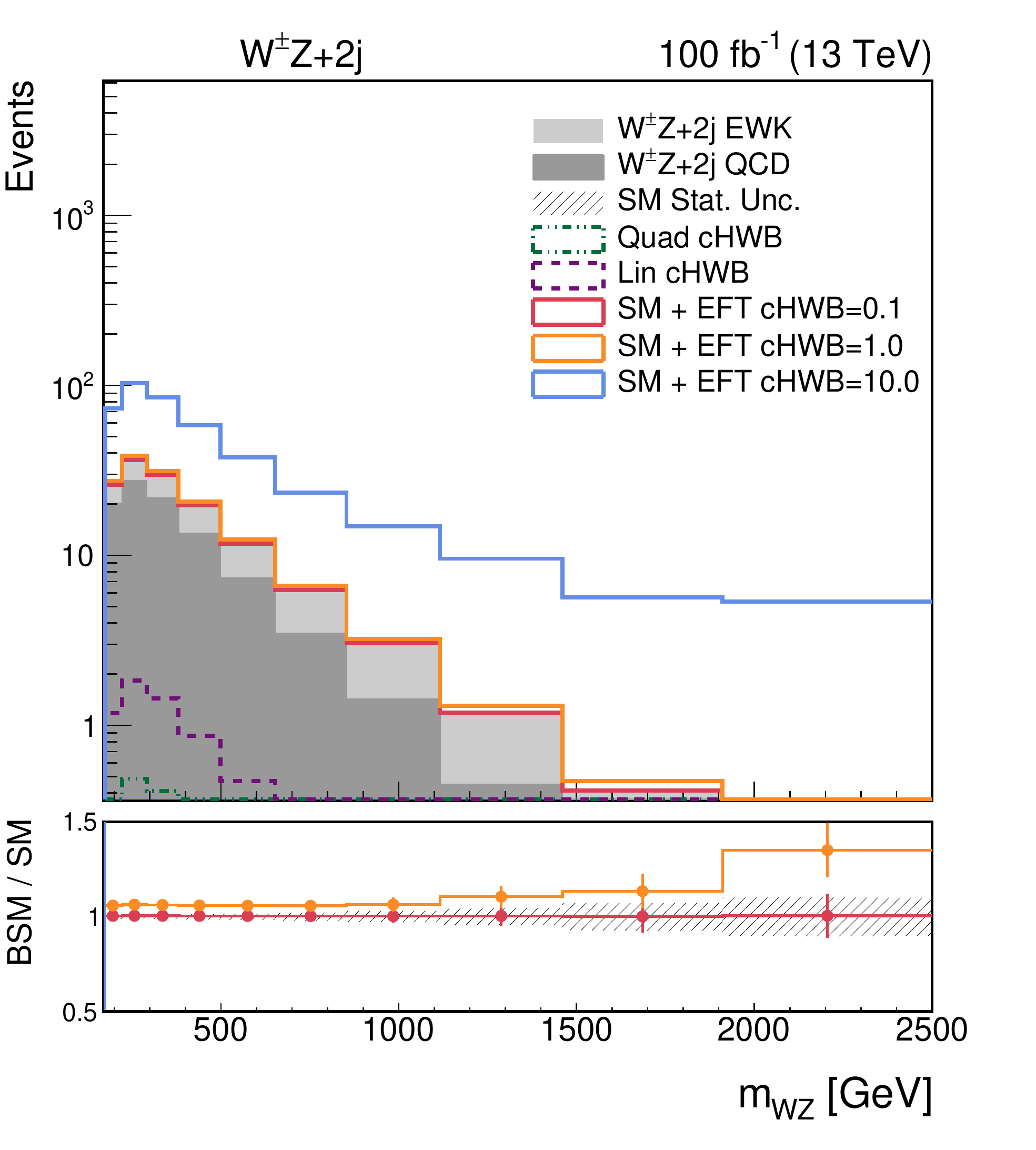}\\
  \includegraphics[width=.49\textwidth]{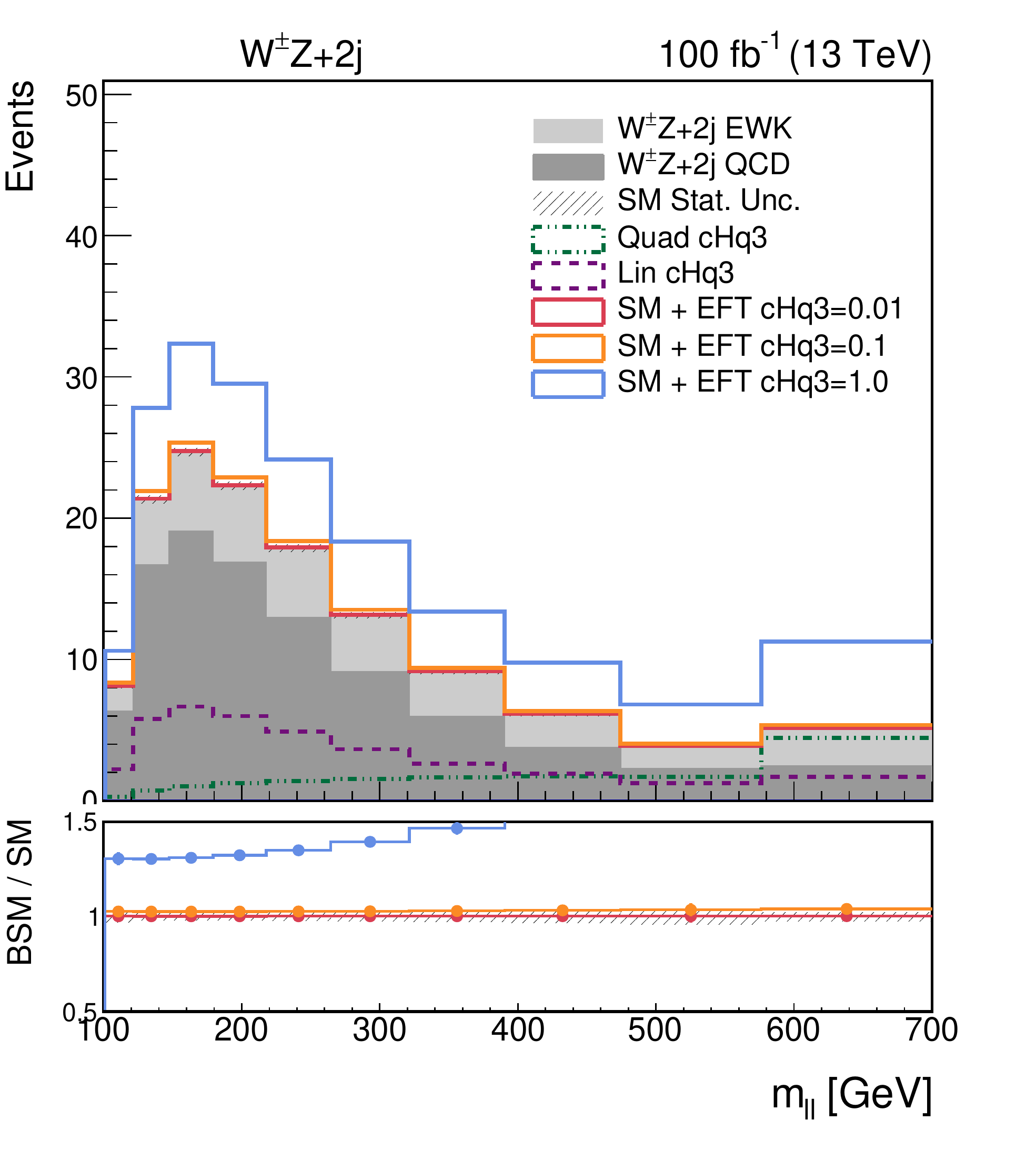}
  \includegraphics[width=.49\textwidth]{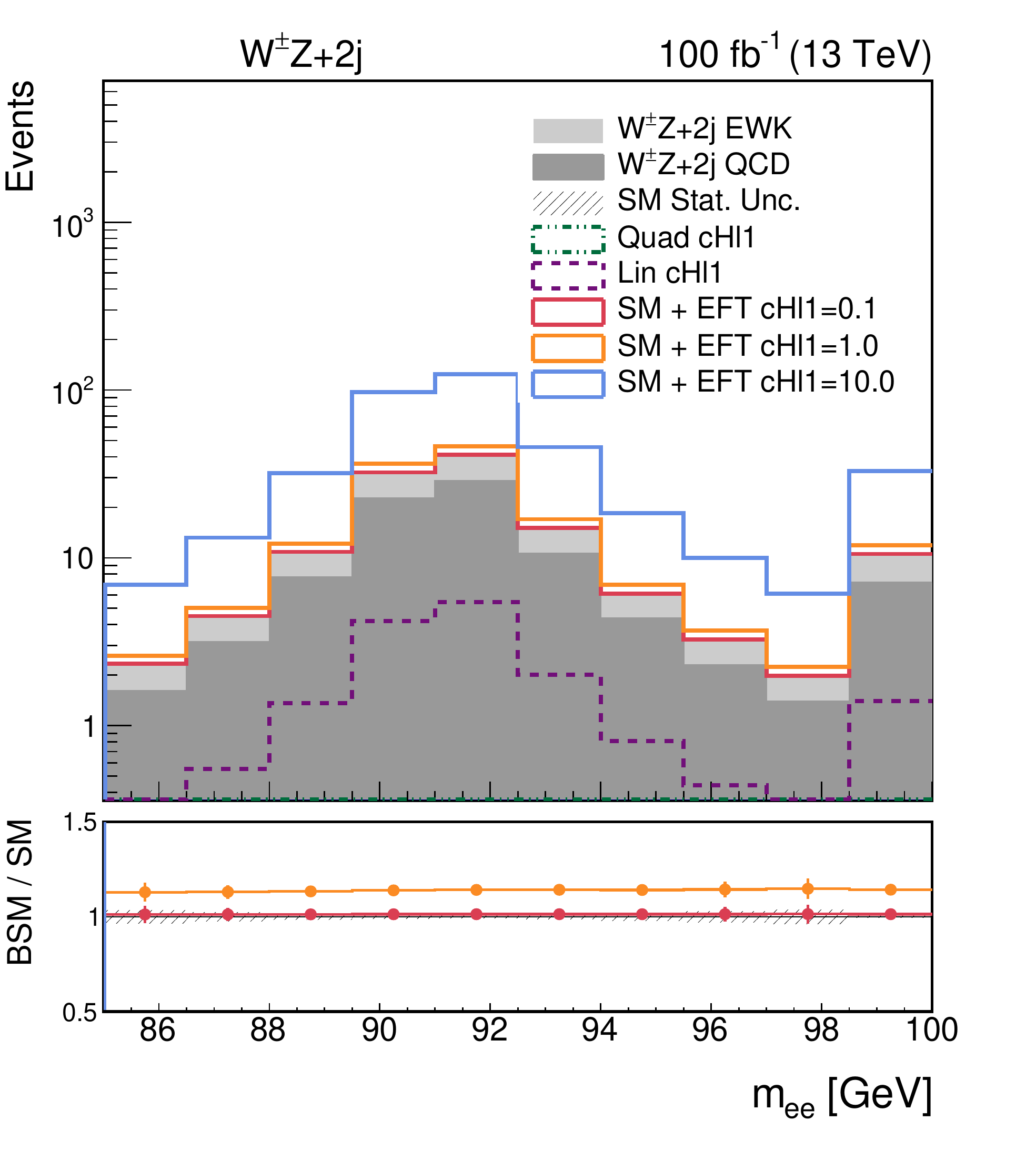}
  \caption{Comparison of SM (filled histograms) and BSM (lines)
      expected number of events 
      after the event selection in Table~\ref{tab:variables},
      for various observables and Wilson coefficients in the WZ+2j-EW and WZ+2j-QCD processes, at an integrated luminosity of $\unit[100]{fb^{-1}}$.
      The SM distribution is represented as a stacked histogram, summing EW (light grey) and QCD (dark grey) components.
      Solid lines show the total prediction for one Wilson coefficient at a time, with $c_\a/\Lambda^2 = 0.01$ (red), $0.1$ (orange) or $\unit[1]{TeV^{-2}}$ (blue).
      The pure interference (quadratic) EFT component, normalized to $c_\a/\Lambda^2=\unit[1]{TeV^{-2}}$, is indicated with a purple (green) dashed line.
      For all distributions, the last bin comprises all the overflow events. 
      \label{fig:Distributions_WZ}  }  
\end{figure}

\begin{figure}[htbp]
  \centering  
  \vspace*{-5mm}
  \includegraphics[width=.49\textwidth]{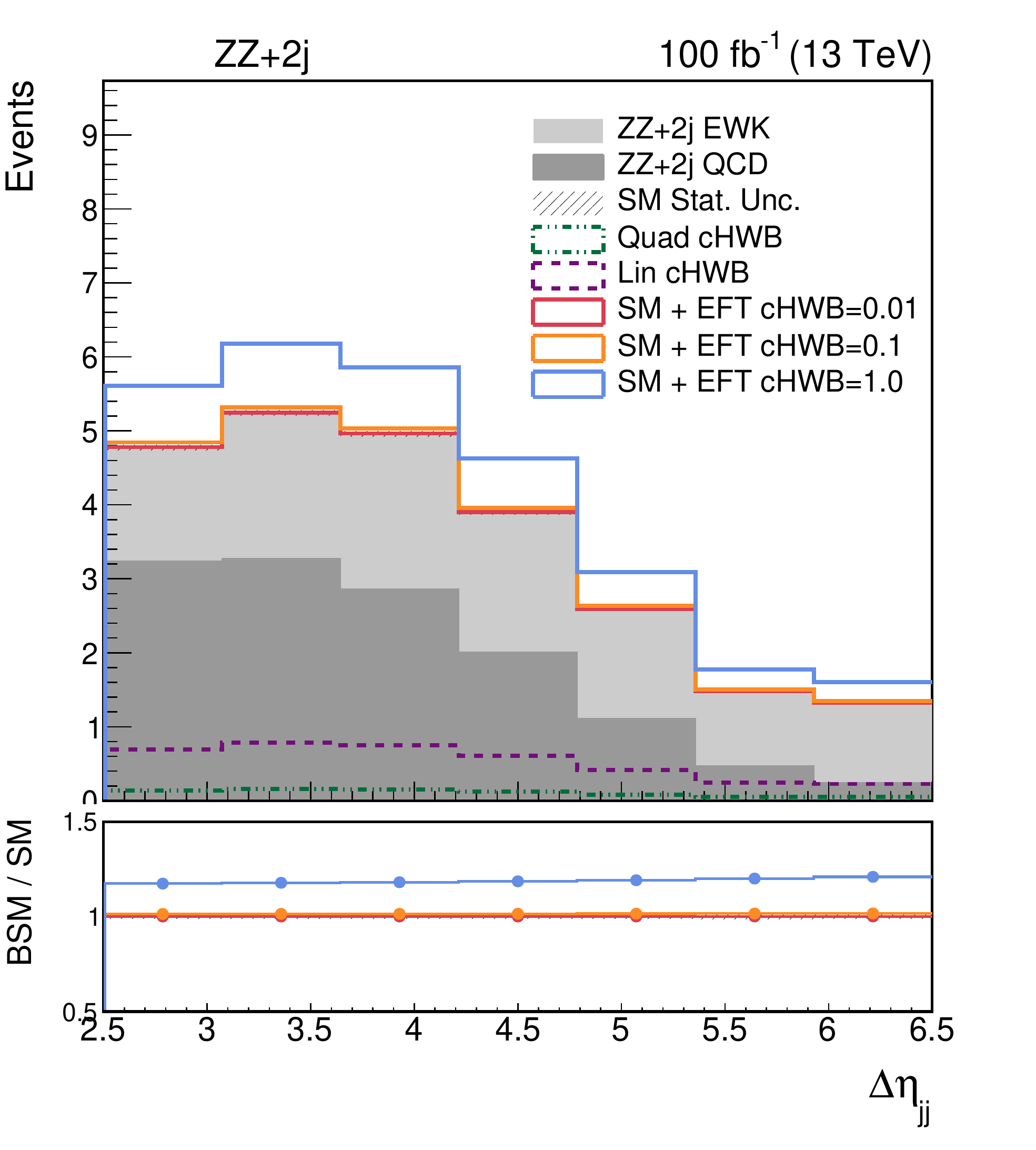}
  \includegraphics[width=.49\textwidth]{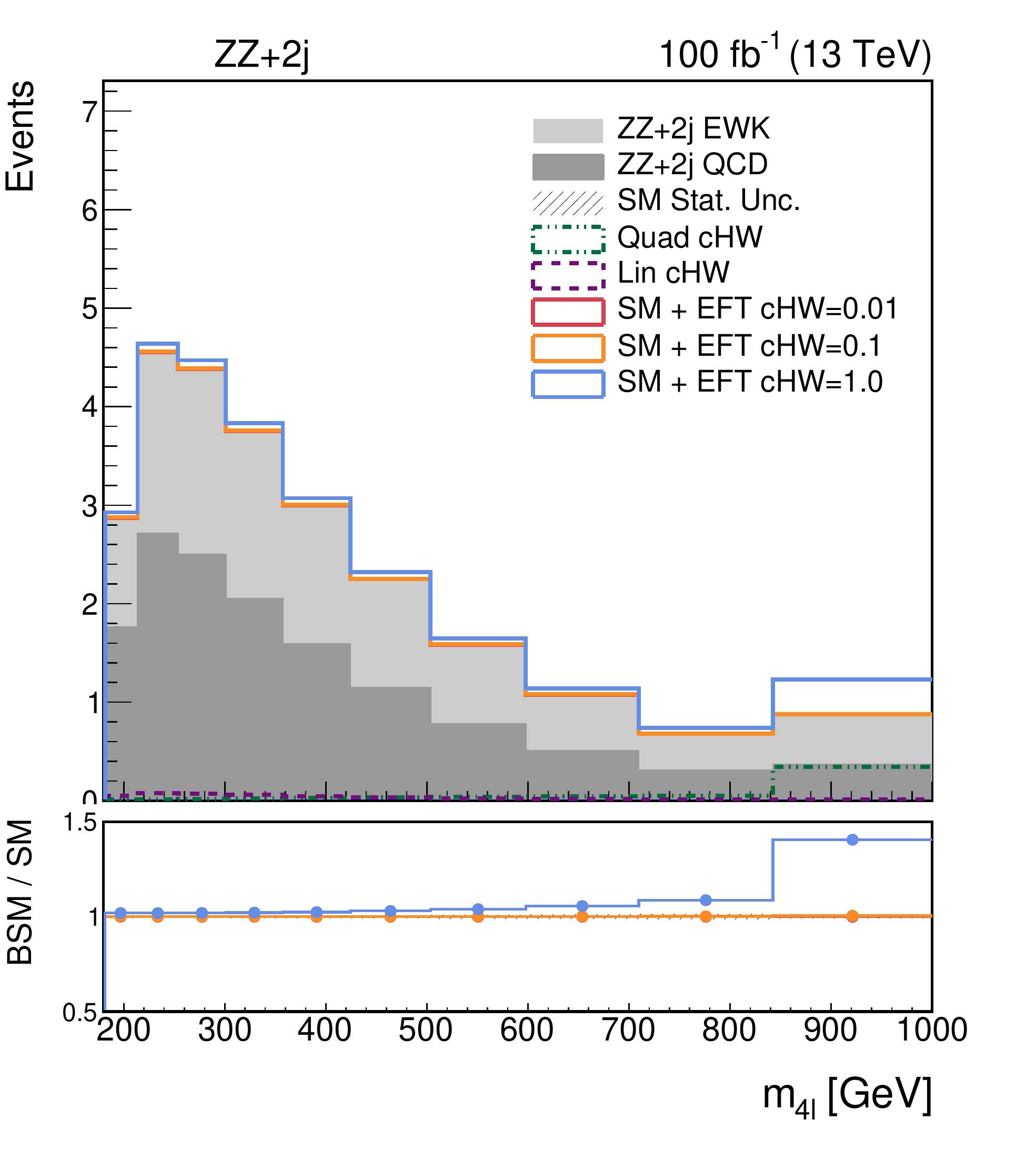}    \\
  \includegraphics[width=.49\textwidth]{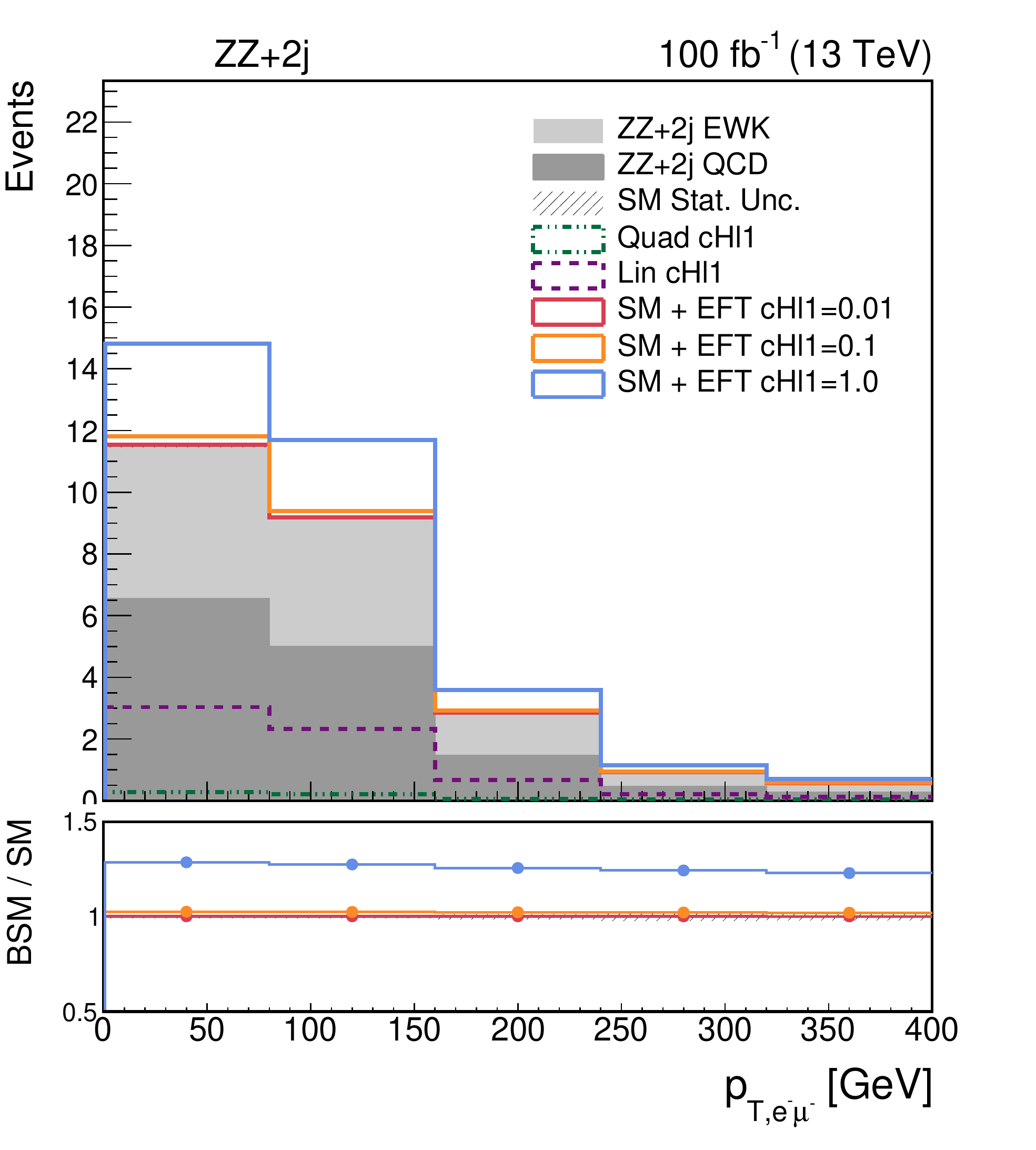}
  \includegraphics[width=.49\textwidth]{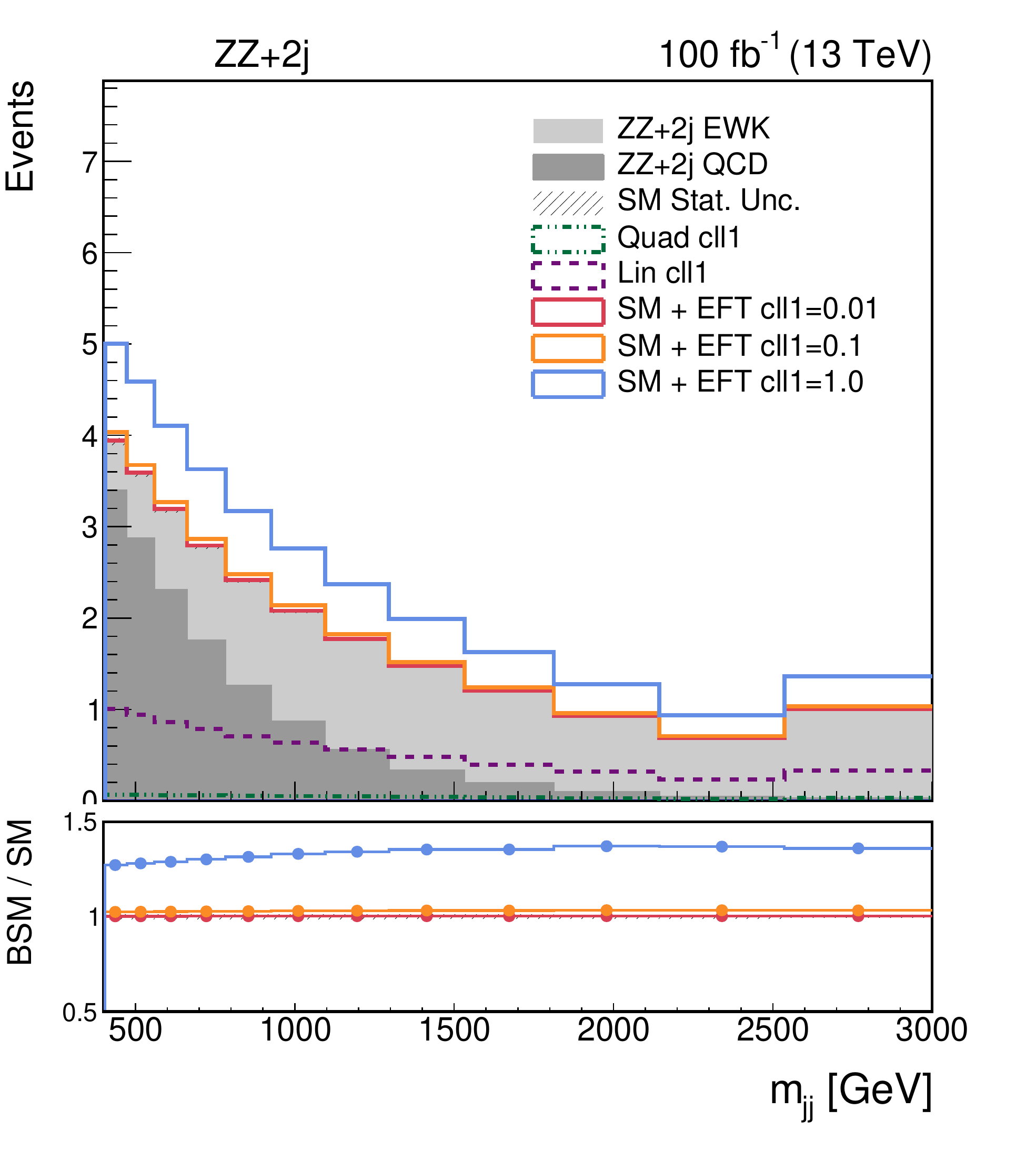}
  \caption{
      Comparison of SM (filled histograms) and BSM (lines)
      expected number of events 
      after the event selection in Table~\ref{tab:variables},
      for various observables and Wilson coefficients in the ZZ+2j-EW and ZZ+2j-QCD processes, at an integrated luminosity of $\unit[100]{fb^{-1}}$.
      The SM distribution is represented as a stacked histogram, summing EW (light grey) and QCD (dark grey) components.
      Solid lines show the total prediction for one Wilson coefficient at a time, with $c_\a/\Lambda^2 = 0.01$ (red), $0.1$ (orange) or $\unit[1]{TeV^{-2}}$ (blue).
      The pure interference (quadratic) EFT component, normalized to $c_\a/\Lambda^2=\unit[1]{TeV^{-2}}$, is indicated with a purple (green) dashed line.
      For all distributions, the last bin comprises all the overflow events. 
      \label{fig:Distributions_ZZ}  }  
\end{figure}

\begin{figure}[htbp]
  \centering \vspace*{-5mm}
  \includegraphics[width=.49\textwidth]{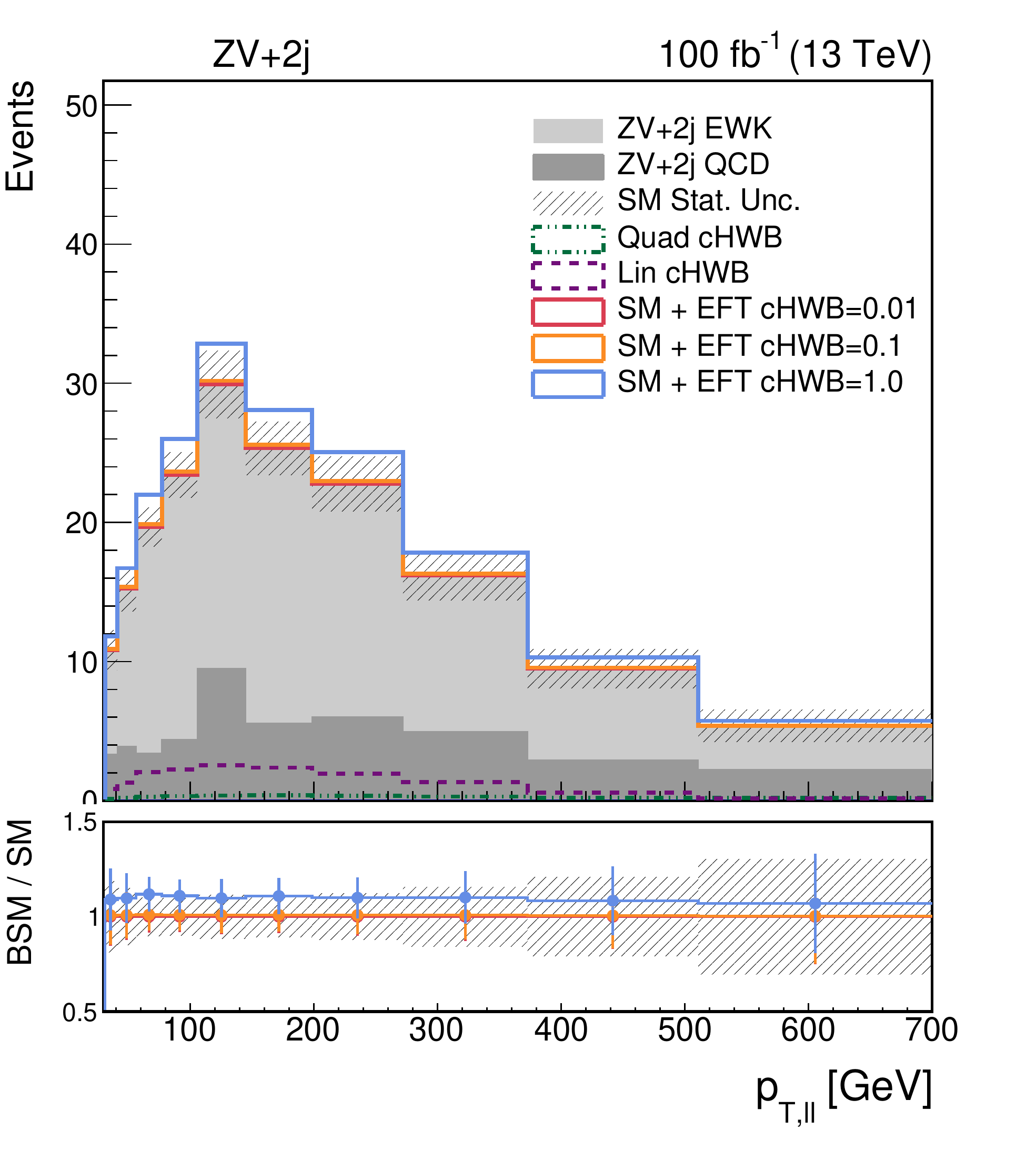}
  \includegraphics[width=.49\textwidth]{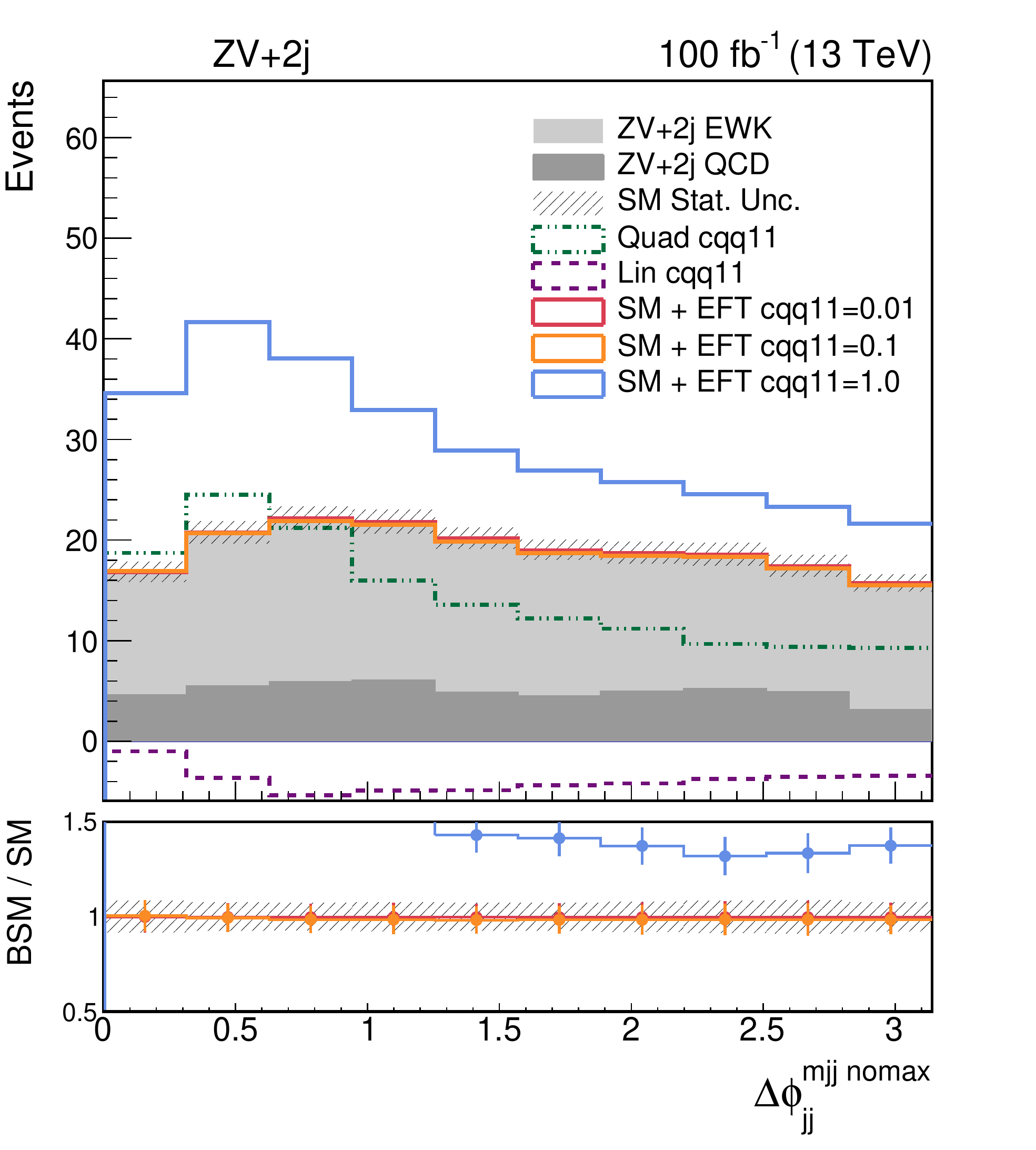}   \\
  \includegraphics[width=.49\textwidth]{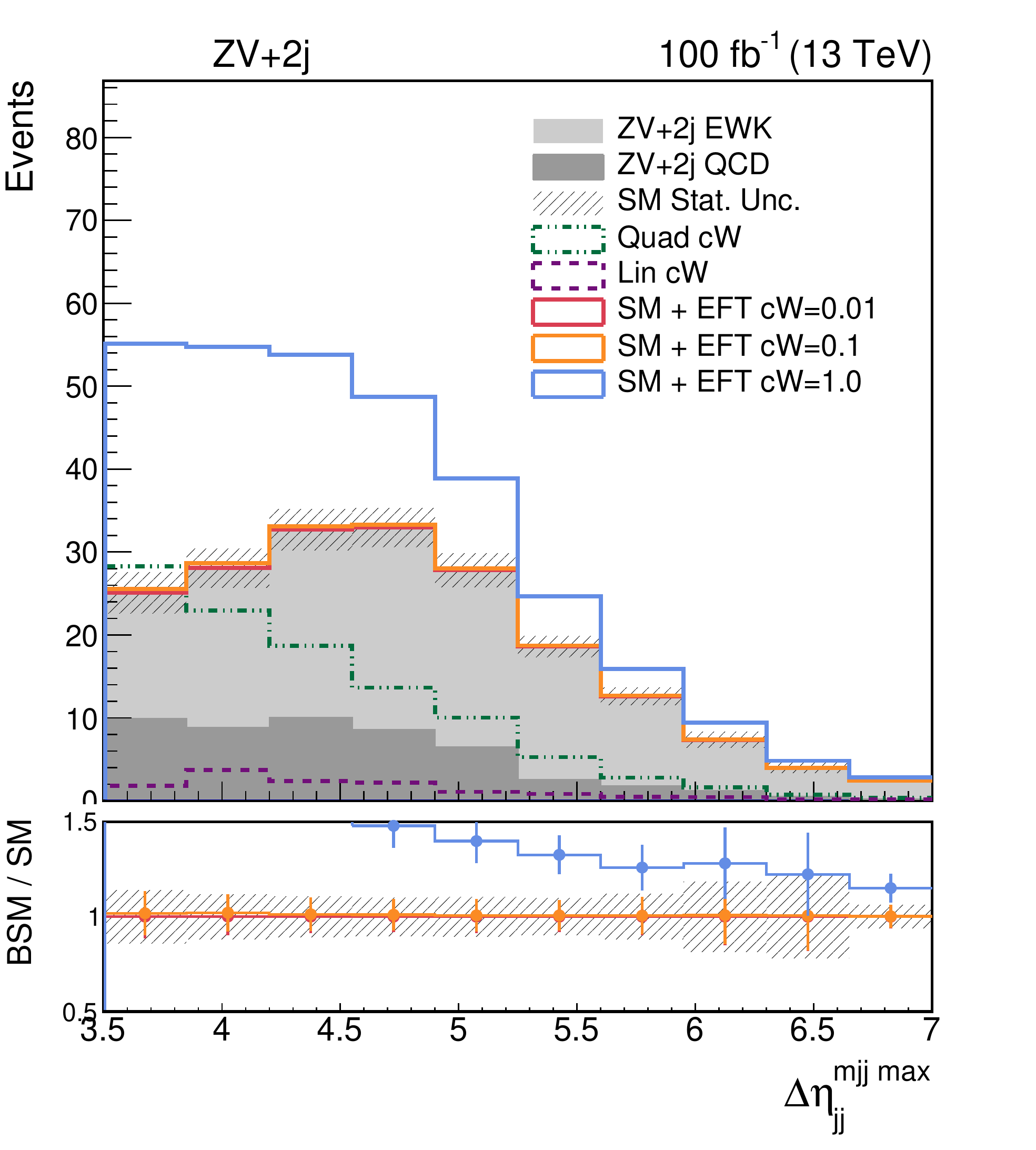}
  \includegraphics[width=.49\textwidth]{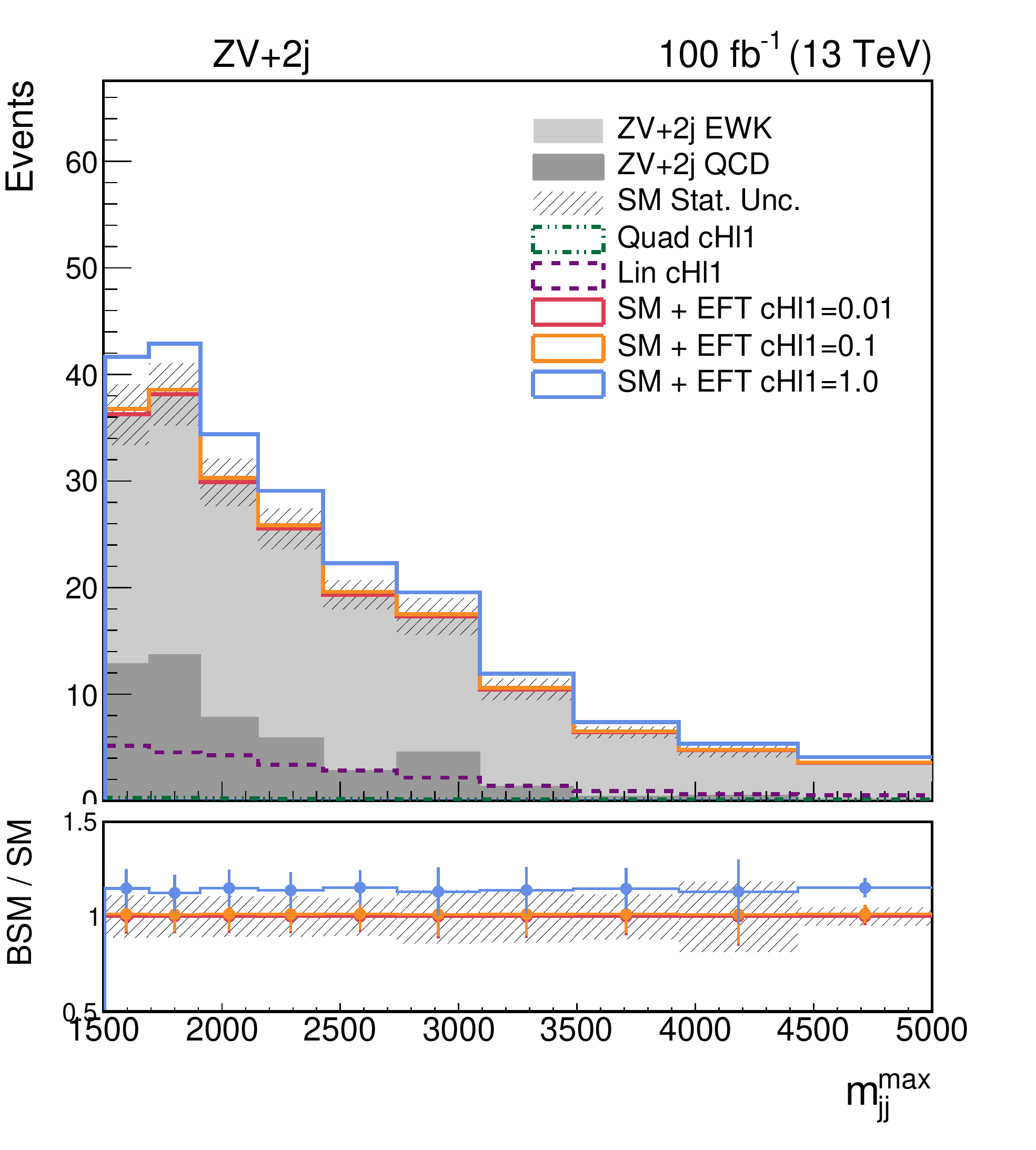}
  \caption{Comparison of SM (filled histograms) and BSM (lines)
      expected number of events 
      after the event selection in Table~\ref{tab:variables},
      for various observables and Wilson coefficients in the ZV+2j-EW and ZV+2j-QCD processes, at an integrated luminosity of $\unit[100]{fb^{-1}}$.
      The SM distribution is represented as a stacked histogram, summing EW (light grey) and QCD (dark grey) components.
      Solid lines show the total prediction for one Wilson coefficient at a time, with $c_\a/\Lambda^2 = 0.01$ (red), $0.1$ (orange) or $\unit[1]{TeV^{-2}}$ (blue).
      The pure interference (quadratic) EFT component, normalized to $c_\a/\Lambda^2=\unit[1]{TeV^{-2}}$, is indicated with a purple (green) dashed line.
      For all distributions, the last bin comprises all the overflow events. 
      \label{fig:Distributions_VZ}  }  
\end{figure}

\clearpage

\clearpage

\bibliographystyle{JHEP}
\bibliography{main}

\end{document}